\newif\iftablet

\iftablet
   \documentclass[14pt,a4paper,twoside]{extarticle}  
\else
   \documentclass[a4paper,twoside]{article} 
\fi

\newcommand{\affil}[1]{$^{\rm #1}$}
\iftablet
\fi

\usepackage[authoryear]{natbib}
\bibpunct{(}{)}{;}{a}{}{,}
\usepackage{color}  

\iftablet
   \relax  
\else
   \usepackage{sectsty} \sectionfont{\large} \subsectionfont{\normalsize}
\fi

\usepackage{graphicx,amssymb}
\usepackage[colorlinks,urlcolor=blue,citecolor=blue,linkcolor=blue]{hyperref} 
\date{\small\raggedright
{\it Updated 9/Jun/2014}. Published  as {\it PASA}, {\bf 30}, e056 (2013). \href{http://adsabs.harvard.edu/cgi-bin/nph-abs_connect?fforward=http://dx.doi.org/10.1017/pasa.2013.34}{doi:10.1017/pasa.2013.34}. A few minor corrections have been made to the journal version.\\
} 
\newcommand{\kms}{\mbox{km\,s$^{-1}$}}
\title{\hrule\bigskip\huge\bf\flushleft The Dawes Review 1: Kinematic studies of star-forming galaxies across cosmic time\bigskip\hrule}
\author{\parbox{\textwidth}{\flushleft
\vspace{-0.5cm}
{\it Karl Glazebrook\,\affil{A,B}}\\
\vspace{0.4cm}
{\small \affil{A}\,Centre for Astrophysics and Supercomputing, Swinburne University of Technology, P.O. Box 218, Hawthorn, VIC 3122, Australia}\\
{\small \affil{B}\,kglazebrook@swin.edu.au}\\
}}
\begin{document}

\twocolumn[
\begin{changemargin}{.8cm}{.5cm}
\begin{minipage}{.9\textwidth}
\vspace{-1cm}
\maketitle
%
%
\small{{\bf Abstract:} The last seven years have seen an explosion in the number of Integral Field  galaxy surveys, obtaining resolved 2D
spectroscopy, especially at high-redshift. These have
taken advantage of the mature capabilities of 8--10m class telescopes and the development of associated technology such as AO. Surveys have leveraged both high spectroscopic resolution enabling internal velocity measurements and high spatial
resolution from AO techniques and sites with excellent natural seeing.  For the first
time we have been able to glimpse the kinematic state of matter in young, assembling star-forming galaxies and learn detailed
astrophysical information about the physical processes and compare their kinematic scaling relations with those in the local Universe. Observers have measured disc galaxy rotation, merger signatures and turbulence-enhanced velocity dispersions of gas-rich discs. Theorists have interpreted kinematic signatures of galaxies in a variety of ways (rotation, merging, outflows, and feedback) and attempted to discuss evolution vs theoretical models and relate it to the evolution in galaxy morphology. A key point that has emerged from this activity is that substantial fractions of high-redshift galaxies have {\it regular kinematic
morphologies} despite irregular photometric morphologies and this is likely due to the presence of a large number of highly gas-rich discs. 
There has not yet
been a review of this burgeoning topic. 
In this first Dawes review I will discuss the extensive kinematic surveys that have been done and
the physical models that have arisen for young galaxies at high-redshift.}


\medskip
\medskip
\end{minipage}
\end{changemargin}
]
\small

\bibliographystyle{yahapj-custom} 

\def\mycite#1{{\color{red} \cite{#1}}}
\def\REF{{\color{red}REF}}
\def\REFS{{\color{red}REFs}}
\def\REFs{{\color{red}REFs}}
\def\XXX{{\color{red}XXX}}
\def\YYY{{\color{red}YYY}}
\def\ZZZ{{\color{red}ZZZ}}
\def\kms{km\,s$^{-1}$}
\def\Ha{H$\alpha$}
\def\Hb{H$\beta$}
\def\TFR{Tully-Fisher relationship}
\def\KSR{Kennicutt-Schmidt relationship}
\def\Msun{{\rm M}_{\odot}}
\def\TODO#1{{\bf TO DO: #1}}
\def\mytopic#1{\smallskip\noindent\textbf{\emph{#1}}}

\iftablet 
   \onecolumn 
\fi

\begin{figure}[t] 
\centering
\iftablet
   \includegraphics[width=12cm]{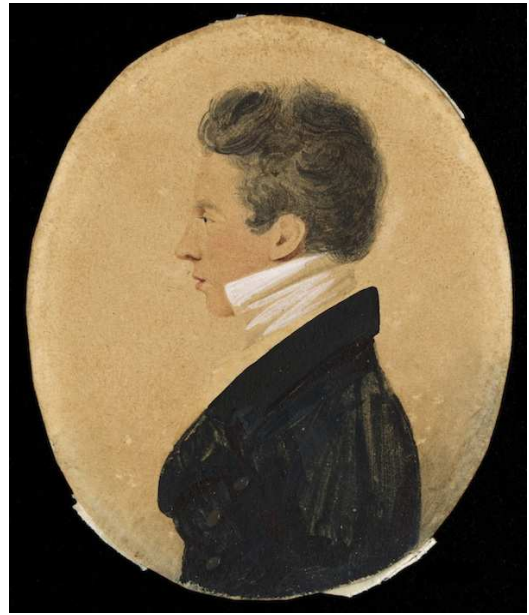}
\else
   \includegraphics[width=7.5cm]{figure-dawes-new.pdf}
\fi
\caption{\small William Dawes was a Royal Marine officer on the `First Fleet' arriving in Australia in 1788. He was a man of many talents: engineer, map maker, botanist, and amateur astronomer. He was one of the first to document the Aboriginal Australian languages spoken in the Sydney region. He was the first person to make astronomical
observations in Australia using telescopes from a place in Sydney Cove now known as Dawes Point \citep{Dawes}. {\it Image Credit: miniature oil painting of Lieutenant William Dawes, 1830s, artist unknown. Collection: \href{http://www.tmag.tas.gov.au}{Tasmanian Museum and Art Gallery}. Reproduced with their permission.}} 
\label{fig:dawes}
\end{figure}

{\it\noindent \textsf{The Dawes Reviews are substantial reviews of topical areas in astronomy, published by authors of international standing at the invitation of the PASA Editorial Board. 
The reviews recognise William Dawes (1762--1836) (pictured in Figure~\ref{fig:dawes}), second lieutenant in the Royal Marines and the astronomer on the First Fleet. Dawes was not only an accomplished astronomer, but spoke five languages, had a keen interest in botany, mineralogy, engineering, cartography and music, compiled the first  Abor\-iginal-English dictionary, and was an outspoken opponent of slavery.}}
\\

\begin{quote}
\it
`Eppur si muove' \\ 
\\
 --  Galileo Galilei (apocryphal)
 \\
\end{quote}

\section{Introduction}
\label{sec:intro}


The advent of new large telescopes coupled with new  instrumentation technologies in the last decade has been extremely powerful in expanding our
view of the high-redshift Universe. In particular, we have seen a flowering of the topic of high-redshift galaxy {\it kinematics} which studies their 
internal motions through high spatial and spectral resolution observations.  The number of papers has exploded and we have seen a variety of surveys of observational approaches, analysis techniques, and theoretical interpretations. This has led to new paradigms of the nature of young galaxies but it has also raised problems in understanding 
as many new techniques have been used making comparison with the local Universe and traditional techniques difficult. 

The \emph{Publications of the Astronomical Society of Australia }has decided to launch this new series of major reviews in honour of Lt. William Dawes. I have chosen to write it on the topic of these exciting new studies of the kinematics of high-redshift star-forming galaxies,
one which has not had a major review and is in need of one. This is the first such Dawes review  and as such there is no tradition to follow, instead one gets to set the tradition. I will choose to write this as a high-level introduction  to the field, perhaps akin to the style of lecture notes, for the new worker in the field (for example an incoming postgraduate student). As such  I will try and favour clarity and simplicity of explanations over totally complete lists of all possible references and ideas on a topic and will discuss analysis techniques in some detail.  I will highlight the main surveys and the main ideas and warn in advance that some things may get left out. I will also allow myself the freedom to give more scientific speculation of my own than would occur in a traditional review, however it will be clearly indicated what is a speculation. Obviously I will use the first person when needed as this seems appropriate for my approach.

\subsection{Background and scope of this review}
The rotation  of the `spiral nebulae' was one of the earliest and most fundamental observations of their nature and the second important discovery from their spectroscopy. Almost exactly 100 years ago in 1912 September, Vesto M Slipher measured the first spectrum and first redshift of a galaxy using a new fast spectrograph he had built \citep{Slipher1913}. This galaxy was M31 and the redshift was actually a blueshift of 300 km$/$s --- this was highly unexpected at the time, it was ten times higher than any previous velocity measured for an astronomical object. Slipher himself thought it good evidence for  the extragalactic model of spiral nebulae \citep{daywefound} and proceeded to embark on a campaign to measure many more velocities \citep{Slipher1917} eventually resulting in one axis of Hubble's famous diagram \citep{Hubble1929}. 

Less well-known is that during this first campaign Slipher also discovered the rotation of galaxies \citep{Slipher1914} --- he noticed the tilt of the spectral lines whilst observing the Sa galaxy M104 and noted the similarity to the same phenomenon when observing planets. Slipher had worked for Lowell for many years measuring the day lengths of various planets. Slipher commented: \emph{`Although from the time of Laplace it has been thought that nebulae rotate, this actual observation of the rotation is almost as unexpected as was the discovery that they possessed enormously high radial velocities'.} 

We now regard galaxies as gravitationally bound extragalactic objects and their internal motions relate to  fundamental questions about their masses and assembly history. In particular the last seven years have seen a wealth of new high-redshift observations measuring for the first time the \emph{kinematics} of galaxies in the early Universe and producing new pictures of star-forming galaxies. These are the topic for this review. I note that I will  favour the term `kinematics' which describes, from observations, the motions of astronomical objects (as opposed to the term `dynamics' which describes the theoretical causes of such motions). 

Large 8-10m class optical telescopes\footnote{The overwhelming majority of kinematic observations at $z>0.5$ have been optical$/$near-infrared utilising nebula emission lines, however  radio/sub-mm  observations will be mentioned and this balance is likely to change dramatically in the next decade with the advent of the Atacama Large Millimetre Array (ALMA).} with their light grasp and angular resolution have been critical for the development of this subject but equally important has been the associated development of astronomical instrumentation sitting at the focal plane. 

\emph{Integral Field Spectroscopy} (IFS) has played a pivotal role due to the complex structures of high-redshift objects. With this technique, it is possible to collect a spectrum of every point in the 2D image of an object, which is contrasted with the classical technique of \emph{long-slit spectroscopy} where spectra are collected along a 1D slice (whose direction must be chosen in advance) through an object. An IFS generally works by reformatting a 2D focal plane, and there are various ways of accomplishing this (for a review of the technology, see \cite{IFSrev})  but a general principle is that because instruments are limited by the number of pixels in their focal plane detectors, an IFS typically has a small field of view with spatial sampling of order 1000 elements\footnote{IFS spatial sampling elements (e.g. lenslets or fibres) are often called `spaxels'. This I mention solely to record for posterity this great
quote: {\it `If spatial bins are spaxels, are spectral bins spexels and time bins tixels? But wait a tixel, those spaxels and spexels are all pixels or voxels! I say, purge the English language of these mongrel wordels!'} (Matthew Colless, 2010, personal communication)} suitable for single object work.
(This is an area that is likely to improve in the future with new instruments and ever large pixel-count detectors).

\emph{Adaptive Optics} (AO) technology which corrects for atmospherical turbulent blurring of images has also become routine on large telescopes 
\citep{AOreview} over the last decade and has allowed the achievement of the angular diffraction limit on 8-10m telescopes --- typically 0.1 arcsec instead of the 0.5--1 arcsec seeing limit imposed by the atmosphere. This is important as 1 arcsec corresponds to 8 kpc for $1<z<3$ which is comparable to the sizes of disc galaxies at these redshifts (e.g. \cite{Ferg04,Buit08,Mos12}). AO observing comes with its own sets of limitations imposed by the requirements to have bright stars or laser beacons to measure AO corrections from and have generally not been possible for all objects in large samples.

It is important when writing a review to carefully define the scope. The topic will be the kinematics of star-forming galaxies at high-redshift (which I will define as $z>0.5$), with a focus on what we have learned  and how we have learned it, from IFS and AO observations. It is not possible to cover, with any comprehensiveness related topics such as (i) general physical properties of high-redshift star-forming galaxies, (ii) the kinematics of star-forming galaxies in the local Universe, and (iii) the kinematics of non-star forming `red and quiescent' galaxies at high-redshift. The first two are already the subject of extensive reviews to which I will refer, and the last is a rapidly burgeoning field which will probably be due for its own review in 2--3 years as the
number of observations increases tremendously with the advent of multi-object near-IR spectrographs.\footnote{I note that {\it spatially resolved kinematic observations} of red galaxies
at high-redshift  will prove  very difficult as it would require the detection and measurement 
of stellar absorption lines at even higher angular resolution in
smaller objects than has been done for the star-forming population.} However, some non-comprehensive discussion of each of these (especially the first two) will be given to set the scene.

The plan and structure of this review is as follows. Firstly, in the remainder of this introduction I will briefly discuss the kinematic properties of galaxies in the modern Universe to frame the comparisons with high-redshift. In Section 2, I will review the earliest kinematic observations of star-forming galaxies at high-redshift from longslit techniques. In Section 3, I will review the most important large high-redshift IFS surveys, how they are selected and carried out and their most important conclusions. In Section 4, I will review the kinematic analysis techniques used by IFS surveys with reference to the surveys in Section 3. In Section 5, I will compare and contrast what we are learning about the physical pictures of high-redshift star-forming galaxies from the various IFS surveys and discuss, in particular, the `turbulent clumpy disc' paradigm that has arisen from these works. In Section 6, I will point to the future, the outstanding questions and the future instruments, telescopes, surveys and techniques that may address them.

This review will adopt a working cosmology of $\Omega_m=0.3$, $\Omega_\Lambda=0.7$,  $H_0= 70$ km\, s$^{-1}$\, Mpc$^{-1}$ \citep{Spergel03}. Since most of the work discussed has been in the last decade, the authors have adopted cosmologies very close to these resulting in negligible conversion factors in physical quantities.   I will adopt the use of AB magnitudes.

\subsection{Kinematics of star-forming galaxies in the local Universe.}

In the local Universe, we see a distinct separation of galaxies in to two types with red and blue colours \citep{Strat01,Baldry04} commonly referred to as the `red sequence' and `blue cloud' reflecting the relative tightness of those colour distributions. The separation is distinct in that there is a clear bimodality with a lack of galaxies at intermediate colours. These colour classes are very strongly correlated with morphology either as determined visually or via quantitative morphological parameters --- a detailed recent review of these properties as derived from large statistical surveys such as the Sloan Digital Sky Survey and exploration of their dependence on other parameters such as environment is 
given by \cite{BM09}. The correlation is sufficiently strong that virtually every massive system on the blue cloud is a rotating star-forming disc galaxy (usually spiral), though there is a rare population of `red spirals` which overlap the red sequence (which is mostly ellipticals) that may arise from truncated star-formation, greater older stellar population
contributions or dust \citep{Masters10,Cortese12}. 

There has been a number of reviews on the topic of the kinematics of local disc galaxies over the years, which should be referred to for a comprehensive discussion. In this section, I will discuss the most important points mostly referencing  recent results whilst noting that the subject has a long history which has been well covered elsewhere. I refer the reader for more depth and history to  \cite{kruit1978}, who review the kinematics of spiral and irregular galaxies and \cite{SofueRubin2001}, which is a more focussed review on the topic of rotation curves. A classic review of the structure of the Milky Way in particular was done by \cite{GWK89}. Recently \cite{vanderKruitFreeman2011}  wrote a very comprehensive recent review of all properties of galaxy discs including kinematics.

For comparison with high-redshift, the most fundamental properties of local star-forming galaxies are their rotation and velocity dispersion, whose most important points I will review below. However, as we will see later in this review, star-forming galaxies at high-redshift show more kinematic diversity than in the local Universe including high fractions which are not dominated by rotation  or which show complex kinematic signatures of mergers. Given evolutionary paths from high-redshift to low-redshift and from star-forming to quiescent are not obvious I will also discuss briefly the kinematics of local elliptical galaxies and mergers.

\subsubsection{Rotation of local star forming galaxies}

The earliest published work on disc galaxy rotation was that of \cite{Slipher1914} but also see \cite{Pease1916}. They measured the rotation of several spirals between 1914 and 1925 including M31 and M104. The review of \cite{SofueRubin2001} gives a historical introduction, so also does the one of \cite{kruit1978}. The early optical work was limited to the central regions of galaxies, the advent of radio telescopes and neutral hydrogen HI observations \citep{VDH57,Argyle65} permitted measurements out at large radii where most of the angular momentum lies. Radio observations led to the well-known and most fundamental scaling of disc galaxies: the `Tully-Fisher Relation'  first reported by \cite{TF77}  between optical luminosity and HI line width. If the HI line width, from an unresolved or marginally resolved single-dish observation, is thought of as tracing the total kinematic shear,  then this becomes a relation between luminosity and rotation velocity, and hence luminosity and a measure of mass. Later, Tully-Fisher work has benefited from greatly increased spatial resolution and 2D kinematic mapping of the rotation field.

In the standard pictures, we now think of galaxies as inhabiting haloes of Cold Dark Matter (CDM), a non-baryonic component that dominates the dynamics and sets the scene for galaxy formation \citep{Blum85,Ostriker93}. The most fundamental of observations supporting this picture is the  `flat rotation curves' of disc galaxies \citep{RF70,Roberts73,Rubin78}. The general picture is of a steeply rising rotation curve in the innermost few kpc followed by the `flat' portion, which really means a turnover and then a slight slow decline in more luminous galaxies or a flatter more constant rotation in lower luminosity galaxies \citep{URC,SofueRubin2001}. This occurs in a regime where the optical surface brightness is exponentially dropping off and the rotation velocity, as traced by HI, stays high past the outer edge of the optical disc. If light traced mass the velocity would drop off more sharply, this is the basic evidence for dark matter haloes (though is not universally accepted, for an alternative paradigm involving 'Modified Newtonian Dynamics' see \cite{MOND}). If a dark matter halo was spherical and isothermal ($\rho¨ \propto r^{-2}$), one expects a perfectly flat rotation curve, in reality simulations predict more complex profiles for dark matter haloes \citep{NFW} and this, together with the stellar contributions, must be carefully considered when fitting rotation curve models \citep{Kent87,BAC01}). As such when defining the `rotation velocity', one must be careful to specify at what radius this is measured. A common convention is to use 2.2 disc scalelengths\footnote{It is useful to also note that 2.2 scalelengths is also 1.3$\times$ the half-light radius for a pure exponential disc.}  (from the surface photometry) as this is the radius where the rotation curve of a self-gravitating ideal exponential disc peaks \citep{F70}. This `$v_{2.2}$' can also be related to the HI line width \citep{Courteau97} which also probes the outer rotation. The typical values for large disc galaxies are in the range 150--300 km$/$s.

The original Tully-Fisher relation displayed a slope of $L \propto V^{2.5}$ (based upon the luminosity from blue-sensitive photographic plates), modern determinations find an increasing slope with wavelength rising to a slope of $V^4$ in the K-band or with stellar mass \citep{BdJ01,Ver01}. This is consistent with galaxies having a roughly constant ratio of dark matter to stellar mass globally\footnote{i.e. if discs form a one parameter sequence of 
constant central surface brightness, then $L\propto r^2$ and with  $G M \propto r V^2 $
one can easily show that if $M\propto L$ then $L\propto V^4$
} --- which is in contrast to the resolved distribution within galaxies where clearly it does not. CDM theory predicts a slope closer to $V^3$ based
on scaling of dark matter halo properties \citep{MMW98}. Some authors have argued that this represents an unreasonable `fine-tuning' of the $\Lambda$CDM model and have proposed an alternative gravity `MOND' mode without dark matter  (e.g. \cite{MOND,McGaugh98,McGaugh12}), however small
scatter can be accommodated within the $\Lambda$CDM framework \citep{Gnedin07,AR08,Dutton2012}.
 MOND does not seem to explain well larger scale structures such as galaxy groups and clusters in the sense that even with MOND there is still a need to invoke dark matter to explain the kinematics \citep{Angus08,Nat08}. This review will only consider the  $\Lambda$CDM cosmological framework.

\subsubsection{Velocity dispersion of local galaxy \\ discs}
\label{sec:local-spirals}

We next consider the vertical structure and pressure support of galactic discs, as this will become quite a significant topic when comparing with high-redshift, where we will see substantial differences. The most obvious visible component of spiral galaxy discs is the so-called `thin disc' which is where the young stellar populations dwell. The stellar component of the thin disc has an exponential scale height of 200--300 pc and a vertical velocity dispersion ($\sigma_z$) of $\sim$ 20 \kms\ \citep{vanderKruitFreeman2011} --- the dispersion is related to the vertical mass distribution by a gravitational equilibrium. This is $\sigma_z^2 = a G \Sigma h_z$ where $\Sigma$ is the mass surface density, $h_z$ is the vertical exponential scale height, and $a$ is a structural constant $= 3\pi/2$ for an exponential disc.
In general, the dispersion of a stellar disc is a 3D ellipsoid $(\sigma_R, \sigma_\theta, \sigma_z)$. The radial ($\sigma_R$) and azimuthal ($\sigma_\theta$) components are
related by the Oort constants (giving $\sigma_\theta \simeq 0.71\, \sigma_R$ for a flat rotation curve) and the radial and vertical components are related to the discs structure and mass to light ratio with a typical value of $\sigma_z / \sigma_R \sim 0.6$ for large spirals (again see van der Kruit \& Freeman and references therein for an extensive discussion of this).

The stellar age range of the Milky Way thin disc is  up to 10 Gyr. Right in the middle of the thin disc is an even thinner layer where the gas collects --- the neutral hydrogen, molecular clouds, dust, HII regions, and young OB and A stars all sit in this thinner layer which has a dispersion of only $\sim 5$--10 \kms\ and scale height of 50 pc in the Milky Way. This thinner disc is where all of the star formation takes place today and in which the characteristic spiral structure of gas and young stars is apparent. In our Milky Way, the youngest stars (OBA spectral types) share the kinematics of the gas disc in which they form, as stellar age increases the velocity dispersion also increases --- this kinematic evolution is interpreted as being due to stars on their orbits encountering `lumps' in the disc, and scattering off them, such as giant molecular clouds (GMC's) and spiral arms. This gives rise to the thin stellar disc having on average a higher dispersion than the gas disc and young stars. The difference in velocity dispersion between different components
gives rise to the phenomenon known as `asymmetric drift'; for example, the rotation of the stellar disc lags behind that of the gas disc due to it's higher radial velocity dispersion which provides additional dynamical support against the galaxy's overall gravitational field.

Many external galaxies have their gas and kinematics observed in the \Ha\ line of ionised hydrogen whose luminosity is generally dominated by
HII regions. In the Milky Way, HII regions and GMCs  share the low velocity dispersion (i.e. between cloud centres, \cite{SB89}) of the gas disc;
however, it should be noted that the \Ha\  line has a thermal broadening due to a characteristic temperature of $10^4$K of $\sim 9$ \kms\ which will increase the observed line width. There is also a turbulent broadening due to internal motions in HII regions of order 20 \kms\ \citep{Metzger67,Shields90}. Adding these in quadrature, we can see the typical dispersion is consistent with the range of 20--25 \kms\  found by observations of external nearby spirals \citep{GHASP,And06}.

The Milky Way also has a  a so-called `thick disc' stellar component \citep{GR83} (though there is still a debate as to whether this is a true dichotomy or a continuous stellar population sequence, e.g. \cite{Bovy12,Bovy12B}.). Thick discs are now thought to be ubiquitous in spirals and may have masses that are,
on average, up to values comparable to the thin disc \citep{Com11}. 
The thick discs contain older, redder, and lower surface brightness populations and negligible on-going star-formation \citep{YD08B}. The thick disc in our Milky Way
has a scale height of $\sim$1400 pc \citep{GR83}. It is 
 low metallicity $\sim 1/4$ Solar,  is $\sim 10$ Gyr old \citep{GWK89} and  has a vertical velocity dispersion of $\sim$ 40 \kms\ \citep{CB2000,Pasetto12}. Other spirals are thought to be similar. The origin of  thick discs is a matter of debate and there are a variety of models --- it may be formed from early merger events, satellite accretion, or secular evolution (see discussion in \cite{vanderKruitFreeman2011} and references therein). A particularly relevant scenario for our later discussion is
the idea that the thick discs form in situ in early gas-rich high-dispersion discs \citep{Bour09}. 

\begin{figure}[t]
\centering 
\iftablet
 \includegraphics[width=11cm]{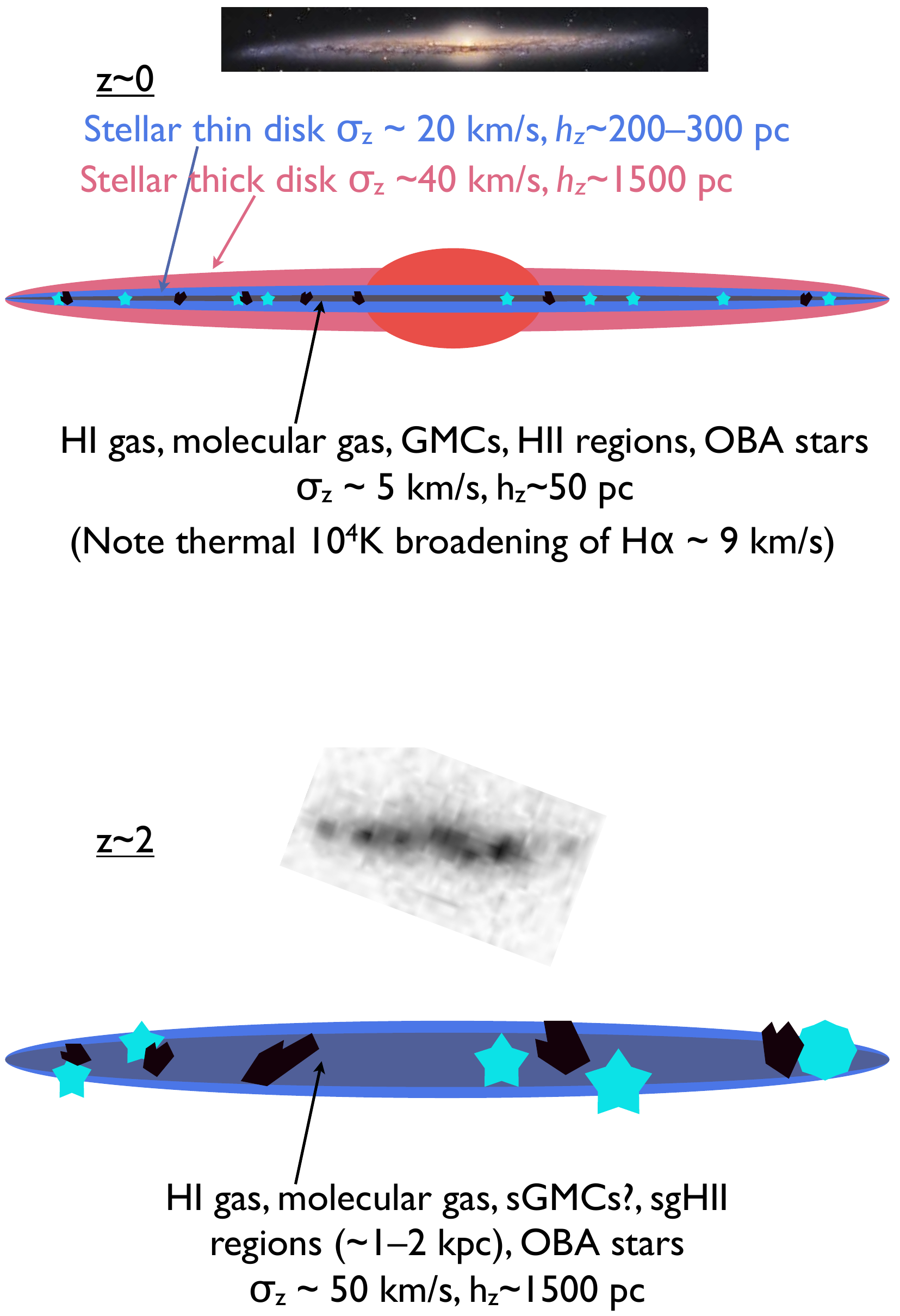}
\else
 \includegraphics[width=6.5cm]{figure-cartoon.pdf}
\fi
\caption{\small Illustrative schematic showing the different structures of low-redshift and high-redshift disc galaxies in an edge-on view. Top: components of the Milky Way and similar
local spirals (see Section~\ref{sec:local-spirals}) containing stellar thin/thick discs and a very thin gas disc in the centre. The latter contains all the Giant Molecular Clouds, HII regions, molecular and neutral gas and young stars. Bottom: a clumpy high-redshift disc (see Section~\ref{sec:turbulent-discs}). This contains
a thick ($\sim 1$ kpc scaleheight) and highly turbulent discs of molecular gas, young stars, super-giant HII regions (kpc scale star-forming `clumps' ) and (presumably)
super-Giant Molecular Clouds. {\it Credit: inset images are of NGC 4565 (top, reproduced by permission of  R. Jay GaBany, \href{http://cosmotography.com}{Cosmotography.com})  and $z\sim 3$ galaxy UDF \#6478 of
\cite{Elm-thick} (their Figure 2, reproduced by permission of the AAS). }}
\label{fig:cartoon}
\end{figure}

Figure~\ref{fig:cartoon}  illustrates these components schematically and also contrasts them with the emerging (but by no means certain) picture of $z\sim 2$ galaxies which we will return to in Section \ref{sec:turbulent-discs}.

In this review, I will use the words `velocity dispersion' frequently. First, I should note that what is measured from spectra is always `line-of-sight velocity dispersion'. Secondly, I note that in the literature it is used in two principal senses:

\begin{enumerate}
\item {\it Resolved velocity dispersion} (sometimes called `intrinsic dispersion' or `local dispersion')  by which we mean the dispersion as measured in line widths of elements of spatially resolved observations. A galaxy disc is a good physical example, in this case the dispersion refers to the random motions of stars and gas around the mean rotation field at each position.
\item {\it Integrated velocity dispersion} by which we mean the dispersion as measured from an integrated spectrum (i.e. spatially averaged). In this case, this will include a (possibly dominant) contribution from any global velocity field such as rotation. The HI line width used in the Tully-Fisher relation is a classic example of this, as are the central `velocity dispersions' measured for elliptical galaxies in long-slit studies.
\end{enumerate}
The measurement difference corresponds to whether we measure the line widths in spatially resolved spectra, and then average or whether we average the spectra and then measure the line width. Physically it is a distinction between different models of internal support against gravity (random motions vs rotational ones).  In practise, any real observation, however fine, will average over some spatial scale and there will always be a contribution from large-scale and random motions to any line width, it is a question of degree and we will return to this point in Section~\ref{sec:dispersion-measures}.
 I will endeavour to be clear about what kind of velocity dispersion is being measured in what context.

\subsection{Kinematic properties of elliptical \\ galaxies.}

While not the focus of this review, it is worth commenting briefly on the major kinematic properties of elliptical galaxies. In particular, one must bear in mind that possible evolutionary processes (such as star-formation `quenching' and galaxy merging) may connect ellipticals at lower redshifts with star-forming galaxies at high-redshift. The historical picture of elliptical galaxies is of large, massive systems with negligible gas and star-formation with small rotation and kinematics dominated by velocity dispersion \citep{DZF91}. The elliptical galaxy analogy of the Tully-Fisher relation is the Faber-Jackson relation \citep{FJ}  relating the integrated velocity dispersion to the luminosity (or stellar mass). It should be noted that what was traditionally measured here is  an integrated velocity dispersion of the brightest central part of the galaxy, usually with a long-slit spectrograph. The Faber-Jackson relation has now been extended to a `Fundamental Plane' \citep{FP} where size, surface brightness, and velocity dispersion (equivalent to size, luminosity, and dispersion) are correlated to define a three parameter sequence with a reduced scatter (see reviews \cite{DZF91,BM09}).\footnote{But see \cite{fund-line} for a contrary opinion where the properties of elliptical galaxies are reduced to a `Fundamental Line.'}

This classical picture has evolved considerably in the last decade with the availability of large-scale IFS observations of nearby elliptical galaxies. In particular, it is now known that a dominant fraction of elliptical galaxies are in fact rotating \citep{Cap2007,Em2007} and one can divide ellipticals in to two classes of  `slow rotators' and `fast rotators' based on angular momentum. The slow rotators tend to be the most massive ellipticals 
(stellar masses $>3\times 10^{11}\,\Msun$) and$/$or the ones found in the centres of rich clusters \citep{Cap2011,DEugenio2013}. The kinematic division may relate to assembly history and the relative role of dissipative (`wet') and non-dissipative (`dry') mergers (e.g. \cite{Burkert2008}) in building the most massive red-sequence galaxies. Detailed kinematics now goes beyond the simple fast/slow overall angular
momentum division
and in particular probing rotation in the outer parts of nearby ellipticals (i.e. well beyond the half-light radii) using IFS and multi-slit techniques provides
detailed information on assembly histories (e.g. \cite{Proctor2009,Arnold2011}).

So far these resolved kinematic observations of local ellipticals are limited to samples of only a few hundred objects, to be contrasted with Tully-Fisher observations of thousands of spiral galaxies, and it is not yet clear how the kinematic classes relate to the classical picture of the Fundamental Plane. This is likely to be an area of fruitful further research.

\subsection{Kinematic properties of local mergers}

As we will see, an important issue in studying galaxies at high-redshift is the kinematic separation of rotating disc galaxies from merging galaxies. At 
z$>$1, the apparent merger rate is high and major mergers typically constitute  up to 20--50\% of observed samples depending on selection details and definition. So trying to systematically identify and classify them is important and critical to issues such as the high-redshift \TFR.

Mergers are much rarer in the local Universe with major mergers being $\sim 1$--2\% of all galaxies \citep{Domingue2009,Xu2012} which is
why {\TFR}s work so well. Departures from the mean relation may be correlated with peculiar velocity structures or recent star-formation
history associated with merging \citep{Kannappan2002,Oliveira2003}.  There is actually a paucity of work systematically examining the kinematics of mergers perhaps
due to this rarity. Typically papers discuss individual objects in detail,  \citep{Colina2005,Dasyra2006,PL2012} rather than trying
to extract characteristic kinematic parameters for statistical analysis.
Sources are generally selected to be major mergers
as Ultra-Luminous IR Galaxies \citep{Arribas2008,AH2009}, or `ULIRGS',\footnote{A note on the terminology: at $z\sim 0$ the `LIRG' / `ULIRG' boundary at $L(IR) \simeq 10^{12} L_\odot$ seems to distinguish normal spirals from major mergers, however this may change to high-redshift in the sense that more galaxies in the LIRGS/ULIRGs are structurally star-forming discs due to the overall evolution in star-formation rates \citep{Daddi2007,Daddi2008,Wuyts2011}.} aided by obvious
morphological criteria (e.g. double-nuclei, tidal tails). Typically active on-going but pre-coalescence mergers display complex kinematic maps (in ionised gas) tracing the discs of each component (with large velocity offsets) plus kinematic disturbances induced by the merger. 
At high-redshift, non-parametric measures such as {\it kinemetry} are being increasingly applied to try and distinguish discs from mergers (see  Section~\ref{sec:merger-disc}). Kinemetry \citep{Kinemetry} was originally
developed to measure the fine kinematic structure of local elliptical galaxies and is the kinematic extension of photometric moments. It has been applied to a small sample of four  local IR-selected merging galaxies by \cite{Bellocchi2012}, who found good consistency with photometric classifications.
There is no publication presenting quantitative or qualitative kinematic classification of a large sample of local mergers, so this would be valuable future work for comparison with high-redshift, where as we will see in Section~\ref{sec:merger-disc} this has been done of necessity.

\section{Early work with long slit \\
spectroscopy}
\label{sec:longslit}

Resolved kinematic work at significant redshifts began with the commissioning of the 10-m W.M. Keck telescope, which was the first optical telescope in this aperture class. Previous 4-m telescope work had studied normal galaxies to redshifts $z\sim 1$ using multi-slit spectrographs --- examples include the LDSS2 redshift survey \citep{KG95} and the Canada France Hawaii Redshift Survey \citep{Lilly95}, but had only attempted integrated spectroscopy due to signal:noise limitations. Early Keck work focussed on integrated velocity dispersions \citep{Koo1995,Forbes1996} using the optical line width in a manner similar to early radio HI line widths. Trends were found of this velocity dispersion with luminosity which was interpreted by Forbes et al. as echoing the local \TFR\ (with 
the large scatter being due to the much broader sample selection and crudity of the method) and by Koo et al. as representing galaxies which might `fade' to become local low-luminosity spheroids. 

The first resolved long-slit work at significant redshift, i.e. constructing true rotation curves, was done by Nicole Vogt et al. \citep{Vogt1996}  again using the Keck telescope. Galaxy rotation curves, with signatures of a turnover towards flatness at large radii,  were measured to radii $\sim 2$ arcsec for galaxies at $0.1<z<1$ in 0.8--0.95 arcsec seeing. An important finding was that high-redshift galaxies have similar rotation curves to low-redshift counterparts and that \emph{`some massive discs were in place by z$\sim$1'}, the first harbinger of the modern picture and in
tension with the $\Omega_m=1$ flat CDM cosmology favoured at the time.  Vogt et al. found evidence for a \TFR\ with only mild evolution.

A key problem in these early studies, and one that remains with us today, is the limited spatial resolution compared to the scale of the objects being studied. In our current cosmology, 1 arcsec corresponds to 6.2--8.5 kpc for $0.5<z<4$. Given a typical spiral disc today has an exponential scale length of only 1--5 kpc \citep{F70}  it can be seen that these high-redshift discs were only marginally resolved in good natural seeing (0.5--1 arcsec). However, the situation is tractable as the exponential is a soft profile detectable
to several scale lengths. 
Because of this, an important development in kinematic modelling was the use of maximum likelihood techniques to fit kinematic models convolved
with the observational Point Spread Function (PSF).

Vogt herself pioneered this technique in her 1996 paper. Another similar approach was that of \cite{SP98}, who applied this to star-forming galaxies
at $z\sim 0.3$ observed with the Canda-France Hawaii Telescope (CFHT) to derive a \TFR. Important conclusions from these early works (that echo later results)
were (i) at least some 
star-forming galaxies at these redshifts displayed clear rotation, (ii) significant fractions (25\% in \cite{SP98}) do not and are `kinematically anomalous', (iii) rotating galaxies appear to follow a \TFR, (iv) the existence of very compact star-forming galaxies at intermediate redshifts, (v) the \TFR\ displays significantly increased scatter compared to the local relation, and (vi) disagreement as to whether the zeropoint of the \TFR\ evolves or not. Note that these early works used a relatively low spectral resolution and could not measure the internal velocity dispersions in the galaxy discs. As we will see at the end of this review the evolution (or not) of the \TFR\ zeropoint is still a matter of debate.

Later, long-slit work built on these. For redshifts $z \lesssim 1$, there was work by \cite{Ziegler01} and \cite{Bohm04} who found evidence for `mass dependent' evolution in the \TFR\ (in the $B$-band, little evolution for more massive galaxies, up to 2 mags in brightening for the fainter galaxies) using the Very Large Telescope (VLT) and the FORS2 spectrograph to study 113 galaxies. Again, spectral resolution was low ($\sigma \simeq 100 $\kms). It is interesting to note that the fraction of anomalous galaxies was $\sim 30\%$ in these papers though that excited negligible comment. \cite{Con05} was the first to look at the {\em stellar mass \TFR\ at significant redshift} using a sample with near-IR photometry and found no evidence for an evolution of the relation from now to $z>0.7$.

At higher redshifts ($z>2$), the earliest kinematic work with long slits focussed on the kinematic follow-up of the so-called `Lyman Break galaxies' (LBGs). These are ultraviolet (UV)-selected star-forming galaxies first characterised by \cite{Steidel1996} at $z\sim 3$. At these redshifts, the galaxies are observed to have low flux (i.e. `dropouts') in the $U$-band from neutral hydrogen absorption bluewards of the Lyman limit together with blue colours (i.e. nearly constant $f_\nu$ flux) in redder filters. \cite{Pett98} and \cite{Pett01} presented near-IR spectra of 15  $z\sim 3$ LBGs. Integrated velocity dispersions were measured from [OII], [OIII] and H$\beta$ emission lines but found to have no correlation with optical or UV continuum properties. In two cases, resolved velocity shear (i.e. tilted emission lines) were detected, but Pettini et al. could not conclude if these were rotating discs.

The UV selection technique has subsequently been pushed to lower redshifts \citep{Steidel2004} ($1.5<z<2.5$) where the galaxies do have $U$-band flux and selection relies on them being bluer in their $U$-band to optical colours than lower redshift galaxies. It is important to note that UV selection does not pick out all galaxies at these redshifts --- in particular it can miss out massive quiescent galaxies (e.g. \cite{Cimatti2004,McCarthy2004,vdK2004}) and populations of dusty star-forming galaxies \citep{Yan2004} which are picked out by red/near-IR colour/magnitude selections (see review on high-redshift red galaxies of \cite{EROs}; an excellent recent review of physical properties and selection
techniques of high-redshift galaxies is given by \cite{Shap11}). \cite{Erb:2006} performed near-IR long-slit spectroscopy of 114 $z\sim 2$ UV-selected galaxies in the \Ha\ emission line. In most cases, resolved information was not measurable and only total line widths were measured. Very little correlation was found between these integrated velocity dispersions, or derived dynamical masses, and stellar mass. A stronger correlation was found between dispersion and rest-frame $V$ luminosity though with a lot of scatter (factors of 3--4 in dispersion at a given luminosity).   In 14 cases (some due to exceptional seeing), resolved velocity shear was measurable and even displayed flat rotation curve tops; the dispersion was well correlated with the rotation velocity suggesting that rotation was the primary contribution to the line widths. Erb et al. also inferred from their sample's star-formation
rate densities that they were gas rich (mean gas fraction $\sim 50\%$) an important point to which I will return later. Finally, I note that
they found that their
sub-sample with shear tended to be the galaxies with older stellar population ages and larger stellar masses leading to the (in hindsight) quite prescient conclusion
that  {\it `the rotation of mature, dynamically relaxed galaxies is a more important contribution to our observed shear than merging, which should not have a preference for older, more massive galaxies'}.

\section{High-Redshift IFS Surveys}
\label{sec:surveys}

\begin{figure*}[t] 
\centering
\iftablet
   \includegraphics[width=18cm]{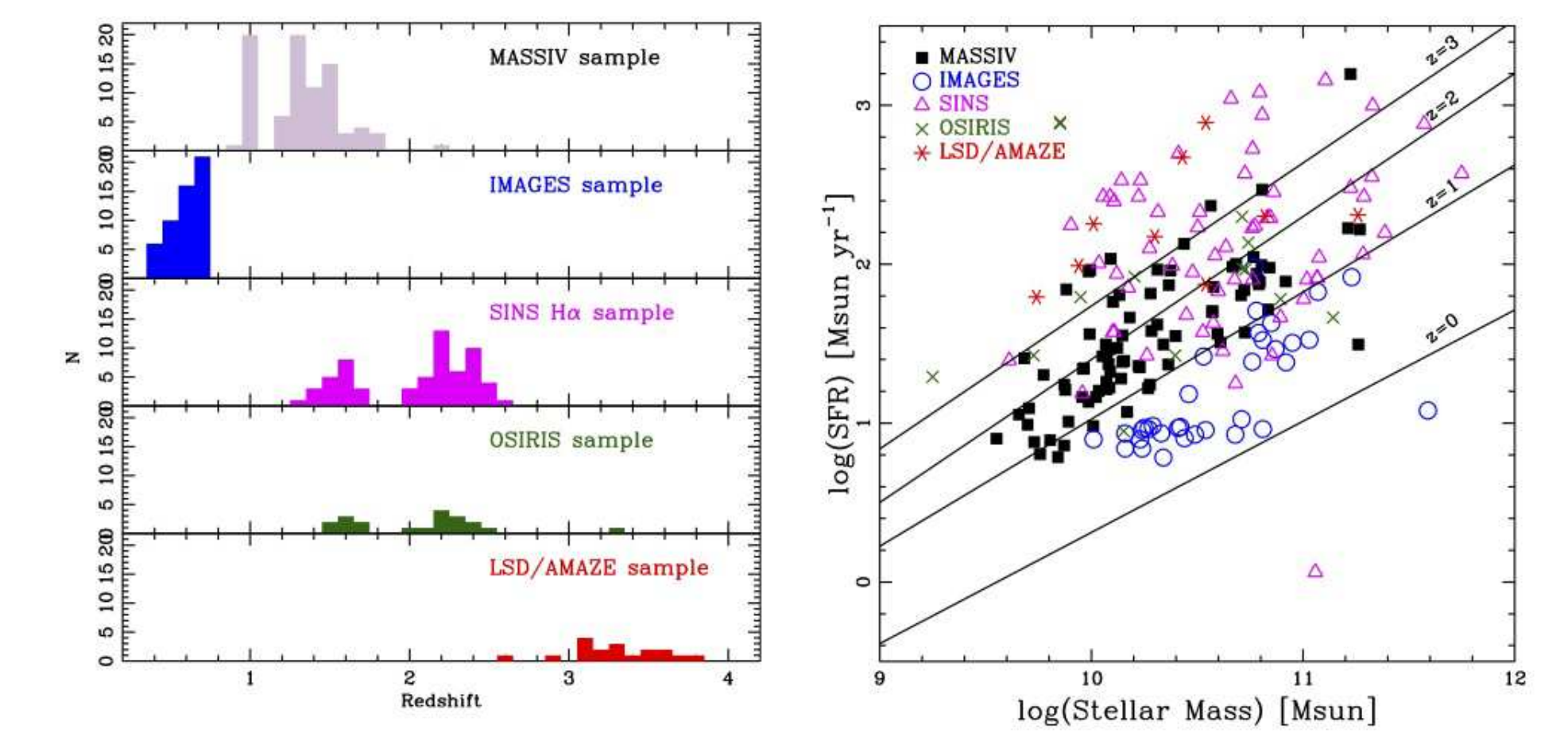}
\else
   \includegraphics[width=15cm]{figure-surveys.pdf}
\fi
\caption{\small The distribution of the principal IFS surveys in the redshift (left) and star-formation rate --- stellar mass (right) space (stellar masses are corrected to the \cite{Salpeter-IMF} IMF). The lines on the right plot are the 
locations of the main galaxy `star-formation main sequence' at different redshifts taken from the models of \cite{Bouche2010}. {\it Credit:  adapted from Figures 10 \& 14 of \cite{MASSIV1}, reproduced with permission \copyright\ ESO. }} 
\label{fig:surveys}
\end{figure*}

The advent of resolved 2D kinematic information coupled with (in some cases) the use of AO to improve spatial resolution has led
to significant new insight. In this Section, I will review the major and most influential surveys, discuss in particular their selection strategies, instrumentation used, and
review the important survey-specific kinematic (and associated) results in their major papers. Figure~\ref{fig:surveys} shows the redshift range and physical parameter
space (i.e. stellar mass and star-formation rates) covered by the main IFS surveys discussed below.

\subsection{The SINS survey}

The SINS  (`Spectroscopic Imaging survey in the Near-infrared with SINFONI') survey was one of the first large IFS surveys of galaxies in the $z\gtrsim 2$ Universe and has been one of the most important for extending our views of early galaxy evolution. 
SINFONI  \citep{SINFONI} is a flexible IFS on the 8-m VLT  capable of both natural seeing and AO modes of operation.
The first results from integral field 
observations in \Ha\ emission, of a sample of 14 
BM/BX galaxies (selected similarly to \cite{Erb:2006}) confirmed the present of a significant
fraction of galaxies with rotation fields characteristic of discs \citep{NFS06} and large enough to be resolved in 0.5-arcsec seeing.
This was one of the first pieces of kinematical evidence for the `clumpy disc' picture (see Section \ref{sec:turbulent-discs}) which I will return to throughout this review.

In the same year, SINS\footnote{Though SINS and SINFONI are often associated as we shall see there are two other large 
high-redshift surveys performed with SINFONI by independent teams, as well as smaller ones.}  published 
one of the very first AO observations of a high-redshift star-forming galaxy,  the $z=2.38$ object `BzK-15504' by \cite{Genzel06}.  The galaxy was a $K$-band selected star-forming galaxy. Redder wavelengths are a good proxy for stellar mass, so being $K=21.1$ meant that this object was selected as a {\it massive} star-forming galaxy (stellar mass
$\simeq 8\times 10^{10}\Msun$). This is an important point because, as we will see throughout this review, the kinematic nature of galaxies trends with stellar mass and in particular we see differences
between $K$-band-selected and UV-selected star-forming galaxies. The galaxy was colour-selected to lie at these redshifts using the $BzK$ colour-selection \citep{BzK} which is one of a family of colour-selection techniques used to select galaxies at high-redshift \citep{Shap11}. It was observed using $K$-band AO in the \Ha\ emission line.

\begin{figure}[t] 
\centering
\iftablet
  \includegraphics[width=12.5cm]{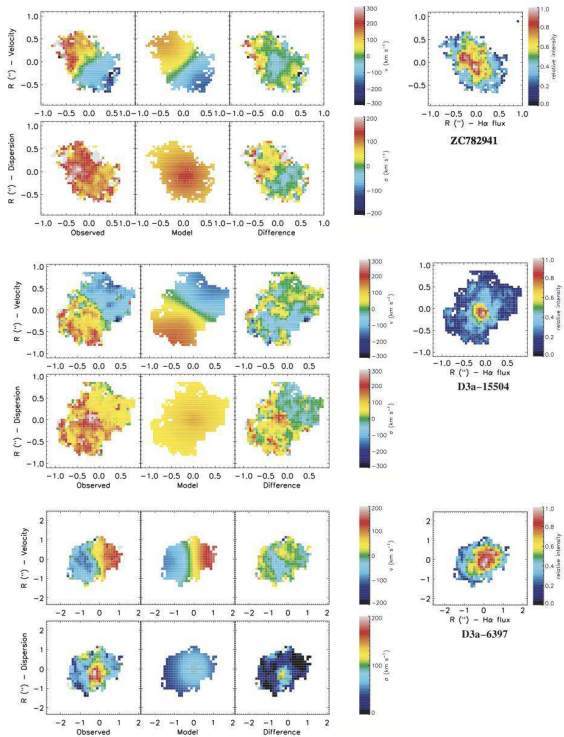}
\else  
   \includegraphics[width=7.8cm]{figure-GenzelObject.pdf}
\fi   
\caption{\small Three selected $z\sim 2$  galaxies from \cite{Cresci09} well fit by kinematic disc models. The middle object, 
 galaxy  D3a-15504, was originally observed by \cite{Genzel06}, here it has  higher signal:noise. These
are \Ha\ emission line maps, top two taken with AO at resolution 0.2 arcsec,  the bottom object  illustrates how these disc kinematics are still
resolved in natural seeing. On the left are the kinematic maps (top row: velocity, bottom row: dispersion) comparing the data and best fit disc models. \Ha\ intensity maps are shown on the top right. Each galaxy is well fit by a rotating disc model  but the velocity dispersion is high. Values reach $>100$ \kms. I call out the  spatial structure in the dispersion maps (see my discussion in Sections \ref{sec:physics}
and  \ref{sec:outstanding}) as a particular striking and unexplained feature, not reproduced in the models.
{\it Credit:  adapted from Figure 2 of  \cite{Cresci09} (selected galaxies), reproduced by permission of the AAS. } } 
\label{fig:GenzelObject}
\end{figure}

This galaxy was the first prototypical case of a galaxy at $z\sim 2$ with clear disc-like kinematics seen at high resolution, as defined by a smooth symmetric velocity gradient with evidence for a turnover to a flat portion and no abrupt discontinuities in velocity as might be expected if it were two objects engaged in a major merger. Subsequent deeper AO observations of this
object (and two others with AO) \citep{Cresci09} have confirmed this picture (Figure~\ref{fig:GenzelObject}). The large star-formation rate and low value of the \cite{Q} $Q$ parameter ($<1$) implied a gas rich disc forming stars {\it in-situ} rapidly and suggested continuous fuelling by cosmological accretion. The value of the local \Ha\ velocity dispersion ($\sigma \sim 50$--100 \kms) was about 2--4$\times$  higher than the thin discs of normal local spirals (see Section~\ref{sec:local-spirals}), however the circular velocity ($v_c$) was quite similar ($\sim 230 $\kms) leading to a much smaller value of $v_c/\sigma$ which \cite{Genzel06} identified as a key kinematic parameter (see later discussion in Section~\ref{sec:turbulent-discs}). 
Genzel et al. pointed out that the dynamically hot disc is more akin to the local thick discs of nearby spirals and there could be a plausible
evolutionary connection. They also identified the energy source supporting the large disc gas dispersion (e.g. star-formation feedback, accretion, etc.) as a key problem to understand, a point to which we will return in Section~\ref{sec:physics}.

The full SINS survey was carried out from 2003--2008  and observed a total of 80 objects (\citep{NFS09}, noting the sample has since been significantly extended \citep{Manc2011}). Sixty-three of the observed galaxies had detected emission-line kinematics and 12 were observed with AO (improving spatial resolution from $\sim 0.5$ to $\sim 0.1$ arcsec). Sample selection is the key to comparing high-redshift IFS surveys, SINS had a range of heterogenous sub-samples and in particular included a large number of $K$-band as well as rest UV-selected galaxies (the latter sub-sample was the focus of the early work of \cite{NFS06}). These formed the majority of the the sample and the various papers focussed on these, in particular with the \Ha\ detected sub-sample with $1.3<z<2.6$ (62 galaxies). A large range of stellar mass was probed ($2\times 10^{9}$ -- $3\times 10^{11}\Msun$ with a median of $2.6\times 10^{10}\Msun$) as the $K$-band and UV selection tended to pick up complementary populations. 

A primary result (echoed in other work) summarised in the survey paper \citep{NFS09} was that around a third of the sample were rotating star-forming discs \citep{NFS06} with large ionised gas dispersions (`turbulent discs') with $v_c/\sigma \sim 2$--4. Another third were objects with no significant kinematic shear but still high dispersion (`dispersion dominated galaxies' in the language of \cite{Law07}) while the remaining third had detectable kinematic structure but no clear disc-like pattern, so they were described as `clear mergers'. This approximately 1/3:1/3:1/3 split of fundamental kinematic classes is echoed in many other surveys we will see in this section though the exact percentages vary. Morphologically, the discs do not resemble local spirals of similar mass, rather they are dominated by giant kpc scale clumps of emission --- and this remains true whether UV, \Ha\ or near-infrared continuum is considered \citep{NFS11}. 

\cite{Cresci09} presented the kinematics of the best quality SINS discs (Figure~\ref{fig:GenzelObject}), mostly those with the highest signal:noise ratio and/or AO observations. These are generally massive star-forming galaxies with $K<22.4$ and quite large (disc scale lengths of 4--6 kpc). The dynamical modelling of the discs required a large component of isotropic velocity dispersion (40--80 \kms), construction of the stellar mass \TFR\ indicated a 0.4 dex\footnote{All dex values 
reported in this review refer are in log mass or log luminosity unless otherwise stated.} offset at $z\sim 2$ lower in stellar mass at a given $v_c$ and is plausibly reproduced by simulated galaxies. \cite{Puech08} raise the question about the choice of
local relation which can have an effect on the amount of evolution; \cite{MASSIV4} argue this makes negligible difference to the results of Cresci et al. as 
the different local relations intersect at $10^{11} \Msun$ which is the mass range of the SINS discs considered. \cite{Bou07} consider the other
velocity--size scaling relation of SINS
galaxies (using half-light radii) and concluded this relation was evolved from $z=0$. 

Clearly distinguishing discs from mergers kinematically is a key issue (to which I will return in the next section), \cite{Shap08} considered this for a sample of 11 SINS galaxies (again highest signal:noise) using the technique of `kinemetry' \citep{Kinemetry}, they find 8/11 are discs by this criterion and classify the rest as mergers, though dispersion-dominated objects were excluded as the sample was biassed towards well-resolved objects.  

The resolved physical properties of SINS discs was addressed in a series of papers, \cite{Gen08} considered possible scenarios for the origin of the turbulence and the evolution of the discs. They argue that the large dispersions applies to cold gas as well as the observed ionised gas and arises from cosmological accretion. There is a correlation of central mass concentration with metallicity (as inferred from the [NII]/\Ha\ line ratio) which which would imply that bulgeless galaxies are younger. 
\cite{NewDD} considered an extended AO sample and compared with non-AO observations, in particular finding that the fraction of `dispersion dominated' galaxies (see Section~\ref{sec:DD}) drops with increasing resolution. 
\cite{Gen11} considers the properties of the giant kpc clumps of five galaxies  in more detail. Key points are that the clumps are entrained in the overall rotation field of the disc (i.e. they are part of the disc not merging external galaxies), that they occur in regions of disc instability as indicated by Toomre $Q<1$ and that they show broad wings\footnote{An important point is that these are even broader wings (several hundred \kms\ width) on a central component which is often confusingly called `narrow' despite being broader than in local disc galaxies.} indicative of star-formation-driven outflows \citep{Newman-outflows}.

\subsection{The OSIRIS survey of UV-selected galaxies}

The IFS survey of  \cite{Law07,Law09} focussed on 13 UV-selected galaxies observed with the OSIRIS IFS \citep{OSIRIS} on the Keck telescope.\footnote{The lack of an acronym is, in my opinion, refreshing.} Twelve of these galaxies are at $z\sim 2.2$ selected using the `BX' colour criteria of \cite{Steidel04} and were a subset with high \Ha\ fluxes from previous
slit spectra \citep{Erb06} or high star-formation rates calculated from rest-frame UV emission. The IFS subset was mostly selected
on the emission line fluxes but also had a subjective selection component
for interesting objects with criteria such as extreme ends of the  young/low mass old/high-mass scales, multicomponent UV morphology, and 
unusual UV spectra. A lower $z\sim 1.6$ sample also selected from the BM/BX catalogue and observed by the complementary project of \cite{Wright07,Wright09}; this is described below along with other OSIRIS work at similar redshifts (Section~\ref{sec:other}).

The Law et al.  sample galaxies are generally of lower stellar mass ($1\times 10^{9} $--$8\times 10^{10} \Msun$ with a median of $1.4\times 10^{10} \Msun$) by a factor of 2 than the SINS discs at the same redshift; however, there is a broad overlap (Figure~\ref{fig:surveys}). All IFS observations where done with laser guide stars (LGS) AO of the \Ha\ line in the $K$-band 
so that spatial resolution was 1--2 kpc, several times better than non-AO observations of other surveys. The price to be paid for this was the lower surface brightness sensitivity for extended emission
due to the finer pixel sampling and the reduced flux from finite Strehl ($\simeq 0.3$).  From the IFS observations,
 6/13 galaxies showed clear velocity shears, though merger interpretations were also plausible in 3--4 of these, with 1--2 being very clear discs. 

Law et al. note the dominance of objects
 with  high  intrinsic dispersions 50--100 \kms\ which were in all cases larger than the maximum velocity shear amplitude (another contrast to the SINS discs), labelling these `dispersion dominated galaxies'. Some objects had no detectable shear whatsoever. In 
the comparison of \cite{NFS09}, it was shown that the typical `circular velocities' (under a disc interpretation) and half-light radii (measured in \Ha) were also a factor of 2--3 smaller than SINS discs, with the smallest Law et al. objects having sizes 
 of $\lesssim 1$ kpc. This seems consistent with a broader picture where UV-selection favours lower stellar mass, smaller star-forming galaxies at these redshifts and is further discussed in Section~\ref{sec:DD}.

 \cite{Law07} looked at the `Toomre parameter $Q$' in three objects as  defined by:
\begin{equation}
Q = { V^2 \over G M_{disc} / r_{disc} } = M_{dyn} / M_{disc}  
\end{equation}
where $V$ is the observed shear. However, this equation does not correspond well to the standard \cite{Q} criterion  --- though one can obtain it by writing $\sigma = V$.  It is better thought of as a ratio of dynamical mass ($r V^2/G$) to visible `disc' mass ($M_{disc}$), the galaxies all had  $Q\lesssim 1$ indicating that the disc mass is unphysical --- i.e. too much mass to be supported dynamically by rotation. I also note that interestingly the equation corresponds very closely to the criterion for exponential disc instability (against bar formation) in a dark matter halo independently identified by \cite{MMW98} (their eqn. 35). The `$Q$' values suggest they may be true dispersion-dominated objects and not stable discs, unless the compactness causes $V$ to be significantly underestimated through resolution effects (and their could also be issues with inclination which is not accounted for).

It is important to note that Law et al. observed a similar number of galaxies which were not detected, there was a tendency for these objects to be observed in sub-optimal conditions (e.g. seeing) but there could be a result of a bias of detections to higher surface brightness. The authors do find a systematic trend in the direction expected for this bias compared to the general galaxy population at this redshift. Interestingly, one of the non-detections was subsequently detected by \cite{Law-Nature} with a five times longer OSIRIS exposure, it proved to be a high-dispersion rotating disc with a spiral pattern (rare at these redshifts, attributed to a minor merger induction). Three more were observed and also detected by \cite{NFS09} in natural seeing and proved to be rotation dominated, clearly the resolution--sensitivity trade of AO observations is playing a role (as did the longer exposures used).

The incidences of possible discs and mergers seem comparable with other work (perhaps with a trend to less of these at lower stellar masses), however the compactness of these galaxies does not lead to unambiguous characterisation and Law et al. caution against over-simplistic classifications in to these two classes.


\subsection{The IMAGES and related FLAMES-GIRAFFE surveys}
\label{sec:IMAGES}

\begin{figure*}[t] 
\centering
\iftablet
  \includegraphics[width=19cm]{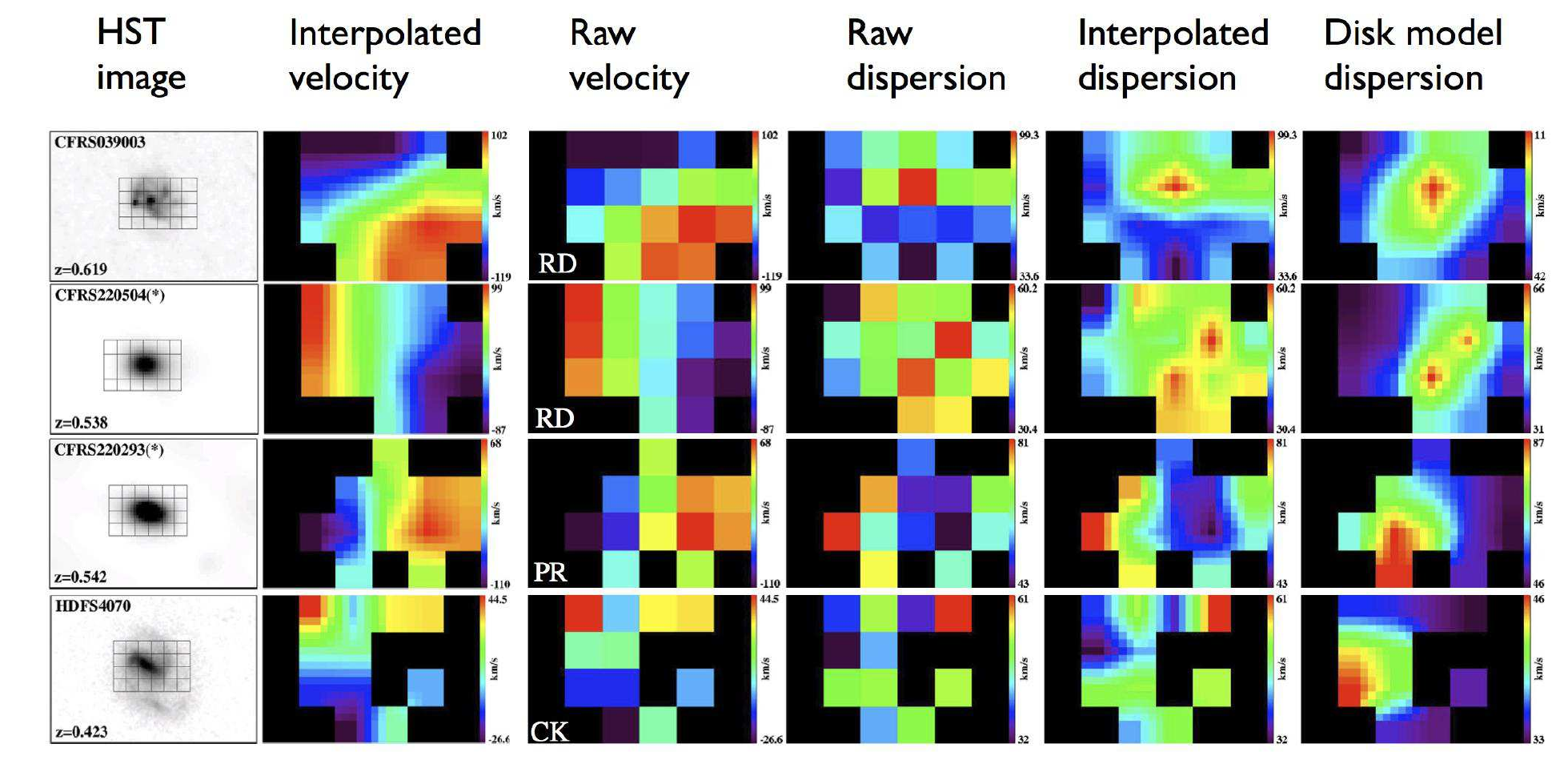}
\else
   \includegraphics[width=15cm]{figure-IMAGES-classes.pdf}
\fi
\caption{\small Images and IFS  maps of galaxies of different kinematic classes from sample FLAMES/GIRAFFE data showing the different kinematic classifications described in the text. Note the rather coarse spaxel scale of 0.52 arcsec (see grid superimposed on higher-resolution
HST image) makes classification challenging and a $5 \times 5$ pixel interpolation scheme was used to smooth the maps. {\it Credit:  adapted from Figures 3 \& 5
(selected galaxies and combined) of \cite{Flores06}, reproduced with permission \copyright\ ESO. } } 
\label{fig:IMAGES-classes}
\end{figure*}

The predominant IFS work at intermediate redshift ($0.3<z<1$) has been done using the VLT's FLAMES-GIRAFFE multi-object integral field facility. This has produced a large sample from the IMAGES (` Intermediate MAss Galaxy Evolution Sequence') VLT Large Program. FLAMES-GIRAFFE \citep{FLAMES}  is an optical facility with 15 separate `Integral Field Units' (IFUs) patrolling a 25 arcmin field-of-view. 

Important early work was done with this instrument by \cite{Flores06} with a sample of 35 objects (a sample of $I<22.5$ emission line galaxies observed at $z\sim 0.6$ and focussing on the \TFR\ and the scatter about that relation identified by the slit based surveys mentioned in Section~\ref{sec:longslit}). An important development was the use of 2D kinematic data to simply characterise/classify the velocity fields of star-forming galaxies. The $I$-band selection at this redshift would be pulling out high stellar mass systems, including objects comparable to the Milky Way. Flores et al. classified galaxies in to three kinematic classes (used extensively in later papers) via inspection of the velocity and velocity dispersion 2D maps:

\begin{enumerate}
\item {\bf Rotating discs (RD)}: These have regular symmetric dipolar velocity fields, aligned with the morphological axis, with symmetric centrally peaked dispersion maps.These objects correspond kinematically most closely to local rotating discs. The centrally peaked dispersion is a product of both the fact that typical discs have a steeper rotation curves in the inner regions combined with the smoothing from the PSF (a.k.a. `beam-smearing'). The unresolved velocity shear appears as an artificial component of velocity dispersion, but the fact that it appears in the middle makes it useful to identify discs.\footnote{Note both of these are required: a purely linear velocity gradient will not have a centrally peaked dispersion, rather the dispersion is uniformly boosted.}
\item {\bf Perturbed Rotators (PR)}: These are similar to the RDs displaying a dipolar velocity field, but the velocity field is not perfectly symmetrical nor aligned with the morphological axis and/or the dispersion peak may be offset from the centre (or absent). Physically, these are identified as disc galaxies with some sort of minor kinematic disturbance (e.g. from a minor merger or gas infall/outflows). 
\item {\bf Complex Kinematic objects (CK)}: This class is everything else, typically a chaotic and/or multipolar velocity field with no symmetry. Physically, these could be identified with systems such as major mergers.
\end{enumerate}

The measured ratio of RD:PR:CK objects comes out as  an almost three way split of 34:22:44 percent. This is a stark contrast to local surveys where virtually all similarly massive galaxies would likely be classified as RD by these criteria, and implies a large amount of kinematic evolution in the galaxy population in the last 6 Gyr. However, star-formation properties also evolve in a similarly dramatic fashion: at these redshifts, nearly half of massive galaxies ($>2\times 10^{10} M_\odot$) are undergoing intense star formation comparable to their past average; this is a significant change from $z=0$ \citep{Bell05,Juneau05}.  Physically, the growth in the CK classification is attributed by the authors to a strong evolution in the major merger rate with the CKs being either in-process mergers or dispersion-supported merger remnants \citep{Puech06}. Such objects represent only a few percent of local massive galaxies  \citep{Domingue2009,Xu2012}. 

It should be born in mind that these classifications are based on natural seeing data of resolution 0.4--0.8 arcsec (2--4 kpc at $z\sim 0.6$) and to make matters worse the IFUs have quite coarse sampling (0.52 arcsec micro lenses). Flores et al. use an interpolation technique to present their IFU maps (see Figure~\ref{fig:IMAGES-classes}) but only about a dozen independent kinematic spatial points are measurable for each galaxy. (HST imaging was available for the entire sample at much better resolution.) The classification was tested using simulated maps of each galaxy. A handicap of working in this redshift range is that the strong emission lines ([OII, \Ha) used to probe the kinematics are in the optical region, where currently AO systems either do not work or deliver negligible Strehl. So it is not even possible to observe sub-samples with AO (as for example SINS did at $z\sim 2$). As AO systems improve and work at bluer wavelengths this may be remedied in the future.

Flores et al. construct a \TFR\ and their most important conclusion was that the large residual scatter identified in slit surveys arose from the new kinematic PR, CK classes. The \TFR\ for the RD class alone shows reduced scatter comparable to the local relation. The RD relation also shows no detectable zero point offset from the local stellar mass \TFR\ of \cite{Ver01}, this is in contrast to previous slit-based work. The authors attribute this to the strong evolution in kinematic classes and the inability of slit surveys to distinguish these classes as the kinematics is only measured along a single slice through the galaxy. The RD class does appear to have a significantly
higher velocity dispersion and consequent lower $v/\sigma$ than local galaxies \citep{Puech07} echoing the trend found in $z\sim 2$ galaxies. The PR class extends this trend to even lower $v/\sigma$ values.

\begin{figure}[t] 
\centering
\iftablet
   \includegraphics[width=12cm]{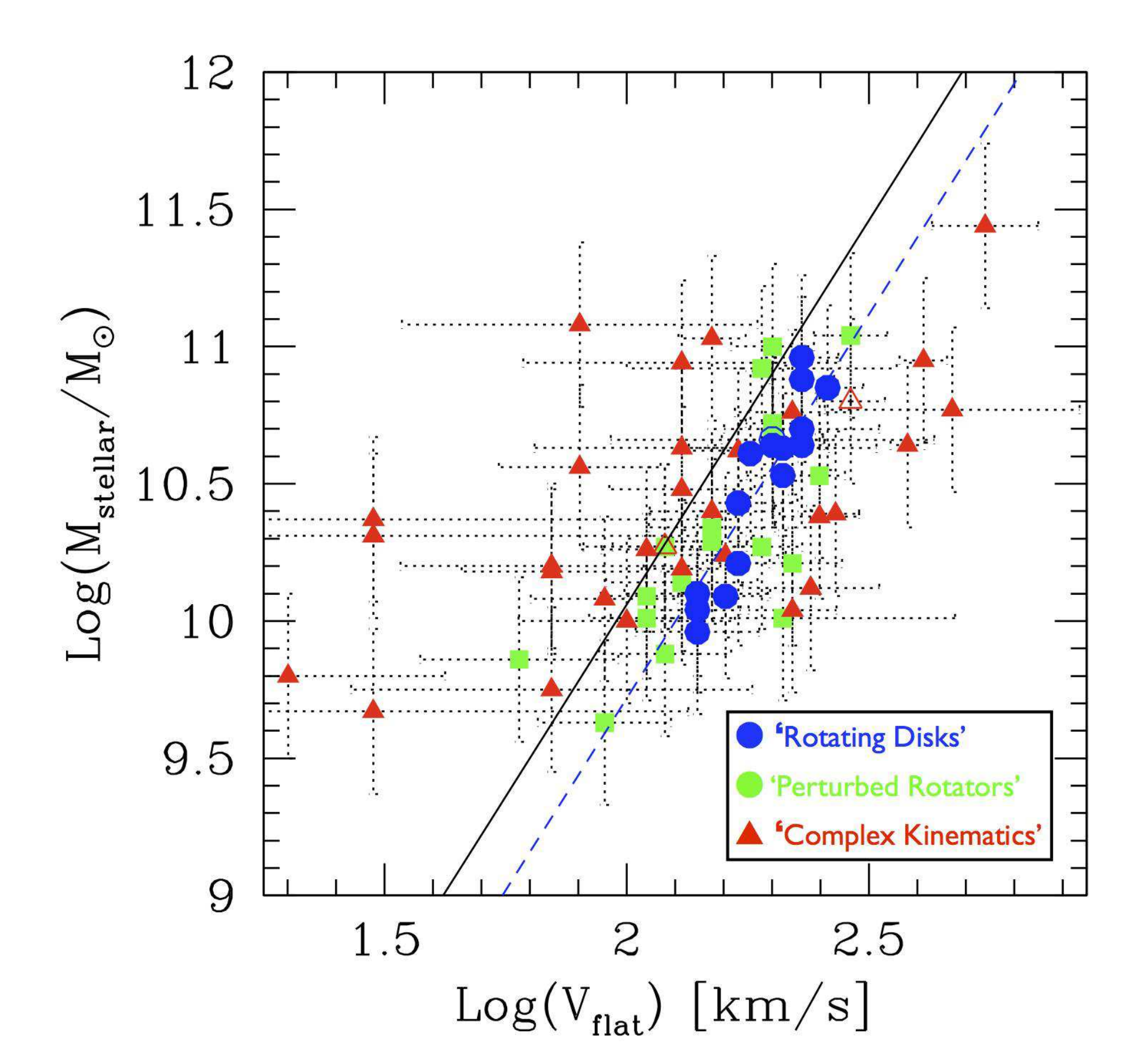}
\else
   \includegraphics[width=7.5cm]{figure-IMAGES-TFR.pdf}
\fi
\caption{\small Stellar mass \TFR\ at $z\sim 0.6$ from the IMAGES survey showing the dependence of the increase of scatter as the kinematic
class goes from regular discs to objects with irregular kinematics. {\it Credit:  adapted from Figure 3 (left panel)  of \cite{Puech10},  reproduced with permission \copyright\ ESO. } } 
\label{fig:IMAGES-TFR}
\end{figure}

The IMAGES large program \citep{Yang08} was an extension of this earlier FLAMES-GIRAFFE work to double the sample size to 63 galaxies over a similar redshift range. From an $I$-band selected input redshift survey of galaxies with [OII] emission, they are down-selected by rest-frame $J$-band luminosity, corresponding to an approximate  stellar mass limit of $>1.5\times 10^{10} \Msun$ at the redshift of the survey.
Yang et al. confirmed the evolution of the kinematic class fractions, with similar values to those quoted above. \cite{Neichel08} examined the relation between morphological and kinematic classes and found a very strong correlation between the RD objects and galaxies that appeared in HST images as spiral discs. The \TFR\ was explored in more detail by \cite{Puech08} who reaffirmed the earlier conclusion that the increase in scatter about the mean relation was due to the `non-relaxed' PR and CK classes (the scatter increases from 0.1 to 0.8 dex from RDs to CKs; shown in Figure~\ref{fig:IMAGES-TFR}). However, with a bigger sample and an improved analysis and a revised local
reference\footnote{The local stellar mass relation was based on the $K$-band one of  \cite{Ham07}, which they derive from the SDSS relation of \cite{Pizagno07}. Hammer et al.
examine the Verheijen relation (which is also the basis of the \cite{BdJ01} relation) and conclude that it is biased and the SDSS relation is more reliable.}, they now found a modest amount of zero point evolution in the K-band \TFR\ (about 0.34 dex or a factor of two in stellar mass at fixed velocity since $z\sim 0.6$). The high star-formation rate of $z\sim 0.6$ galaxies implies they are likely to be much more gas rich than local spirals. \cite{Puech10} tried to incorporate this gas in to the mass budget by inverting the Kennicutt-Schmidt relationship \citep{KSR} between gas and star-formation surface density\footnote{A recent review of such 'Star-Formation Laws' in nearby galaxies is presented by \cite{SFL-review}.} and construct a {\it baryonic} \TFR. They find that the zero point of this relation does not evolve, that galaxies in their sample have approximately equal stellar and gas masses, and hence 
conclude that the evolution of the stellar mass \TFR\ simply reflects the conversion of this gas in to stars since $z\sim 0.6$.

\subsection{The MASSIV survey}

The {\it Mass Assembly Survey with SINFONI} (MASSIV) sample is an IFS survey at $0.9<z<1.8$ of 84 galaxies, 11 with AO-LGS \citep{MASSIV1}. Selection is from the VVDS redshift survey \citep{VVDS} either using the [OII] emission line strength or rest-frame UV luminosity at the higher redshift end, and with a hierarchical selection scheme (`wide', `deep' and `ultra-deep' VVS parent samples). Early
results from preliminary samples were presented on kinematic classification \citep{Epinat09}.
The full survey description of Contini et al. shows a comparison in the star-formation rate-stellar mass main sequence plane with other samples (reproduced in Figure~\ref{fig:surveys}). The distribution of star-forming galaxies at $1<z<2$ is reasonably sampled by MASSIV, though of course their might be biases (e.g. against dusty star-formers without UV or line emission) and there is a deficit of the very massive 
($>10^{11}\, \Msun$) star-forming galaxies sampled by SINS at $z>2$.

\cite{MASSIV2} presents an analysis of the kinematical distribution. After considering multiple possible classification parameters (strength of velocity shear, 
kinematic/morphological alignment, residuals to disc fits,  velocity dispersion maps, presence of companions --- B. Epinat, 2013, private communication), the team
settled on two principal classification dimensions. The  first was between `rotators' (44\%) and `non-rotators' (35\%) with the remaining 21\% not having sufficient signal:noise to classify.  The second was between isolated and merging  / interacting galaxies, the latter make
up  29\% of the entire sample but 
it is important to note  that there is some overlap (e.g. some rotators are interacting). This categorisation is rather different to the classifications done in the other surveys (e.g. SNS, IMAGES) where for example rotators and mergers are exclusive categories. This partly arises from the fact that the identification of mergers in MASSIV comes from the presence of multiple components (separated
spatially and kinematically) in their IFS images, this is different from the approach of identifying irregular velocity maps. That said, there is a considerable overlap between the non-rotators and mergers (about half of non-rotators are classified as interacting vs only 20\% of rotators) and the isolated non-rotators tend to be smaller. Thus, if one were to think of this in terms of the disc:merger:dispersion-dominated trichotomy of other surveys, the fractions
are similar --- a roughly three way split.  \cite{MASSIV5} present a more detailed analysis of the merger rate in the sample, taking advantage of the wide-field of the SINFONI IFS ($\sim 70$ kpc at $z\sim 1.3$) to systematically define the close pair fraction by spatial proximity and separation
in redshift. This is a unique IFS science application; imaging surveys can not determine the association along the line of sight and long-slit observations
do not cover enough sky area to find non pre-selected secondary objects. Of course the IFS approach does require the companion to have emission lines above a 
detection limit, as such they are only sensitive to gas-rich mergers. They found a merger fraction of $\simeq 20\%$ across a range of redshift; using a time-scale model this was translated in to a merger rate and cumulative merger number for massive galaxies over $0<z<1.5$. I discuss 
the merger rate and the comparison with other techniques in more depth in Section~\ref{sec:merger-rate}.

The `rotator' classification is made by considering fractional residuals from a fitted disc model vs alignment between kinematic and morphological axis (I discuss this further in Section~\ref{sec:merger-disc}). Rotating galaxies are found to be larger and have higher stellar masses and star-formation rates (typically by 
a factor of two in each), a result similar to other surveys.  The typical disc velocity dispersion is found to be $\sim 60$ \kms. Comparing with the SINS/AMAZE/LSD samples at higher redshift, and the lower redshift IMAGES and the local GHASP \citep{GHASP} samples, evidence is
found for a smooth evolution in disc local velocity dispersions. Interestingly, similar dispersions are found for rotators and non-rotators,
with the latter having a strong {\it anti-correlation\/} between size and dispersion. 

\cite{MASSIV4} presents the \TFR\ and size-velocity scaling relations and again compare with the IFS samples at different redshifts. The rotators at $<z>\, \simeq 1.2$ show consistency with a small scatter stellar mass \TFR, whilst the non-rotators depart radically from this. The question of evolution 
depends on which local \TFR\ is assumed (an issue also highlighted by \cite{Puech08}), but the comparison with \cite{Pizagno07} suggests a $-0.36$
dex evolution of the zeropoint fairly similar to that found by SINS \citep{Cresci09} at $z\sim 2$, consistent with the idea of discs increasing their stellar mass with time at a fixed $v_c$. Consistent gas fractions were found using both the \KSR\ and the difference between dynamical and stellar mass. The 
baryonic \TFR\ does not appear evolved since $z=0$ similar to the findings of Puech et al. Size-velocity evolution in MASSIV appears modest (at most 0.1 dex smaller sizes  
at high-redshift at a given stellar mass).

\subsection{The AMAZE/LSD surveys}

The AMAZE (`Assessing the Mass-Abundances Z Evolution') and LSD (`Lyman-Break Galaxies Stellar Populations and Dynamics') are two related surveys (by substantially the same team, and usually analysed jointly) using SINFONI of galaxies at $z>3$, a substantially higher redshift than the other large surveys. AMAZE \cite{AMAZE1} targeted UV-selected galaxies (classical LBG selection) mostly at $3<z<3.7$ ($U$-band dropouts) with a
few at  $4.3<z<5.2$ ($B$-band dropouts) from deep spectroscopic surveys in the Chandra Deep Field South and performed observations in
natural seeing (0.6--0.7 arcsec PSF). LSD \citep{LSD1} employed a similar LBG selection at $z\simeq 3$ and focused on natural guide star AO observations, drawing on the large catalog of \cite{Steidel2004} to find objects near suitable AO stars. Typical magnitudes were $R\lesssim 24.5$ corresponding to a mass range of $10^{10-11} \Msun$ at $z\simeq 3$. Some lensed galaxies were also included but their analysis has not been published.

\cite{AMAZE-TFR} presented the kinematic analysis of the AMAZE/LSD samples, in particular 23 AMAZE and 9 LSD galaxies all
in the range $2.9<z<3.7$ apart from one object at $z=2.6$. They presented a
two-stage approach to identifying rotating disc galaxies: first, they fitted a simple linear velocity shear model to the IFS maps. Galaxies with 
statistically significant shear were classed as `rotating' and then subject to full disc model fitting. An advantage of this approach is that fitting a shear
requires substantially less model parameters and is more robust in low signal:nose data. This gave some quite interesting results
which shed light on the comparisons of other surveys: 10/23 of the AMAZE galaxies but only 1/9 of the LSD galaxies were rotators. In my
view, this is quite a significant difference given the very similar and comparable selection and observations of the AO and non-AO samples
and the previous comparisons of the SINS (mostly non-AO, larger fraction of rotation-dominated discs) and OSIRIS (all AO, mostly dispersion-dominated) samples. 
It is likely
that at least part of this is due to the greater sensitivity of the non-AO observations to the low surface brightness extended, rotating, outskirts of disc galaxies as
Gnerucci et al. note. There was no overlap between AO and non-AO samples.

For the rotators, estimates of dynamical mass were constructed from the modelling and were consistent with large gas fractions (up to 90\%) when compared to the stellar mass. Gnerucci et al. compared this to the gas masses inferred another way by inverting the \KSR\ and found a plausible 1:1 correlation.
The $v/\sigma\sim 2$ values of the rotators were a factor of two less than that of SINS discs at $z\sim 2$, which seems consistent with the higher gas
fractions compared to $z\sim 2$ in the framework of the `turbulent gas-rich disc' model (see Section \ref{sec:turbulent-discs}). The \TFR\ of the discs was consistent
with a large $-1.0$ dex decrease in the stellar mass at a given velocity relative to local galaxies, substantiality more than at $z\sim 2$,
but with a very large scatter ($\sim 0.5$ dex).

Other AMAZE/LSD papers considered the evolution of the stellar mass:global metallicity relation \citep{FMR, LSD1} and the discovery of positive metallicity gradients (i.e. metal poor galaxy centres)
in a sub-set of galaxies \citep{Cresci10}. 

\subsection{Other optical/near-infrared IFS surveys at $z>1$}
\label{sec:other}

As well as the large surveys, several smaller projects should be mentioned, these tend to probe complementary parameters spaces. 

In particular in the $1<z<2$ regime AO is possible, but difficult, since one must target \Ha\ in the near-infrared $H$-band where it has reduced Strehl. \cite{Wright07,Wright09} present OSIRIS AO kinematics of seven galaxies at $1.5<z<1.7$ UV-selected and with prior optical
spectroscopy using the BM/BX criterion. They find  
four of these to have kinematics consistent with disc systems and high intrinsic velocity dispersions ($> 70$ \kms) in at least two of these. \cite{Wis11}
presents OSIRIS AO kinematics of 13 galaxies a $\sim 1.3$, again selected by rest-frame UV emission and optical spectroscopy, but selected from a wider area survey 
probing higher UV luminosities and star-formation rates higher than more typical $z \sim 1.3$ galaxies (but comparable to $z>2$ SINS disc). They again find that around half the objects have disc-like kinematics and high intrinsic velocity dispersions and clumpiness. The resolved star-formation properties and clump scaling
relations were examined further in \cite{Wis12}.
An interesting difference between the Wright et al. and Wisnioski et. al. samples
lies in the nature of the non-disc candidates --- in the first they are extended objects with multiple sources of resolved \Ha\  emission and irregular kinematics whereas
in the latter they tend to be single compact sources of \Ha\ emission with mostly dispersion-dominated kinematics (i.e. similar to the Law et al. objects at $z>2$ illustrated in Figure~\ref{fig:DD}). It is not clear if this reflects the different selection, luminosity and/or space density, or simply the signal:noise of the data. More luminous \Ha\ objects are easier to map; however, the actual effect seems reversed in that the fainter non-disc sources of Wright et al. tends to have more extended \Ha\ morphologies.

An alternative selection technique to the broad-band-selected surveys mentioned above is via the use of narrow-band imaging which has the advantage that the 
galaxies are already known to have the strong emission lines required for successful IFS observations. \cite{SHiZELS1} observed with SINFONI 14 \Ha\ selected emitters
at redshifts 0.8, 1.5 and 2.2 corresponding to the wavelengths of their narrow-band filters of the parent imaging sample ('HiZELS', \cite{Sobral2009,Sobral2013}), and with a stellar mass range similar to the SINS survey. AO IFS maps were obtained for nine of these galaxies, five of which were classified as discs ($+$ two mergers and two compact galaxies), again very similar fractions to other surveys. The stellar mass \TFR\ was examined showing a factor of two evolution in mass at a fixed velocity since $z\lesssim 2$. The discs themselves 
were physically very similar to SINS discs in that they had high dispersion, low $v/\sigma$ values as well as  clumpy star-formation and high gas fractions \citep{SHiZELS2}. One particularly interesting point was that two of the objects with AO observations were at $z=0.84$, though the reported Strehl was low ($\simeq 10\%$) as is normal with current systems for $J$-band observations. Notably this is the only AO IFS observations I know of for galaxies at $0.3<z<1$.

\begin{figure}[t] 
\centering
\iftablet
   \includegraphics[width=12cm]{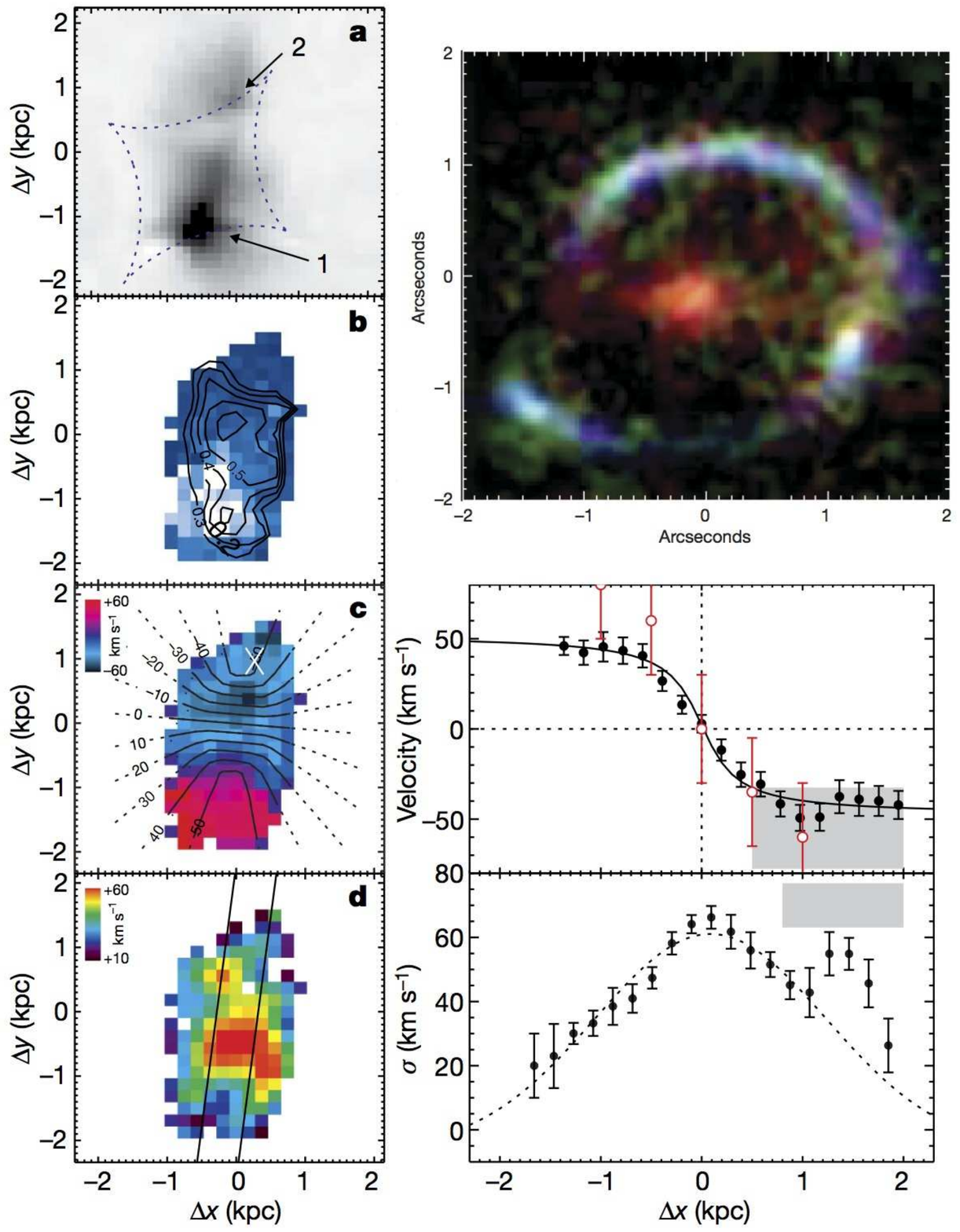}
\else
  \includegraphics[width=7.5cm]{figure-stark.pdf}
\fi
\caption{\small A beautiful example of a small disc galaxy at  $z=3.07$ with dynamical mass $\sim 2\times 10^9 \Msun$ and star-formation rate 
$\sim$40 $\Msun$ yr$^{-1}$
from \cite{Stark08} lensed 28-fold and demonstrating a
near-complete Einstein Ring
observed at $\sim$100 pc resolution with the assistance of gravitational
lensing and AO. Maps on the left show (a)  lens reconstructed  rest-UV continuum ($\sim$1500\AA) emission, (b) [OIII] 5007]\AA\ line emission 
(with contours showing \Hb), (c) velocity map and rotation curve showing a characteristic `spider diagram' and (d) dispersion map and curve (tilted
lines show extraction axis).  (See Stark et al. for full figure details.) The galaxy is clumpy in 
continuum and line emission but is a clear disc with a turnover and high dispersion in the kinematics. 
The top-right panel shows the original sky plane image (composite
red: $K$-band, green: [OIII], blue: HST $V_{606}$ filter) known as the `Cosmic Eye' with the red central source being the $z\sim 0.7$ lens.
{\it Credit: adapted from Figures 1 \& 2 of Stark et al. (selected panels and combined), reprinted  by permission from Macmillan Publishers Ltd: \href{http://www.nature.com}{Nature}, 455, 775 \copyright\ 2008.  } } 
\label{fig:stark}
\end{figure}

An especially powerful combination has been to combine AO IFS observations with the gravitational strong lensing effect of giant clusters at intermediate redshift which can often magnify background galaxies by factors of up to 10--50 (see the review of \cite{strong-lensing}). Such strong lensing only occurs over limited sky areas, and the objects most magnified tend to be the faint but numerous objects not probed by other surveys. A key question is: do the sensitivity and coarser resolution limits of the non-lensed surveys give us a biased view of the high-redshift population? The lensed surveys also allow us to probe $z>3$ and smaller spatial scales. \cite{Stark08} (Figure~\ref{fig:stark}) and \cite{Jones10} consider a sample of six lensed sources magnified up to 50$\times$ at $1<z<3$. They find a much higher-incidence of rotating, high-dispersion discs (4/6) than in the more luminous sources probed by unlensed surveys and in all cases the galaxies are resolved in to multiple emission line clumps. \cite{Yuan11,Yuan12} present observations of two more objects 
which again seem consistent with the picture of high dispersion, low $v/\sigma$ clumpy discs. The highest redshift examples to date are two lensed $z\sim 5$ galaxies \citep{Swin07,Swin09} observed in [OII] and with very low dynamical masses ($10^{9-10} \Msun$) and  velocity shears ($<100$ \kms) compared to the other surveys but still  relatively large velocity dispersions ($\sim 80$ \kms). 
A particular benefit of the gravitational lens observations is the use of the high linear magnification to probe the size of star-forming clumps. Currently, only lensing can deliver $\sim$100 pc spatial resolution of high-redshift galaxies, resolution is critical for accurate size measurements  and hence
testing the  picture of large high-Jeans mass clumps in unstable discs \citep{Jones11,Liv12}. I will return to this scenario in Section~\ref{sec:turbulent-discs}.
Two more notable lensed objects are presented in (a) \cite{Nes06} of a giant arc at $z \sim 3$ whose de-lensed kinematics suggests a rotating disc and 
(b) \cite{Nes07} a lensed sub-mm galaxy with merger-like kinematics.

In addition to the MASSIV survey, other samples of galaxies selected from the VVDS sample have been observed with SINFONI (non-AO).
\cite{Lem10a} present a sample of ten $1.0<z<1.5$ galaxies selected on their [OII] emission, finding eight rotating high-dispersion discs, 
one clear merger, and one object with no kinematic 
variation interpreted as face-on. They split the discs almost equally between `rotation dominated' and `dispersion dominated' around $v/\sigma = 1.66$ which
appear to follow relatively offset stellar mass \TFR\ relations (and both evolved from the local relation). \cite{Lem10b} select three intermediate
stellar mass (1--$3\times 10^{10}\,\Msun$)  $z\sim 3$ galaxies from VVDS
based on their rest-frame UV VVDS spectra and observed in the near-IR in \Hb, [OIII] lines. They have very high star-formation rates for this redshift  --- as a result
of being selected as $I<24$ in VVDS they are brighter in the rest frame UV than typical $z\sim 3$ galaxies. All three have high dispersion and small shears ($v/\sigma \lesssim 1$), one was tentatively classified as a merger based on anomalous kinematics, and the other two were consistent with rotating disc models. However, interestingly, both of the latter displayed secondary components consistent with close companions. The typical velocity shears are small ($< 50$ \kms)
and they argue the properties of the sample are very similar to those of the Law et al. objects at $z\sim 2$. Another $z\sim 3$ LBG observed with SINFONI is presented by
\cite{Nes08}, this is interpreted as a merger.

\subsection{Sub-mm line surveys}

All of the surveys presented so far, and a majority of the discussion, has focussed on 2D kinematics measured using rest-frame optical emission lines observed in the near-infrared. A change from this and an interesting development has been the first kinematic measurements at high-redshift using sub-mm wavelength lines, so far of the CO molecule. 

High-redshift star-forming galaxies are rich in molecular gas and dust. In particular, among the massive star-forming galaxies, we see a population of `sub-mm galaxies' \citep{Blain2002} with strong emissions at these frequencies due to star-formation rates up to 1000 $\Msun$/yr per year (e.g. \cite{SMG-SFRs}). A strong correlation of star-formation approximately
proportional to stellar mass is observed at high-redshift (the `star-forming main sequence'  \citep{main-sequence,Daddi2007} but the 
classical `sub-mm galaxies' may lie above this relation and may represent rare events
such as major mergers \citep{Daddi2010}. 
Main sequence massive star-forming galaxies at $z\sim 2$ have up to 50\% gas fractions \citep{Tacconi10,Tacconi12} several times higher than local massive spirals. 

There are now over 200 total molecular gas measurements at high-redshift (e.g. see review of \cite{CarriliWalter13}), although the spatial resolution is usually rather coarse (0.5--1.0 arcsec) due to the baseline limitations of current sub-mm interferometers; however, this does allow some kinematic measures for larger galaxies. Early work by 
\cite{Gen03} using the IRAM Plateau de Bure Interferometer  modeled the CO kinematics of a sub-mm-selected galaxy as a large rotating disc. \cite{Daddi2008}  observed a more normal main-sequence galaxy and showed that it was disc-like.
The `PHIBSS' CO survey of 52 main sequence star-forming galaxies  \citep{Tacconi10,Tacconi12} at $z\sim 1.2$ and 2.2 
found that 60\% where kinematic discs and that the CO velocity dispersions were high and agreed with the \Ha\ values.
This is an important point as molecular gas is likely to dominate the mass budget with ionised gas being only a small fraction.
A detailed spatial comparison of \Ha\, optical, NIR and CO data was performed for one of these galaxies by \cite{Gen13}. They found that  \Ha\ and CO traced the same rotation curve and also evidence for variable dust extinction, an important caveat to be considered when
interpreting optical maps. 

\begin{figure}[t] 
\centering
\iftablet
   \includegraphics[width=12cm]{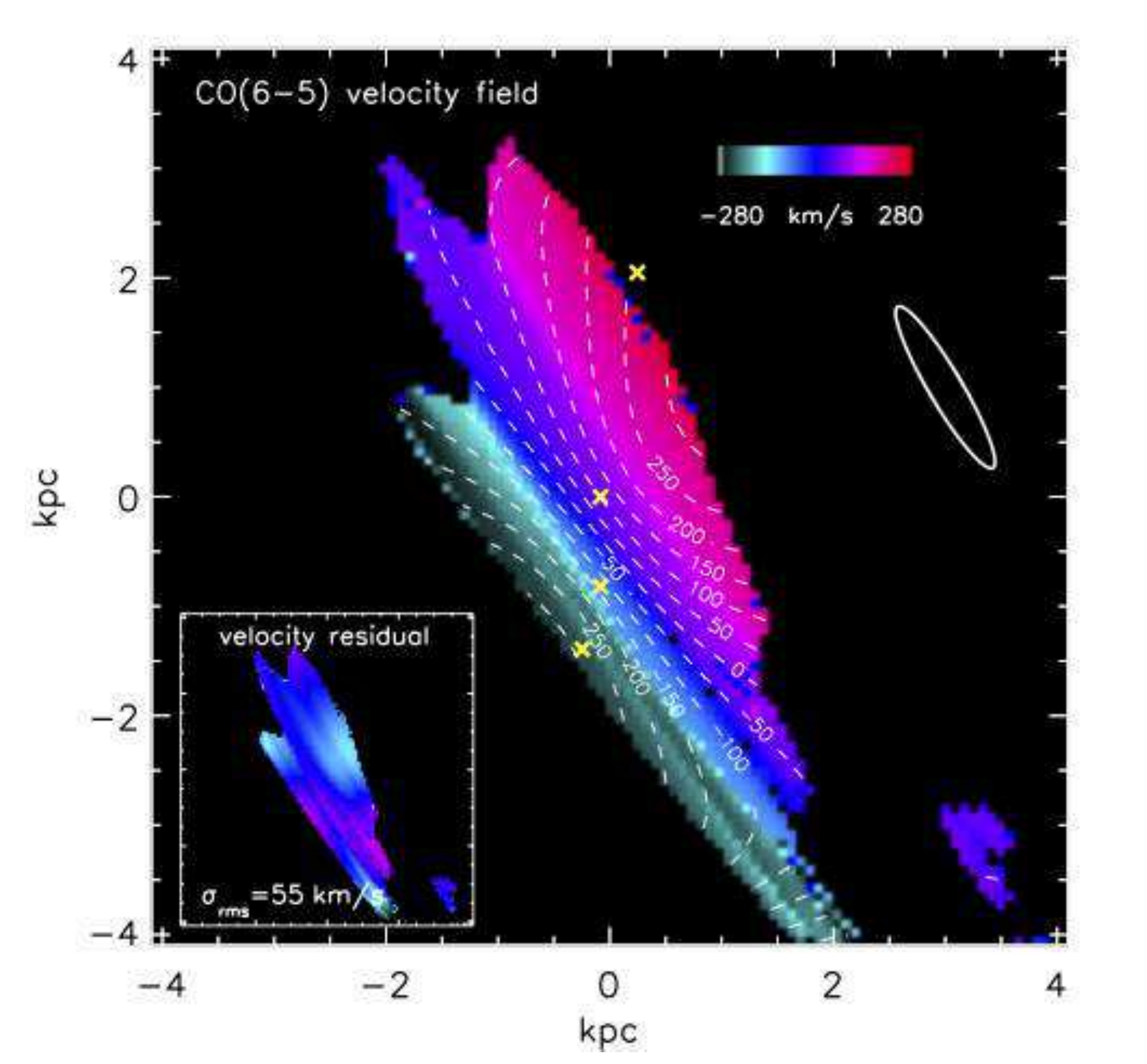}
\else
   \includegraphics[width=7.5cm]{figure-COmap.pdf}
\fi
\caption{\small Resolved CO velocity map of lensed $z=2.32$ sub-mm galaxy SMM J1235-0102 reconstructed in the source plane. This is one of only two published well-resolved molecular line velocity maps of a high-redshift disc galaxy. The effective lensing PSF (which is
anisotropic) is shown as the white ellipse at the top right.  Contours are of velocity and the yellow crosses are the locations of the star-forming clumps. The galaxy is well fit by a disc model, the inset shows the residuals. {\it Credit: from Figure 4 (top panel) of \cite{Swin11-CO}, reproduced by permission of the AAS.} } 
\label{fig:COmap}
\end{figure}

There are only a handful of cases in the literature with kpc resolution and these are mostly objects with a gravitational lensing boost to the resolution. 
\cite{Swin11-CO} presented 
CO line observations of the $z=2.32$ lensed sub-mm galaxy SMM J1235-0102 which has a 
total star-formation rate of $\sim 400\,\Msun$/yr (about 10$\times$
the `main sequence' value for it's mass). The beam was $\sim 0.5$ arcsec  and with the 30-fold lensing boost resolution of 100 pc was obtained. Despite
the extreme star-formation rate the CO kinematics showed that the molecular gas was distributed in a turbulent rotating disc (see Figure~\ref{fig:COmap}) with
$v/\sigma \sim 4$ consistent with the picture inferred of other massive $z\sim 2$ star-forming galaxies from ionised gas.  
\cite{Hodge12} analysed a single $z=4.05$ very bright sub-mm galaxy where the CO lines are redshifted down to 
the higher radio frequencies.  Using a wide Very Large Array spacing and 120 h
of integration, they made a map at $0.2$ arcsec resolution which revealed a clear disc of dispersion $\sim 100$ \kms\ with clumpy molecular gas (clump masses $\sim 10^9\Msun$). 
In contrast to these results,  \Ha\  kinematics of $z\sim 2$ sub-mm galaxies have instead found that they mostly have  complex velocity fields with multiple components showing distinct kinematic offsets (\cite{Nes07,SMG-Kinemetry} and notably \cite{SMG-clumps} the first with AO). The origin of this difference between kinematics in CO vs \Ha\ is not clear but is likely due to 
the small numbers of objects involved and heterogenous selection. Of course the optical/near-IR selected general star-forming galaxy populations are also diverse but are somewhat
better characterised.

With the ongoing deployment of ALMA \citep{ALMA}, we can expect such observations to become routine, and expand to non-lensed samples of non-extreme objects, in the next few years. 

\subsection{Recent multislit surveys}
\label{sec:mslit}

While my main consideration in this review is IFS kinematic surveys in the last decade, it is necessary to also mention some of the important
kinematic results from the more recent contemporaneous slit-based surveys as these have sampled much larger numbers of high-redshift galaxies, albeit in 1D. 
These have mostly come from the DEIMOS multi-object spectrograph on Keck \citep{DEIMOS} due to its relatively large slit mask area and high spectral
resolution for kinematics. \cite{Weiner06a,Weiner06b}
examined the \TFR\ of $\sim 1000$ galaxies at $z\lesssim 1$ in the `Team Keck Redshift Survey' using both {\em integrated} velocity dispersion (i.e. a
similar idea to \cite{Forbes1996}) and resolved rotation curve fits for the larger galaxies ($\sim$ a third of the sample) and found strong evolution in the $B-$band
(fading with time) but little in the near-infrared, with large scatter (0.3 dex) attributed to dispersion-dominated galaxies.

\cite{Kassin07} looked at the stellar  mass \TFR\ of 544 galaxies ($0.1<z<1.2$) with resolved kinematic modelling from the DEEP2 redshift survey \citep{DEEP2}, as in earlier work they found a large scatter ($\sim 1.5$ dex) at higher redshifts dominated by the more disturbed morphological classes and the lower stellar masses and echoing the results from the IFS-based IMAGES survey at similar 
redshifts discussed earlier. Kassin et al. and Weiner et al. introduce a new kinematic measure $S_{0.5} = 0.5\, v^2 + \sigma^2$,
combining rotation and velocity dispersion  and found the $S_{0.5}$ \TFR\ of all
kinematic classes showed considerably reduced scatter ($\sim 0.5$ dex) and no evolution in intercept nor slope (see Figure~\ref{fig:lowzTFR}). 
The conclusion was that at higher redshifts, 
the star-forming galaxies are increasingly supported by dispersion arising from disordered motions \citep{Weiner06a}. 
The $M$--$S_{0.5}$ relation was also found to agree with the local Faber-Jackson relation for elliptical
galaxies suggesting a possible evolutionary connection. Also using DEEP2, \cite{Lorenzo09,Lorenzo10} considered the evolution of the \TFR\ to $z\gtrsim 1$ using integrated line widths to estimate rotation
velocities of visually selected spirals; they found evolution in the $B$-band consistent with other studies but not in the $K$-band. However, when they consider the required
evolution in $K$-band mass:light ratio with time, they conclude that stellar mass may have doubled at fixed velocity in the last 8 Gyr. 

The recent DEIMOS survey of \cite{Miller11} has provided a different, possibly conflicting, perspective on \TFR\ evolution. They observed
 only 129 $0.2<z<1.3$ galaxies 
but unlike previous surveys, which were typically 1--2 h spectroscopic exposures,  they took much longer 6--8 h exposures and took care to align
the slits to within 30$^\circ$ of the HST-derived galaxy major axis. Like previous studies, they find evolution in the blue but little in the stellar mass \TFR, but 
interestingly they report a smaller scatter of only $\sim 0.06$ dex in $\log_{10} v$ (0.2 dex in stellar mass), a factor of two less than in previous surveys. This they attribute to their longer exposures which, for what they call {\it `extended emission galaxies'}, means they can reach the flat-portion turnover in 90\% of their galaxies and place all of them on a tight \TFR\ and without requiring an extra parameter such as $S_{0.5}$. This seems in 
contradiction to the IFS results (primarily of the IMAGES survey) in the same
redshift range. Nearly half the IMAGES sample are the CK class which contribute $\sim$0.8 dex of scatter and do not have regular disc-like kinematics. It 
also seems to conflict with the kinematic fractions in the larger, but shallower, survey of Kassin et al. with the same spectrograph.

If the samples are broadly comparable, there is definitely a contradiction. It is important to note that Miller et al. only recover rotation velocities for 60\% of their targeted sample (the remaining 40\% are too compact in emission or have no emission) and that they did not target 20\% of their input sample as, again, being too compact. It seems unlikely though that pure sample effects can explain the discrepancy completely, as many of the Miller et al. galaxies have the peculiar/disturbed 
morphologies characteristic of other samples. Perhaps the explanation is that  deeper  observations of `CK objects' show large-scale rotation? (And it can not simply be deeper observations revealing shear from merging components as one would not then expect them to lie on the \TFR). The IMAGES
survey also used 4--15 h exposures, though it is expected that an IFS instrument may have less throughput and the FLAMES-GIRAFFE sampling
was relatively coarse. Interestingly, Miller et al. did observe three of the actual CK 
galaxies from the \cite{Flores06} sample, noting the velocities were consistent within the IFS area. Comparing the tabulated properties of the objects in common,
I note that Miller et al. report masses 0.8--0.9 dex less for these same three objects, with velocity agreement for two. These  particular CK objects lie fairly close to the IMAGES \TFR\  compared to other CK objects, and all show distinct velocity shears in the maps of \cite{Yang08} and so may be mis-classified. They may not be comparable
with the other CK objects; a proper comparison would require more overlap. 

\cite{Miller12} extends their sample to $1.0<z<1.7$ taking advantage of a newly installed extra-red sensitive CCD in the LRIS spectrograph on Keck \citep{LRIS-Red}. 
They successfully
detected extended emission and measure rotation curves in 42 galaxies (out of 70 observed) and report a virtually non-evolving stellar  \TFR\ at these redshifts (in conflict with \cite{MASSIV4}), again with small scatter (\cite{Miller13} attributes residual scatter to bulgeless galaxies at $z>1$ following an offset |\TFR). Again there seems to be a conflict in kinematic classification, half the MASSIV sample at
similar redshifts were classified as non-rotating based on 2D IFS data (albeit several times shorter exposure times). 
One must consider that in these slit surveys, the slit angle must {\it a priori} be chosen from imaging data, it seems unlikely that one could choose this correctly
to align with the rotation axis as would be required to make a tight \TFR\ relation, given that MASSIV \citep{MASSIV2} reports that at least half of their
sample are highly misaligned. Is it possible that kinematic and photometric alignment could only be revealed at low surface brightness on large scales? Is it really
possible to determine kinematic axes photometrically from the clumpy morphologies of galaxies at these redshifts? The implications of the disagreements apparent in the literature are not yet clear. 

These questions aside, there does seem to be  general agreement between IFS and slit surveys on the increasing contribution of internal velocity dispersion to disc kinematics. \cite{Kassin12} (based on the same DEEP2 sample) report  a continuous increase in dispersion to $z=1$, matching with the IFS samples at the same, 
or higher, redshift and consequent decrease in $v/\sigma$. Interestingly, by defining a `disc settling criteria' of $v/\sigma>3$  (a value which they claim
correlates with normal vs disturbed physical and kinematic morphologies), they find a `kinematic downsizing'
trend with stellar mass in the sense that high-mass galaxies `settle' at earlier times (for example 50\% of $10.3<\log(M/M_\odot)<10.7$ galaxies are settled
at $z=1$ compared with 90\% at $z=0.2$). 

\subsection{`Local analogue' samples}
\label{sec:analogues}

A couple of groups have published IFS kinematics of rare samples of nearby galaxies that are possible analogues of high-redshift populations.

The `Lyman-Break Analogues' (LBAs) are galaxies at $z\sim 0.2$ selected as Lyman dropouts from space-UV observations from the GALEX satellite. In particular, \cite{LBA} define a population with very similar UV luminosity, stellar masses, and star-formation rates to $z\sim 3$ LBGs and divide them in to `compact' and `large' categories based on UV size and surface brightness. The large LBAs have stellar masses of $\sim 10^{11}\,\Msun$, lower surface brightnesses and sizes of 
up to 10 kpc, the compact LBAs are typically a factor of ten less massive and sizes $<2$ kpc.  They display similar colours and metallicities to the LBGs
\citep{LBA-O} and similar morphologies dominated by large clumps of star-formation \citep{LBA-morphs, LBA-clumps}.

At these modest redshifts, it is possible to do AO observations using the Paschen-$\alpha$ line in the $K$-band, which is not possible at zero redshift as it falls
in the absorption trough between the $H$ and $K$-bands. For Case B recombination \citep{CaseB} Pa-$\alpha$ is  $12\%$ the intensity of H$\alpha$ (in the absence of dust --- any extinction would make the ratio more favourable) however it is easily detectable in such nearby galaxies with excellent spatial resolution. A high-luminosity subset of the compact population (dubbed `supercompact') has been followed up by AO IFS using
OSIRIS \citep{LBA-BZ,LBA-G} and reveal themselves to be excellent analogues to the dispersion dominated galaxies studied by Law et al. at $z\sim 2$. They have high ionised gas dispersions (50--130 \kms), some evidence of small rotation/kinematic shears and low $v/\sigma \lesssim 1$ all very similar to the properties of the Law et al. sample. This was
confirmed  by carrying out an `artificial redshifting' computation to simulate the appearance of the galaxies to OSIRIS and SINFONI at $z\sim 2$.
\cite{LBA-morphs} and \cite{LBA-G} concluded that LBAs are mainly mergers based on HST morphology and OSIRIS kinemetry.

More recently, \cite{Green10} and \cite{Green13} analysed an IFS sample of nearby ($z\sim 0.1$) but rare galaxies selected on their high \Ha\ luminosity from SDSS spectra. In galaxies with $L(H\alpha)>10^{42}$ erg/s they made kinematic maps at $\sim 2$ kpc resolution in \Ha\ (natural seeing observations) and identified galaxies with high ionised gas dispersion($>$ 50 km/s), about two-thirds of which were discs. This high-incidence of rotation, 
the large stellar masses (up to  $10^{11} \Msun$ and large sizes (2--10 kpc) suggest that they could be more similar to $z\sim 2$ discs than LBAs; 
however, further work and higher spatial resolution observations (see discussion in \cite{Davies11}) are required to confirm this.

Another interesting set of local analogues are `tadpole' galaxies which have a  `single clump $+$ tail' morphology. 
These were first identified at high-redshift by
\cite{vdB96} where
their incidence is higher. A handful have since been identified locally in the SDSS survey \citep{Straughn06,UDF-tadpoles}. \cite{local-tadpoles}  found these to constitute 0.2\% of UV bright surveys (compared 
to 6\% of high-z galaxies; \cite{Straughn06}) and have stellar masses 
$\lesssim 10^9\, \Msun$; they attribute the morphology to lop-sided star-formation. The clumps have masses of $10^{5-7}\Msun$; the galaxies appear to resemble
scaled-down high-redshift tadpoles. The tadpoles have high \Ha\ velocity dispersion and show evidence for marginal rotation dominance \citep{tadpole-kinematics}. Yet another class of  rare low mass galaxies which might be similar to high-redshift objects are the `green peas' \footnote{The name denotes their compact, unresolved, green appearance in SDSS with the colour arising from the particular combination of strong emission lines, 
redshift and SDSS filter set.} first discovered by public volunteers inspecting SDSS images in the
Galaxy Zoo project \citep{green-peas}.  These are very compact (2--3 kpc) low mass ($10^8$--$10^{10}\Msun$) but with high star-formation rates ($>10$--$30\, \Msun$ yr$^{-1}$), low metallicities
and have complex kinematics with velocity dispersions of 30--80 \kms\ \citep{Amorin12} suggesting similarities (apart from the substantially lower stellar masses) to the `dispersion-dominated' objects seen  at high-redshift (see Section~\ref{sec:DD}). Only a limited amount of high-resolution HST imaging has been done but reveals clumpy morphologies. IFS 
observations are needed (for example to compare $v/\sigma$).

A final point to remember in considering such `local analogue' samples is that one is inherently selecting rare and unusual populations nearby, which are then being compared to the bulk galaxy population at high-redshift. It is quite possible that physical processes that are rare locally,
such as mergers, may dominate such selections and make a comparison misleading. The advantage of course is that a much greater wealth of 
multi-wavelength and high spatial and spectral resolution follow-up observations are available than at high-redshift to test physical models. A
simple example is using deep imaging to test for tidal tails from mergers, which could be too low surface brightness to be seen at high redshift.

\section{Analysis Techniques in IFS surveys}
\label{sec:techniques}

In this section, I will review some of the primary analysis techniques employed in IFS kinematic surveys at high-redshift. As can be seen from the discussion in the previous section,
some of the key issues the surveys are tackling are:

\begin{enumerate}
\item Extraction of kinematic maps.
\item Measuring the rotation curve and circular velocities (ideally the near flat post-turnover portion by some quantitative definition) of disc galaxies.
\item Objectively classifying discs from mergers.
\item Measurements of intrinsic velocity dispersion and higher-order moments of spectral lines.
\item Calculation of dynamical mass.
\item Identification of sub-galactic structures (e.g. star-formation complexes or merging galaxies) and measurement of their physical properties.
\end{enumerate}

Quantitative measurements are of course desirable and a variety of numerical techniques, many of which are new, have been devised to reduce IFS kinematic maps to a few basic parameters. All high-redshift observations are subject to limited signal:noise and spatial resolution, the best techniques allow for possible biases from such effects to be measured and corrected for --- or be built in to the methodology. 

Before launching in to the discussion of techniques, it is worth making some specific points about AO vs non-AO observations. While AO
offers greater spatial resolution, a price is paid in the loss 
of  light and signal:noise (for fixed integration times) through several principal effects:
\begin{enumerate}
\item AO PSFs are divided in to two parts: a sharp `core' and a broad `halo', where only the sharp core is corrected and contributes high-resolution information. The faction of light in this core is given by the Strehl factor
which is typically 0.3--0.4 in the $K$-band and 0.1--0.3 in $J$ and $H$ with current technology. 
\item AO optical systems have a substantial number of additional optical elements which reduces throughput. 
\item AO optical systems are usually located in front of the instrument in a non-cryogenic environment and hence generate extra thermal background which reduces signal:noise. 
 \item AO observations necessitate finer pixel scales which introduces additional read noise in to the system which cannot be removed by post-binning. 
 \end{enumerate}

Of course, AO observations reveal more about detailed structure resolving higher surface brightness features  and for brighter more compact objects this can be critical for kinematic modelling and classification. Ideally one would use AO and natural seeing observations on the
same objects and compare the results (e.g. \cite{Law-Nature,NewDD}, discussed further in 
Section~\ref{sec:DD}). Future work may combine both datasets, for example one can imagine a joint maximum-likelihood approach to disc fitting where the natural seeing data was used for faint, diffuse galaxy outskirts and complementary AO data  used for the 
bright central regions.

\subsection{Making maps}
\label{sec:maps}

As a first step (after data reduction to calibrated cubes), almost all analyses start out by making 2D maps of line intensity, velocity, and dispersion from 3D data cubes, this is a type of {\it projection}.
The basic technique overwhelmingly used is to fit Gaussian line profiles in the spectral direction to data cube spaxels. The mean wavelength gives the velocity and the standard deviation the dispersion. The integral gives the line intensity. Typically, this can be done robustly when the integrated signal:noise (S/N) per resolution element being fitted is greater than a few, for example \cite{NFS09} used S/N$>$5, \cite{Flores06} used S/N$>$3.

In order to estimate the dispersion map, it is necessary to remove the contribution from the instrument's spectral resolution. For resolved kinematics resolutions $R\gtrsim 3000$ are normally 
considered suitable (noting this is independent of redshift). Normally the instrumental resolution is subtracted `in quadrature', meaning:
\begin{equation}
\sigma_{gal} = \sqrt{ \sigma_{obs}^2 - \sigma_{instr}^2}
\end{equation}
(e.g. \cite{SHiZELS1}) which is formally correct as the two broadenings are independent. However, in the case of low signal:noise and/or low dispersion this becomes problematic, if the best fit has
$\sigma_{obs} < \sigma_{instr}$ due to noise then the quadratic subtraction can not be done, and when this happens it is ill-defined. A better approach that avoids
this problem \citep{NFS09,Davies11,Green13} is to start with a model instrumental line profile, 
and broaden it by convolving it with Gaussians of different $\sigma_{gal}$ (constrained to be $>0$) 
until a good fit to the observed profile is achieved. This can also handle non-Gaussian instrumental profiles and gives more realistic errors.

Of course, making projections necessarily loses some of the information in the original cube, for example asymmetries and non-Gaussian wings on line profiles which can convey 
additional information on instrumental effects (such as beam-smearing) as well as astrophysical ones (such as infalls and outflows of material). At high-redshift, lack of signal:noise
means these higher order terms can't currently be measured well anyway for individual spatial elements, however with future instruments and telescopes this will not remain true.

\subsection{Measuring rotation curves}
\label{sec:disc-fitting}

For galaxies identified as discs from kinematic maps, perhaps the most important kinematic measurement is to fit a model velocity field to extract rotation curve parameters. This allows construction of the Tully-Fisher relation at high-redshift for comparison with models of disc galaxy assembly.

In samples of nearby galaxies with long-slit optical observations, rotation curves with high-spatial resolution are constructed piecewise (i.e. binned velocity vs coordinate along the slit axis); the maximum velocity can be either read of directly (e..g. \citep{MAT92}) or a 
rotation curve model is fitted to it (i.e. a $V(r)$ function) (e.g. \cite{SS90,Courteau97,Catinella2005}). There is a necessary assumption that the slit is along the principal kinematic axis. For 2D IFS observations (radio HI or Fabry-Perot emission line cubes in the early days) the `tilted ring' approach has become standard (e.g. \cite{tilited-rings,Schommer1993}), where each ring measures the velocity at one radius and also represents a piecewise approach. At high-redshift, different approaches have been used primarily due to two factors (i) the lower
signal:noise does not allow complex models with numerous parameters to be fit and (ii) the limited spatial resolution (often 5--10 kpc for a natural seeing PSF at $z>1$) means `beam smearing' effects are more severe and comparable to the scale of the underlying galaxy itself.

A standard approach to fitting `rotating disc models' to 2D IFS data of high-redshift galaxies has emerged and been adopted by different groups and the essentials consist of:

\begin{enumerate}
\item Model the rotation curve assuming some simplified parametric $V(r)$ function, with essential parameters being a kinematic spatial scale (in kpc) and velocity (usually corresponding to the flat part of the curve). 
\item Allow the rotation curve parameters, and galaxy inclination, and orientation (PA) to vary.
\item Model the galaxy photometric profile --- almost invariably as an exponential discs with the scale length as a free parameter.
\item Combine the kinematic and photometric parameters to make a model galaxy. 
\item Convolve the model galaxy with the PSF.
\item Minimise with respect to the data using some metric such as $\chi^2$ or maximum likelihood on the 2D velocity map and search for a best fit solution and errors on parameters. For IFS data, one would normally use the velocity maps and for slit data the velocity profile; other maps can provide additional constraints.
\item Extract a $V_{max}$ parameter from the best fit de-projected disc model.
\end{enumerate}

This approach was originally developed for fitting long-slit data of high-$z$ galaxies (see Section~\ref{sec:longslit}) and was an outgrowth of similar techniques being used in 2D galaxy photometry with the Hubble Space Telescopes \citep{Schade95}.

\begin{figure*}[t] 
\centering
\iftablet
   \includegraphics[width=19cm]{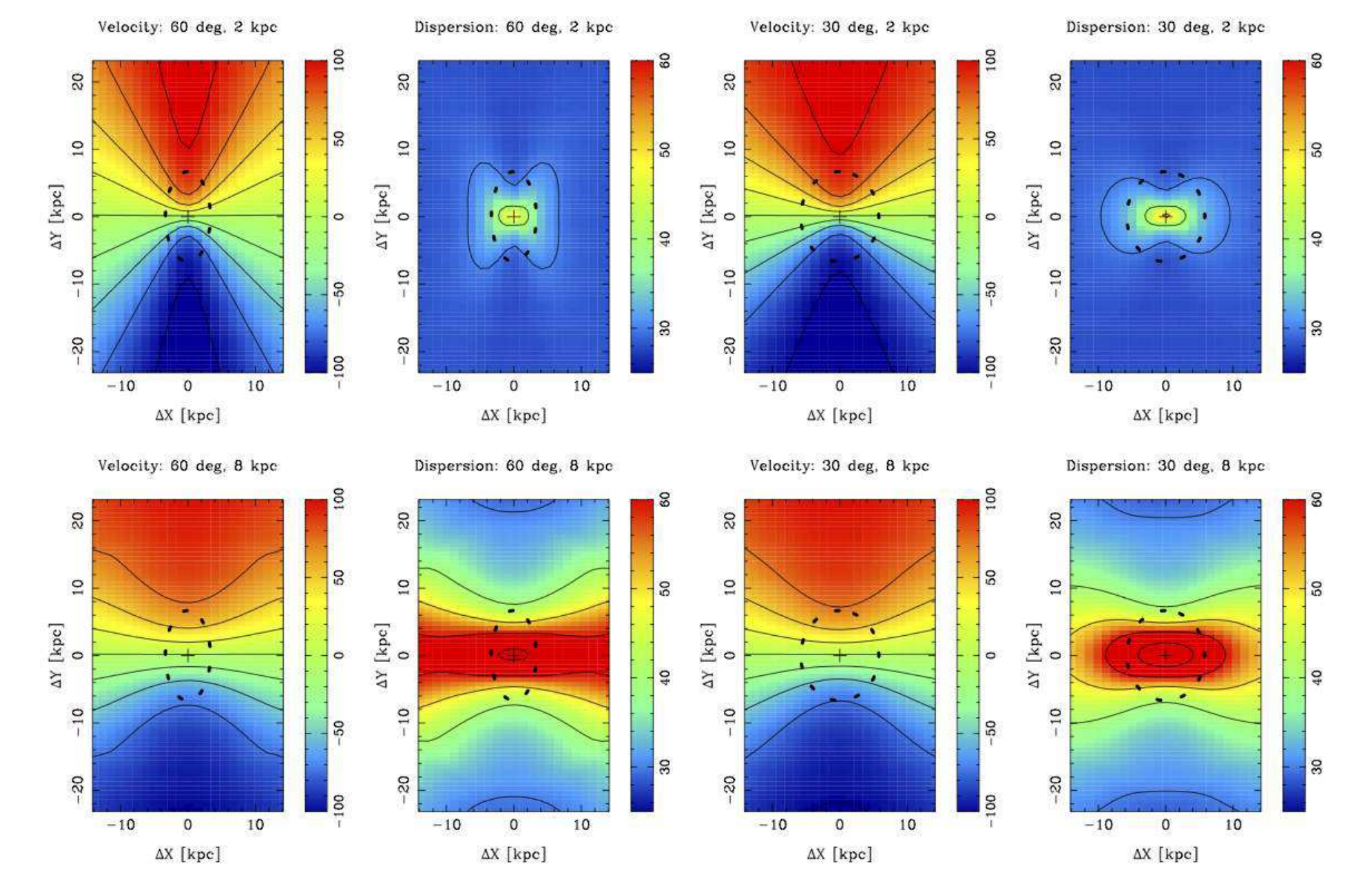}
\else
   \includegraphics[width=16cm]{figure-diskmodels.pdf}
\fi
\caption{\small Model disc galaxy velocity and dispersion fields at inclinations of 30$^\circ$ and 60$^\circ$. The assumed galaxy model is an exponential disc in \Ha\ emission with scale length $h=3$ kpc (the heavy dashed ellipse shows the extent at 2.2$h$) and a rotation curve taken from Equation~\ref{eq:arctan} with $r_d  = 1$ kpc.
The top row is for a spatial resolution of 2 kpc (i.e the FWHM of a Moffat PSF) and the bottom row is for 8 kpc (a coarse resolution representing typical $z>1$ natural seeing observations) and the intrinsic spectral resolving power is 7000.
Models at different inclinations are  defined to have constant $V_{max} \sin i = 110$ \kms\  to illustrate this approximate degeneracy in velocity maps and the intrinsic dispersion is 20 \kms. Contours run linearly from  $-100$ to $+100$ \kms\ in velocity (25 \kms\ steps)  and 30 to 70 \kms\ in dispersion (10 \kms\ steps). Note that the maps are projected from the underlying 3D  disc model by fitting Gaussians to the spectral line profile as is standard for IFS observations, the high dispersion central peak is the result of beam-smearing, which is significantly worse at 8 kpc resolution, and the elongated high-dispersion bar arises from the Gaussian being a poor representation of the beam-smeared line shape. This can be
accounted for in 3D disc fitting (i.e. summing $\chi^2(\hbox{RA},\hbox{DEC},\lambda)$) and this is in fact done by the code used to produce this figure.  {\it Credit: kindly provided by Peter McGregor (2013). }} 
\label{fig:disc-models}
\end{figure*}

The key advantage of this approach is the explicit inclusion of beam-smearing via the PSF convolution step, this leads to more unbiased best fit parameters. However, it is 
necessary to assume an underlying photometric model because the convolution with the PSF will mix velocities from different parts of the galaxy according to their relative luminosity.  

The largest disc galaxies at high-redshift have effective radii of 5--8 kpc \citep{Labbe2003,Buit08}, comparable to typical natural seeing PSFs, however the exponential profile is relatively slowly declining so that useful
signal:noise is obtainable on scales of 2--3 arcsec, this is why the method works. AO data provides higher spatial resolution but at a considerable cost in signal:noise. Natural seeing data has proved surprisingly more successful than AO in revealing discs at high-redshift and it is thought this is due to its greater sensitivity to extended lower surface brightness emission at the edges of galaxies. The most successful AO projects have used very long exposures ($>5$ h per source). 
That said, it can been seen that there are numerous galaxies at high-redshift that do not show rotation, and these may simply be instances whereas they are too small compared to 
natural seeing to resolve and too faint to be accessible to AO.

Variations on this core technique abound and is useful to review them. First, there is the choice of $V(r)$ function. The two most commonly used analytic choices 
for high-redshift analyses are (i) the `{\tt arctan}' function:
\begin{equation}
\label{eq:arctan}
V(r) = V_{max} \frac{2}{\pi} \hbox{arctan}\left( \frac{r}{r_p} \right) 
\end{equation}
of \cite{Courteau97} which is an analytic form that expresses a profile initially rising smoothly with a `kinematic scale radius' $r_p$ \citep{Weiner06a,Puech08} and
smoothly transitioning to a flat top; and (ii) the linear ramp function:
\begin{equation}
V(r) = \frac{1}{2} \, V_{max} \times  \left\{ \begin{array}{rl}
r/ r_p &\mbox{ if $r<2 r_p$} \\
1 &\mbox{ if $r\ge 2 r_p$}
\end{array} \right.\end{equation}
which has a sharp transition\footnote{In the way I have expressed both of these at $r_p$ the velocity is half of the maximum value.}
\citep{Wright09,MASSIV2,SHiZELS1,Miller11}. Both have exactly two free parameters though the ramp model
better fits artificially redshifted simulations of high-redshift galaxies \citep{GHASP} as it reaches its asymptote faster. 
Usage of the ramp function does make
it clearer if the presence of any turnover (i.e $V_{max}$) is well constrained by the data. 

The other common approach to $V(r)$ is to assume a mass
model, usually a thin exponential disc, and integrate this up \citep{Cresci09,AMAZE-TFR}. The solution for an ideal infinitely thin 
exponential disc  is given by \cite{Freeman1970} (his Equation 12 and Figure 2) in terms of modified Bessel functions of the 
first ($I_n$) and second kind ($K_n$):
\begin{equation}
V(r) = \left( \frac{2 G M}{h_r} \right)^\frac{1}{2}  x \left[  I_0(x) K_0(x) - I_1(x) K_1(x)  \right]^\frac{1}{2} 
\end{equation}
where $M$ is the disc mass, $h_r$ is the disc scale length, and $x=r / 2 h_r$. This
has a maximum velocity peak (with a shallow decline at large radii) at  $2.15 h_r$ which is commonly called `$V_{2.2}$' and $V_{2.2}/2$
occurs at $0.38 h_r$. More complex functions can arise from multiple components, for example at low-redshift the `Universal Rotation Curve' formula
of \cite{URC} approximates an exponential disc $+$ spherical halo and they arrive at a luminosity dependant shape.\footnote{This dependence gives rise to circularity issues in Tully-Fisher applications \citep{Courteau97}.}
The large velocity dispersions at high-redshift also motivates some authors to include
this support in relating the mass model to the rotation curve (e.g. \cite{Cresci09}). 

These variations can cause issues when trying to consistently compare {\TFR}s, for example some authors may choose the {\tt arctan} function but then evaluate it at 2.2 scale lengths (e.g. \cite{Miller11}). Finally I note that it is often useful to fit a pure linear shear model (i.e. like a ramp functions but with no break) 
\citep{Law09,Wis11,MASSIV2}. Because it has one less free parameter, it does not need a centre defined, and is not changed by beam-smearing it can be very advantageous
for low signal:noise data and as a means of at least identifying  candidate discs (see Section~\ref{sec:merger-disc}). 

The choice of underlying photometric model is also important, because of the PSF convolution this affects how much velocities in different parts of the galaxy are mixed in the observed spaxels. An accurate PSF is also obviously vital and this can be problematic for AO data. Since the kinematics is being measured in an emission line such as \Ha\ then one needs to know the underlying \Ha\ intensity distribution to compute this correctly --- however, this is not known as one only observes it smoothed by the PSF already so this is essentially a deconvolution problem. Most authors simply assume the profile is exponential (which may be taken from the IFS data or separate imaging)
which is potentially problematic as we know high-redshift
galaxies have clumpy star-formation distributions  and possibly flat surface brightness profiles \citep{ELM-SB2,ELM-SB,Wuyts2012}. It may not make much of a difference if the galaxy is only marginally resolved  as the PSF (which is much larger than the clump scale) dominates for natural seeing data (but see \cite{Gen08} who investigate and compare more complex $M(r)$ mass models with the best resolved SINS galaxies). A different, arguably better, approach is to try and interpolate the intensity distribution from the data itself \citep{MASSIV2,Miller11}. The photometric profile is also usually used to estimate the disc inclination and orientation for the deprojection
to cylindrical coordinates, this is ideally done from HST images but is sometimes done from the projected data cube itself. Trying to determine the inclination from the kinematics is particularly problematic
(though \cite{Wright09} and \cite{SHiZELS1} do attempt this) as there is a strong near-degeneracy between velocity and inclination. One
can only measure the combination $V \sin(i)$  except in the case of very high signal:noise data where the curvature of the `spider diagram' becomes apparent
(a well known effect, e.g. \cite{Beg89} Section A2). I demonstrate this explicitly in Figure~\ref{fig:disc-models}.   
The final key choice is the matter of which data is used for the fitting process. Obviously one must use the velocity map and associated errors, and most
authors simply use that with a $\chi^2$ or maximum likelihood solver. One also normally has a velocity dispersion map and can also use this \citep{Cresci09,Weiner06a}, the dispersion maps contains information which constrains the beam-smearing (via the PSF convolution); however, one must make additional assumptions about the intrinsic dispersion (e.g.
that it is constant). This case arises naturally if using a dispersion-supported component in a mass model. 

Fitting disc models of course requires well-sampled high signal:noise data and of course the model needs to fit. With noisier data, fitting shears is a simpler approach, another  simple parameter is to simply recover some estimate of $V_{max}$ from the data cube \citep{Law09,NFS09}. If we expect discs to have a flat rotation curves in their outskirts then the outer regions will reproduce this maximum value with little sensitivity to exact aperture. Often the maximum pixel
or some percentile is used. One also knows that for a random distribution of inclinations $\left<V \sin i\right >$ $=\left<V\right> \left<\sin i\right >$ and $\sin i$ 
is uniformly distributed with an average of $\pi/4$ \citep{Law09}. However, the use of a maximum may be subject to pixel outliers and the use of a limiting 
isophote may not reach the turnover (this is true for model fitting too but at least one then knows if one is reaching sufficiently far out). 
An unusual hybrid approach to fitting adopted by the IMAGES survey \citep{Puech08} motivated by their relatively coarse IFS sampling
was to estimate $V_{max}$ from essentially the maximum of the data within the IFU, but use model fitting to the velocity map to calculate a correction from the data maximum to $V_{max}$ which they justified via a series of simulations of toy disc models.  

All the papers in the literature have fit their disc models to 2D projections such as the
velocity map; however, in principle it is possible for perform the same fitting in 3D to the line cube. This may provide additional constraints on aspects such as beam-smearing via
the line profile shape (e.g. beam-smearing can induce asymmetries and effects in projection as is shown in Figure~\ref{fig:disc-models}). This last approach has been tried in radio astronomy on HI data of local galaxies but is computationally
expensive \citep{TiRiFiC}.
Fitting algorithms also require a good choice of minimisation algorithm to find the lowest $\chi^2$ solution given the large number of free parameters. Commonly steepest descent type algorithms are used; \cite{Cresci09} used the interesting choice of a genetic algorithm where solutions are `bred' and 
`evolved'. \cite{Wis11} used a Monte-Carlo Markov Chain approach which allows an efficient exploration of the full probability 
distribution (and marginalization over uninteresting parameters). In my view, this approach, which is common in for fitting cosmological parameters,
is potentially quite interesting for future large surveys as in principle it could
allow errors from individual galaxies to be combined properly to compute global quantities such as the circular velocity distribution function. 

Given the variety of choices in disc fitting approaches by different authors,
it is desirable to compare these using simulated galaxies and/or local galaxies (with well-measured kinematics) artificially degraded
to simulate their appearance at high-redshift and explore systematics such as PSF uncertainty. This has not been done comprehensively, but
a limited comparison was done by \cite{GHASP} using data
from a local Fabry-Perot survey in \Ha\  of UGC galaxies and simulating their appearance at $z=1.7$ in 0.5 arcsec seeing; however, this was primarily focussed on
evaluating beam-smearing effects. They did conclude that the galaxy centre and inclination are best fixed from broad-band high-resolution imaging and
that using the simple ramp model statistically recovered reliable $V_{max}$ values more often than other techniques for large galaxies (size $>3\times$ seeing). 
They also argue that the velocity dispersion map adds little constraining power to the disc fit.

\subsection{Dynamical masses}

In the absence of 2D kinematic data and modelling, the `virial estimator' for dynamical mass:
\begin{equation}
M_{dyn} = C \sigma^2 r / G
\end{equation}
where $\sigma$ is the integrated velocity dispersion and $r$ is some measure of the size of the object has often been used \citep{Erb:2006,Law09,NFS09,Lem10a}. In this case, $\sigma$ represents unresolved velocity contributions from both pressure and rotational support and $C$ is a unknown geometric factor of $O(1)$ ($C=5$ for a uniform rotating sphere, \cite{Erb:2006}). 
This is cruder than 2D kinematics in that kinematic structure and galaxy inclination are ignored, in a sense the goal of 2D kinematics is to reliably measure $C$. However it can be applied to larger samples.

A related novel method is the use of the technique of `spectroastrometry' to measure the dynamical masses of
unresolved objects. This technique was originally developed to measure the separations of close stellar pairs \citep{spectroastrometry} and allows relative astrometry to be measured
at accuracies very much larger than the PSF limit. In the original application, it relies on measuring the `position spectrum', i.e. the centroid of the light along the long-slit as a function of wavelength. As one crosses a spectral line, with different strengths and shapes in the two unresolved stars, one measures a tiny position offset. Because this is a differential technique as
very close wavelengths systematic effects (e.g. from the optics, the detector and the PSF) cancel out and the measurement is essentially limited by the Poisson signal:noise ratio. Accuracies of milli-arcsec can be achieved in natural seeing. \cite{Gner10} developed an application of spectroastrometry 
for measuring the masses of black holes and extended this \citep{Gner11} to galaxy discs in IFS measurements.
The technique here now involves the measurement of the position centroid (now in 2D) of the blue vs red half of an emission line as defined by the integrated spectrum and mean wavelength. 
In the presence of rotation, there is a small position shift.
Unlike stars galaxies are complex sources and there can be systematic effects, for example if the receding part of the disc has a different clumpy \Ha\ distribution than the approaching part.
The classical Virial mass estimator  ($M_{dyn} \sim r \sigma^2 / G$) requires 
a size measurement $r$ which is difficult for unresolved compact objects and is often taken from 
associated HST imaging for high-redshift galaxies; this gets replaced by the spectroastrometric
offset $r_{spec}$. Gnerucci et al. find the spectroastrometric estimator gives much better agreement ($\sim$ 0.15 dex in mass) than the virial estimator for high-redshift galaxies with good dynamical masses from 2D modelling. They also argue from simulations that it ought to work well for unresolved, compact galaxies. It is certainly a promising avenue for further work and could help, in my view, resolve the nature of dispersion-dominated compact galaxies. 
Spectroastrometric offsets do require modelling to interpret, however the presence of a position offset is a robust test 
for the presence of unresolved shear irrespective of a model.

\subsection{Velocity dispersion Measures}
\label{sec:dispersion-measures}

A key discovery is that is has been consistently found that high-redshift galaxies have higher intrinsic (i.e. resolved) velocity dispersions than local galaxy discs.\footnote{I re-emphasise that in this section I am {\em not} talking about dispersions of integrated spectra.} Thus, it is necessary to reliably measure this quantify, and in particular derive some sort of `average' from the kinematic data. In fact, one approach has been to define a simple average:
\begin{equation}
\sigma_a = \frac{ \Sigma_i \, \sigma_i } {N_{pix}}
\end{equation}
over the pixels, as used in \cite{AMAZE-TFR,MASSIV2}. One can also define a flux or luminosity weighted average:
\begin{equation}
\sigma_m = \frac{ \Sigma_i \, f_i \sigma_i } {f_i}
\end{equation}
\citep{Law09,Green10} which is less sensitive to low S/N pixels and exact definition of outer isophotes. 

The observed dispersion will include a component of instrumental broadening which must be removed (see Section~\ref{sec:maps}) but will also include a component 
from unresolved velocity shear (such as might be caused by systematic rotation). 
The PSF of the observation will cause velocities from spatially nearby regions to be mixed together and if these are different
then this will show up as an increased velocity dispersion called `beam-smearing'. 
This will get worse for spatial regions with the steepest velocity gradients and for larger PSFs. Brighter spatial regions will also dominate
over fainter ones so the effect also depends on the intrinsic flux distribution. One method to correct for this beam smearing is to
try and compute a `$\sigma$ from beam-smearing' map from the intensity/velocity maps (interpolated to higher resolution) and then
subtracting this in quadrature from the observed map \citep{AMAZE-TFR,MASSIV2,Green10}. Use of $\sigma_m$ and $\sigma_a$ has the advantage that they 
are non-parametric estimators and have meaning even when the dispersion is not constant (i.e. they are averages).

Another approach to calculating the intrinsic dispersion is to incorporate it in to the disc modelling and fitting approaches discussed in Section~\ref{sec:disc-fitting}. For example,
in their dynamical modelling, \cite{Cresci09} incorporated a component of isotropic dispersion (they denote this the `$\sigma_{02}$' parameter), which they then fit to their velocity and velocity dispersion maps jointly. In this way, 
the PSF and the beam-smearing are automatically handled as it is built in to the model. Their model maps are effectively constant and $\simeq \sigma_{02}$ 
except in the centre where there is a dispersion peak due to the maximum velocity gradient in the exponential disc model (e.g. see Figure~\ref{fig:GenzelObject} examples). 

\cite{Davies11} compared these different approaches to calculating dispersion using a grid of simulated disc galaxies observed at different spatial resolutions (and inclination etc.) 
similar to high redshift surveys. 
In particular, they concluded that the method of empirically correcting from the intensity/velocity map is flawed as that map has already been smoothed by the PSF,which leads to less apparent shear than is really present, 
and that the $\sigma_m$ and $\sigma_a$ type estimators are highly biassed even when corrected. They also concluded that the disc fitting approach
was the least biased method for estimating $\sigma$. However, I note that the underlying toy disc model used for the simulated data 
closely agrees with the model fitted to the data by construction, so this is not in itself
surprising. A parametric approach such as fitting a disc with a constant dispersion may also be biased if the model assumptions are wrong --- for example, if the dispersion is not in fact constant or the rotation curve shape is incorrect. It is clear though that use of $\sigma_m$ in particular should be with extreme caution as it is one of the most sensitive of
the dispersion estimators to beam smearing in high-shear galaxies. The parametric approaches tend to underestimate the dispersion at low signal:noise and the non-parametric ones
to over-estimate it.  The dispersion measures of \cite{Green10}, \cite{AMAZE-TFR} and \cite{MASSIV2} may be biased by the effect Davies et al. discusses, the degree to which will depend on how well the parameters of the galaxies
in question reproduce those chosen in the simulations of Davies et al.  and this is yet to be quantified.

\subsection{The merger/disc classification}
\label{sec:merger-disc}

One early goal of high-redshift IFS surveys was to try and kinematically distinguish modes of star-formation in high-redshift galaxies. It was already known that star-formation rate was
typically factors of ten or more higher at $1<z<4$ than locally \citep{Madau96,Lilly96}, and that massive galaxies ($\sim 10^{11}\Msun$) in particular were much more actively forming stars \citep{Bell05,Juneau05}. Models of hierarchical galaxy assembly a decade ago were typically predicting that mergers (as opposed to in-situ star-formation)
were the dominant source of mass accretion and growth in massive high-redshift galaxies \citep{Som01,Cole00}. Is it possible that the increase in cosmic star-formation
rate with lookback time is driven by an increased rate of merger-induced starbursts?

In photometric surveys on-going mergers have been identified by irregular morphology \citep{Con03,Con08,Bluck12}, however this is not always definite because as we will see in Section~\ref{sec:turbulent-discs} it is now established that many disc-like objects at high-redshift appear
photometrically  irregular as their star-formation is dominated by a few large clumps embedded in the discs. Thus,
it is desirable to additionally consider the kinematics. Another popular technique has been to count `close pairs' in photometric surveys and/or redshift surveys (i.e either `close' in 2D or 3D) and then calibrate how many of them are likely to merge on a dynamical time scale via simulations \citep{Lotz08,KW08}. This technique may fail for a late stage merger when the two components are not well separated any more. 

A number of techniques have been used to try and differentiate between galaxy-galaxy mergers and discs in kinematic maps. As can be seen from Section~\ref{sec:surveys},
some high-redshifts surveys have found up to a third of their targets to have merger-like kinematics so this is an important issue. 

The first and most widely used technique is
simply visual classification using either the velocity and/or dispersion maps. One expects a disc to have a smoothly varying clear dipolar velocity field along an axis, and to be symmetric about that axis. At high signal:noise, one would see a `spider diagram' type pattern. The
dispersion field would also be centrally peaked if the rotation curve was centrally steep and beam-smearing was significant. One might expect a merging second galaxy
component to distort the motions of the disc, one would also expect to see a discontinuous step in the velocity field when one transitions to where the second galaxy
dominates the light. A good example of a $z=3.2$ merger with such a step is shown in \cite{Nes08} -- their Figure 3.\footnote{One may also expect the spectral line ratios to change abruptly if, for example, the galaxies had different metallicities.} Such visual classifications have been used in \cite{Yang08}, \cite{NFS09} and \cite{Law09}. Of course,
such visual classifications are subjective and also susceptible to signal:noise/isophote levels.  In imaging surveys, the equivalent `visual morphologies' have often been exhaustively tested by comparing different astronomer's classifications against each other and against simulations as a function of signal:noise, this has not yet been done for kinematic
classifications.

Turning to algorithmic methods to classify galaxy kinematics, the most popular technique has been that of `kinemetry' (the name is an analogy of `photometry') which tries to quantify asymmetries in velocity and dispersion maps. Originally developed by 
\cite{Kinemetry} this was used to fit high signal:noise local elliptical galaxy IFS observations, but has been adapted to high-redshift  $z\sim 2$ SINS discs by \cite{Shap08}. By analogy to surface photometry kinemetry proceeds to measure the 2D velocity and velocity dispersion maps using
azmimuthal kinematic profiles in an outward series of best fitting elliptical rings. The kinematic profile as a function of angle $\theta$ is then
expanded harmonically. For example:
\begin{eqnarray}
K(a,\theta) =&  A_0(r) +  A_1(a) \sin(\theta) + B_1(a) \cos(\theta) \nonumber \\
& +  A_2(a) \sin(2 \theta)  +  B_2(a) \cos(2 \theta) +\cdots
\end{eqnarray}
where $a$ would be the semi-major axis of the ellipse (which defines $\theta=0$). This is of course equivalent to a Fourier transformation, the terms
are all orthogonal. When applied to an ideal disc galaxy we expect
(i) the velocity field should only have a single non-zero $B_1$ terms since due to its  dipolar nature  it goes to zero at $\theta = \pm \pi/2$, with $B_1(a)$ representing
the rotation curve (ii) similarly the symmetric dispersion map should
only have a non-zero $A_0(a)$ term representing the dispersion profile. When higher terms are non-zero, these can represent various kinds of disc
asymmetries (bars,warps, etc.) or just arise from noise. The primary difference between the high signal:noise local application and the low signal:noise high-redshift
application of Shapiro et al. is that in the latter a {\it global\/} value of the position angle and inclination (ellipticity) is solved for instead of allowing it to vary in each ring. These are found
by searching over a grid of values to find those which essentially minimise the higher-order terms.

\begin{figure}[t] 
\centering
\iftablet
   \includegraphics[width=14cm]{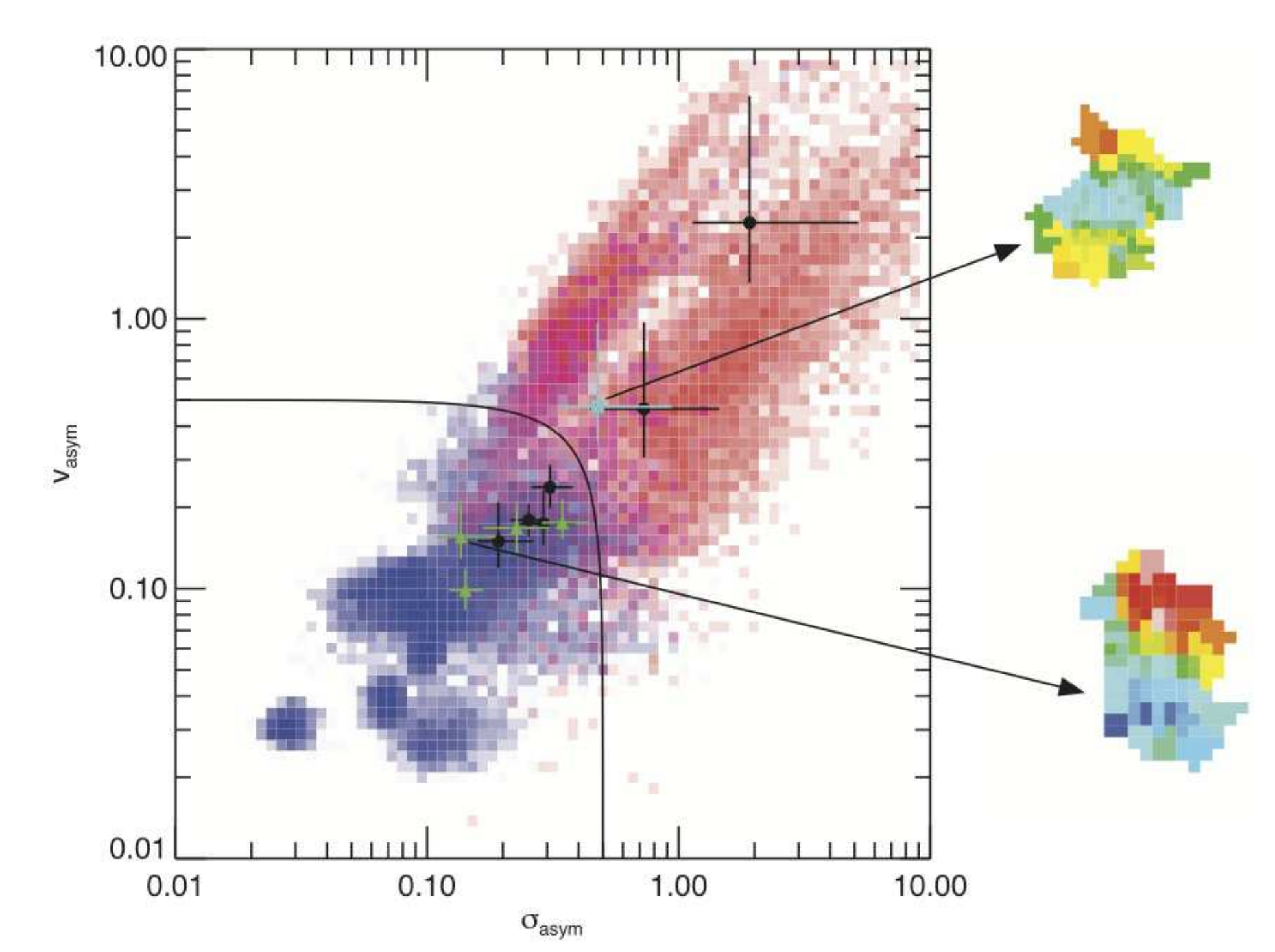}
\else
   \includegraphics[width=7.5cm]{figure-kinemetry.pdf}
\fi
\caption{\small Kinemetry diagram classifying SINS galaxies. Axes are the velocity and dispersion asymmetry (as defined
in the main text) with the line showing the proposed disc/merger boundary  $K_{asym}=0.5$. The points
are the SINS objects classified by Shapiro with the outset velocity diagrams  showing two sample objects classified as a disc (bottom object) and a merger (top object). Note how the disc shows a  dipolar velocity field whereas the merger is more complex. The  red/blue colour scale shows the probability distribution of  {\it simulated\/} merger/disc objects at $z\sim 2$ (see Shapiro et al. for details). {\it Credit:  from Figure 7 of  \cite{Shap08}, reproduced by permission of the AAS. } } 
\label{fig:kinemetry}
\end{figure}

Shapiro et al. expanded
their kinemetry to fifth order and in particular defined an average power in higher-order coefficients (which should be zero for a perfect disc) as 
\begin{equation}
k_{avg} = \frac{1}{4} \sum_{i=2}^5  \sqrt{A_i^2 + B_i^2}
\end{equation}
then they defined velocity and dispersion asymmetry parameters as 
\begin{eqnarray}
v_{asym} & = \displaystyle \left< \frac{ k_{avg,v}(a) }{ B_{1,v}(a) } \right> \\
\sigma_{asym} & =  \displaystyle  \left< \frac{ k_{avg,\sigma}(a) }{ B_{1,v}(a) } \right>
\end{eqnarray}
(where the second subscripts denote the relevant maps) that are normalised to the rotation curve (representing mass) and
averaged across radii. The use of these particular parameters was justified by a series of simulations of template galaxies
artificially redshifted to $z \sim 2$, these templates (13 total) included toy models of discs, numerical simulations of cosmological discs
and actual observations  (15 total) of local discs and ULIRG mergers. Figure~\ref{fig:kinemetry}\ shows the location of these in the
$v_{asym},\sigma_{asym}$ diagram and Shapiro et al.'s proposed empirical division, represented by
\begin{equation}
K_{asym} = \sqrt{ v_{asym}^2 + \sigma_{asym}^2 }  = 0.5
\end{equation}

Using this approach, Shapiro et al. successfully classified 11 of the highest signal:noise galaxies in the SINS samples (see Figure~\ref{fig:kinemetry}), and concluded that $\simeq$ eight were
discs and $\simeq$ three were mergers (both $\simeq \pm 1$), agreeing with visual classification. This kinemetry technique was also applied by \cite{SHiZELS1} to their
high-z sample and they also did an independent set of simulations to verify the $K_{asym} <0.5$ criteria and found a 55\% disc fraction. \cite{LBA-G} applied this to their
sample of $z \sim 0.2$ compact dispersion-dominated `LBAs' (see Section \ref{sec:analogues}), both as observed and when artificially redshifted to $z=2.2$ (specifically simulating SINFONI in natural 0.5 arcsec seeing). 
They found a `merger' fraction ($K_{asym} >0.5$) of $\sim 70\%$, predominately for galaxies with stellar
masses $<10^{10}\Msun$, but this dropped to $\sim 40\%$ for their high-redshift simulations, i.e. a large number of mergers were misclassified as discs,
due mainly to the loss of visibility of outer isophotes and PSF smoothing. Of course these issues also affect visual classifications.

\cite{SMG-Kinemetry} applied kinemetry to a sample of nine sub-mm galaxies at $2<z<2.7$ observed with IFS. They found that essentially all of these were mergers with high asymmetries in 
both velocity and dispersion maps. \cite{Bellocchi2012} found that local LIRGS (two mergers and two discs) were reliably classified by kinemetry and simulated their appearance at high-redshift.
They advocated a modified version of kinemetry where the ellipses were weighted by their circumference which gives more weight to the outer
regions of the galaxies. An analysis of a larger number of 38 galaxies in this latter sample is forthcoming.

Some caveats are warranted, in my view particularly in the application of galaxy simulations to calibrate kinemetry. In the Shapiro et al. figure
(Figure~\ref{fig:kinemetry}) 
and the other papers which use this classification diagram
all the simulations are lumped
together; however, it would be desirable to obtain a deeper understanding of the range of applicability by separating these out. For example, considering real
galaxies and model galaxies separately, understanding the effects from the different kinds of simulation, how parameters degrade with signal:noise, the effects
of choice of radial binning, inclination and resolution and so on. A paper exploring these in detail would be of great value to the literature. 

A different approach to quantitative classification was used in the MASSIV survey \citep{MASSIV2}. They considered two classification parameters they derive from their disc fits. (i)
the mean amplitude of the velocity residuals from the disc fits (normalised by the maximum velocity shear) and, (ii) the alignment between kinematic
axes and photometric axes (determined from broad-band imaging). They identify an isolated cloud of points near the original with aligned axes 
(agreement $<20^\circ$) and 
velocity residuals $<20\%$ which they label `rotators' and constituting half their sample. They argue the results are consistent with kinemetry, however no extensive set of simulations were done to establish the reliability. Dispersion information was not considered, they argued that the velocity and dispersion
asymmetries  are well correlated anyway (this is indeed evident in Figure~\ref{fig:kinemetry}) and of course Shapiro et al. did in fact combine these in to a single parameter.  

Finally, I note that all these  approaches are more or less parametric model-fit approaches\footnote{Noting the key difference between parametric and non-parametric 
approaches is the assumption of an underlying parameterised model whose residuals are minimised with respect to the data.}  based on 2D projections, disc fitting plays a key role in that rejecting it is a basis for potential merger classifications. Kinemetry has similarities to non-parametric measures used in quantitative image morphology, 
especially in its use of an asymmetry measure which is
similar to that used in morphology \citep{Abraham96,CAS}. However, model-fitting is still performed in order to  determine a best fit inclination and PA before calculating the 
kinemetry coefficients. Soler \& Abraham (2008, private communication) investigated the use of the Radon Transform  to compute a statistic to distinguish model discs from mergers with some success. However, again this relies on statistics measured from 2D projections of the intrinsically 3D data. In principal, one can imagine deriving statistics from
the 3D intensity emission line data cubes --- a disc model makes a characteristic pattern of shapes in 3D position-velocity when viewed from different angles. However, to my knowledge 
no such approach has yet been undertaken in the literature; this is quite a contrast to 2D morphology where we have seen the use of a variety of statistics such as concentration, asymmetry,  clumpiness, \citep{CAS} `M20' and Gini \citep{Gini,Lotz04} has contributed to quantitative  study of morphological evolution.

Regardless of the techniques used, it is clear that a substantial amount of further work is required to calibrate the quantitative application of these techniques at high-redshift. Also, it is desirable to move away from the simple `merger vs disc' dichotomy which some authors have reasonably argued is an oversimplification \citep{Wis11,Law09} of continuous mass assembly by competing processes. It would be desirable to be able to apply quantitative techniques to estimate merger mass ratios and merger evolutionary stages (e.g. 
first approach, fly-by, coalescenceÉ as \cite{Puech12} attempts visually) from IFS data,  each of which have their own timescales, in order to test galaxy formation models.

\subsection{Properties of substructures}

As we will see in Section \ref{sec:turbulent-discs}, the `clumpy turbulent disc' model is emerging as a key paradigm to understanding the physical structures of at least some 
high-redshift galaxies as revealed by high-resolution imaging and IFS data \citep{ELM-clump04,ELM-clump05,Genzel06,Elm09,Bour09,Gen11,Guo12}. In this scenario, large  `clumps' which are peaks of local emission are distinct physical 
structures in a galaxy disc and tests of the clump model involve measure of their resolved spatial and kinematic properties. Additionally, we have seen that some high-redshift 
objects are thought to be advanced stage mergers, in this case the sub-components may be distinct galaxies. I will briefly review some of the techniques used
to define the physical properties of such substructures, the physical models will be discussed further in Section~\ref{sec:physics}. It will be seen that the techniques used so far have 
been very basic and unlike the total galaxy measurements, absolutely require AO (or HST observations in the case of pure morphological work) as the 1--2 kpc scales need to be resolved. The extra resolution further provided by gravitational lensing (100--200 pc) has been especially critical in developing this area.
I do not strictly consider IFS data in this section, but also imaging data as the techniques are in common.

A necessary starring point is identification of clumps. Often this is done by simple peak-finding codes, visual inspection or validation \citep{Swin09,Gen11,Wis12,Guo12} as the 
number of individual clumps per
galaxy is typically $\sim$ 2--5. Of course the problem of identifying compact blobs against a background is a long studied one in astronomy and there has been considerable
borrowing of well-established algorithms. In local galaxies, flux isophotes in \Ha\ are often used to identify the numerous HII regions in nearby galaxies
 \citep{KEH89}; this has also been applied at high-redshift
\citep{Jones10,Wuyts2012}. More sophisticated techniques allow for a variable background, e.g.  \cite{NFS-clumps} applied the star-finding software {\it daofind\/} \citep{Stetson87} to HST infrared images, looking for local maxima above a background threshold and validated visually, to find 28 clumps in six $z \sim 2$ galaxies. If clumps are resolved then star-finders may
not be appropriate, especially if they assume point sources with a particular PSF, because of this \cite{Liv12} adopted the {\it clumpfind} program \citep{clumpfind} (albeit in a 2D mode for their HST images) originally developed for the analysis of molecular line data in the Milky Way. This proceeds by thresholding
at a series of progressively fainter isophotes to try and deblend overlapping clumps. Resolution effects are an issue --- if we looked
at a grand design spiral with resolution of only 1 kpc, would we see the numerous HII regions in the spiral arms merge together to make only a few larger single objects? The
answer so far appears to be subtle: \cite{Swin09} did simulations of this effect using local galaxies and found such `region merging' did indeed result in few regions of greater size and
luminosity. They argued the vector of this change was nearly parallel to the existing size-luminosity relation and did not result in an {\it offset\/} relation as they found at high-redshift. A similar effect was found in \cite{Liv12}, noting that the magnitude of the effect (vector A in their Figure 6) is approximately a factor of two in size and luminosity. Both of these particular studies were being compared to lensed high-redshift sources where the resolution was $\sim 300$--600 pc and the clump radii up to a kpc.

The method of measuring clump sizes is also an issue. The traditional approach in local galaxy studies is after finding HII regions through an isophotal selection to simply define the radius as $\sqrt{{\rm Area}/\pi}$, i.e. the radius of a circularised region.  The other approach is to fit profiles to the regions and then use the half-width at half maximum (HWHM) as the size, this is known as the `core method' \citep{Kennicutt79}. Isophotal sizes are problematic in that they depend on the exact isophote chosen, and often it is not a defined physical surface brightness in \Ha\ but simply a signal:noise level (e.g. \cite{Jones11}).  In this case, just taking deeper data will result in larger sizes. Compared to a faint region a region with a higher luminosity but the same core radius will have a larger isophotal radius, this is particularly problematic
when comparing different redshifts as there will be a degeneracy between luminosity and size evolution in region properties. \cite{Wis12} and \cite{Liv12} both considered
the effect of the choice of core vs isophotal radii. \cite{Wis12} found the isophotal radiii in their local galaxy comparison samples were up to three times larger than core radii (determined by fitting 2D gaussians) and attributed this to the inclusion of diffuse emission in isophotes; in contrast, the respective luminosities were much more in 
agreement as they are dominated by the brighter inner parts. They argued that core radii were a more robust choice for high-redshift comparisons.
\cite{Liv12} found good agreement between clump sizes at high-redshift from {\it clumpfind} isophotes and core sizes (for sizes $>100$ pc).

Both types of radii are subject to resolution effects
which clearly need to be simulated and this has been done by several groups \citep{Elm09,Swin09,Liv12}. Even for nearby galaxies this may be critical, for example \cite{Pleuss00} studied resolution effects in M101 comparing HST data of this nearby galaxy  to simulated natural seeing at a distances several times greater. They found the effect of changing the resolution from 4 pc up to 80 pc (still much better than typical high-redshift data) was to merge regions due to their natural self-clustering and boost their isophotal sizes by factors of 2--4 . They even hypothesised that the `break' in the HII region \Ha\ luminosity function at $\sim 10^{39}$ ergs s$^{-1}$ \citep{Rozas96} could be entirely due to resolution effects in typical local data.  
The  largest and most luminous HII regions with sizes of
up to 300 pc were the least effected by the degradation, as might be expected, this conclusion echoes the earlier work of \cite{KEH89}. 
More systematic studies of the effect of resolution on size measurements at high-redshift is clearly needed; existing work only treats this topic briefly on the way to the high-redshift 
results of interest. There is clearly a problem: for example Figure~6 of \cite{Liv12} suggests there is a factor of 10 vertical offset between the nearby and $z\sim 1$ luminosity-size diagram of clumps (see also \cite{Swin09,Jones10}). However, \cite{Wis12} (their Figure 6) argues that there is a single relation. This large difference seems to arise from the use of isophotal  vs core  sizes.  Another potential issue is that a 
significant number of the size measurements in the
literature exploit the extra magnification due gravitational lensing  \citep{Swin09,Jones10}. This of course allows smaller physical scales to be
resolved but it should be noted that the magnification is very anamorphic and the extra
resolution is only attained in one spatial dimension.

The topic of {\it clump\/} velocity and velocity dispersion has cropped up in a few papers. Generally, dispersion is normally measured from the integrated spectrum in an aperture at the position of the clump and can be used to estimate scaling relations and derive Jeans masses  \citep{Swin09,Gen11,Wis12}. Clumps share the velocity field of the underlying 
galaxy disc; this in fact is a key test of the clump disc model. (If they were external merging galaxies one would expect a kinematic discrepancy and
this is seen in these cases, e.g. \cite{SMG-clumps}). They also seem to share
the dispersion of the disc; at least distinct features (such as a peak or trough) are not apparent in dispersion maps at clump locations (for example see Figure 3 of \cite{Wis11},
Figures 3--6 of \cite{Gen11} or Figure~\ref{fig:GenzelObject}). One novel technique to investigate clump formation physics is to calculate spatial maps of the Toomre $Q$ parameter under the expectation
that clump locations might correspond to $Q(\hbox{RA},\hbox{DEC})<1$). This requires a disc model velocity field and an inclination; I discuss the physical basis for this and potential problems 
of $Q$-maps further in Section~\ref{sec:turbulent-discs}. Clumps may also rotate internally and have significant dynamical support from this rotation and this is suggested by some simulations \citep{Bour07,Cev12}. This
has been looked for by searching visually for apparent shears in residual velocity maps (after subtracting the best fitting disc model) with perhaps a tentative detection of
small signals in some cases \citep{Gen11}; however they are small at the $\sim 15$ \kms\ kpc$^{-1}$ level. 
One can also consider the application of resolved rotation curves to derive resolved mass profiles of galaxies at high-redshift. Kinematics of course can be sensitive to unseen components, for example evolved central bulges in disc galaxies may have no \Ha\ emission but reveal themselves through their effect on rotation curves.
There has been little of this in the high-redshift literature, probably 
because to do this properly requires AO observations and the number of AO samples is small and they only contain handfuls of galaxies. \cite{Gen08} present an 
application of this to the SINS survey (five galaxies which were the best observed, two with AO) where they extract a `mass concentration parameter' defined as the ratio of total dynamical mass within the central 0.4 arcsec (3 kpc at $z=2$) to the total (limited at 1.2 arcsec); the technique to derive this was to add $M_{dyn}(0.4'') / M_{dyn}(1.2'')$ as an extra free parameter to the mass modelling of the {\it 1D} rotation curves along the major axis, holding the previously determined 2D disc fit parameters fixed. They do find an interesting correlation with emission line ratios in the sense that (their interpretation) more concentrated galaxies are more metal rich. One of their galaxies (BzK6004) shows high concentration
and a beautiful detection of a central red bulge in the K-band continuum, surrounded by a clumpy \Ha\ emitting disc. However, it is not clear in my view 
if `mass concentration' in the sense defined  on average means presence of a bulge or simply a more concentrated disc; and this would be a fruitful area to examine further with larger AO 
samples particularly looking at this type of modelling in more detail, with greater numbers and correlating with bulge presence (e.g. as revealed by HST near-infrared observations). 

Finally, using IFS data one can measure other spectral, non-kinematic properties of galaxy sub-structures. Line luminosities and ratios can be measured by
standard aperture photometry techniques. These can be used to derive physical properties such as star-formation rate
and  gas-phase metallicity in much the same way as for integrated spectra. These physical conversions can be complex and are beyond the
scope of this review's discussion,  for a thorough discussion of star-formation
indicators see \cite{Hop03} and for gas phase metallicity measurements, see \cite{KE08}.


\section{Physical kinematic pictures of star-forming high-redshift galaxies}
\label{sec:physics}

The surveys outlined in Section \ref{sec:surveys} have transformed our pictures and physical understanding of the nature of high-redshift star-forming galaxies. The development of resolved
kinematic measurements at $z>1$ to complement photometric ones has allowed deeper evolutionary connections to be made between galaxies in the early Universe and locally.
Whilst the story is by no means complete, some clear physical pictures, which one might call useful `working models' to prove further (or refute),  of the nature and structure of these galaxies have emerged which I will attempt to summarise here. I will defer outstanding observational and physical questions to the final section.

\subsection{Turbulent disc galaxies}
\label{sec:turbulent-discs}

An important early question was whether disc galaxies existed at all at high-redshift  \citep{BCF96,Weil98,Mao98}. The existence of a disc presupposes some degree of gas settling, 
the fact that most high-redshift star-forming galaxies at high redshift showed much higher star-formation rates than those locally \citep{Bell05,Juneau05} and also exhibited lumpy, somewhat irregular morphologies  \citep{KGMDS-95,Drv95,Abraham96,Abraham-HDF} led some to hypothesise that perhaps they were all mergers: after all the highest star-formation rate objects locally are merger-driven ULIRGS and early versions of the Cold Dark Matter model predicted high merger rates at high-redshift from hierarchical growth \citep{BCF96,Weil98}. Of course not every galaxy could be seen in a merger phase, but if imaging surveys were mostly sensitive to high star-formation rate galaxies this could be interpreted as a selection
effect.  Is it possible that the cosmic star-formation history is merger driven \citep{Tiss2000}?

An alternative viewpoint is that a typical massive galaxy's star-formation history could be dominated by continuous star-formation, with a higher value than today as the galaxy would be more gas rich in the past. In this scenario, we would expect the gas and young stars to have settled in to a rotating disc.  In the more
modern $\Lambda$CDM model, the different expansion history tends to produce a lower merger rate than flat $\Omega_m = 1$ CDM models
and the late time
evolution of large galaxies is less rapid \citep{Kauf99}. Further to this
new analytic arguments
and hydrodynamical simulations have suggested mechanisms where galaxies sitting in the centre of haloes can continuously accrete new gas at significant
rates of 
via `cold cosmological flows' \citep{Dek09,DSC09}. Observationally, the revelation of a tight star-formation rate -- stellar mass `main sequence' whose locus
evolves smoothly with redshift is also more in accord with a continuous accretion process dominating the star-formation; stochastic merger-driven bursts would introduce too much scatter in this main sequence \citep{main-sequence,Daddi2007,Rod2011}. The merger rate has been derived from close pair counts in high-redshift data (see Section ~\ref{sec:merger-rate}, at $z>1$ it is about 0.1--0.2 per Gyr (for typically mass-ratios $>1/4$)). If all star-forming galaxies were undergoing mergers (with a duty cycle of 1--2 Gyr) then the rate would have to be 3--4$\times$ higher. 
Direct comparisons can be made of observed galaxy growth vs those predicted by mergers, e.g.  \cite{Bundy07} who compared the rate of production
of observed `new spheroids' in each redshift bin with merger rates from simulations and 
\cite{Con12} who compared empirical merger growth from pair-counts vs the `in-situ' growth calculated from their measured star-formation rates. These analyses favour in-situ type processes for galaxy star-formation and quenching. 

Kinematic studies have been motivated by these results and as we have seen in Section~\ref{sec:surveys} a large fraction ($\sim 30\%$ or larger) of galaxies seen
at high-redshift are clearly rotating discs, i.e. while the broad-band with HST appears photometrically irregular the objects appear {\em kinematically regular}
\citep{UDFz1.6,Starken08,Puech-clumps,Jones11,NFS11,Gen11}, a key point. Generally, the rest-frame UV and \Ha\ from AO IFS trace each other 
\citep{Law09} whereas
the stellar mass is smoother  (but still clumpy) \citep{NFS11,Wuyts2012}. 
The fraction of discs seems to increase towards higher stellar masses \citep{NFS09,Law09}. The typical rotation velocities are
100--300 \kms\ so very similar to local galaxies \citep{Cresci09,AMAZE-TFR,MASSIV4}. The big surprise has been the high values of the {\it velocity dispersion\/}
found in galaxy discs. First observed by \cite{NFS06} and \cite{Genzel06} typical dispersion values (in all surveys) range from 50--100 \kms. It is helpful to frame this as $v/\sigma$, the
ratio of circular rotation velocity to dispersion. For the larger discs (stellar masses $>5\times 10^{10} M_\odot$) $v/\sigma$ typically ranges from 1--10 at $z\sim 2$ \citep{Starken08,Law09,NFS09,AMAZE-TFR,Gen11}. There are also a number of objects that appear not to be dominated by rotation with   
$v/\sigma \lesssim 1$; this class has been called `dispersion-dominated objects' \citep{Law09,Kassin12}. These values compare with a value of $\sim 10-20$ for the Milky Way
and other similar modern day spiral discs \citep{GHASP,DISKMASS}. If these measured values of the dispersions correspond to the dynamics of the
underlying mass distribution these high-redshift discs are `dynamically hot'. 

A simplified physical picture of such objects was first described by \cite{Nog98,Nog99} and is nicely summarised by \cite{Gen11}. (For a more detailed theoretical treatment see \cite{DSC09}). The arguments goes as follows. The classical \cite{Q} parameter $Q$ for stability of a gas disc  is: 
\begin{equation} \label{eq:Q}
Q_{gas}=\frac{\kappa \sigma} {\pi G \Sigma_{gas}}
\end{equation}
where $\Sigma$ is the mass density and $\kappa$ is the epicyclic frequency. 
 $\kappa = a \, v/ R$ where $a$ is a dimensionless factor
$1<a<2$ depending on the rotational structure of the disc, $v$ is the circular velocity, and $R$ is some measure of the radius (for an exponential disc
the scalelength). The  $Q$ parameter
can be understood by considering a gas parcel large enough to collapse under self-gravity despite it's velocity dispersion, i.e. larger than the `Jeans length' $L_J \simeq \sigma^2 / G \Sigma$.
However, as gas parcels rotate around with the disc in their reference frame they also experience an outward centrifugal acceleration $\simeq L_J \kappa^2$; if this
is larger than the gravitational acceleration $G \Sigma$ then the disc is stable. Local spiral discs tend to have $Q\sim 2$ \citep{Q-spirals}.

Following Genzel, if we express the total dynamical mass as $M_{dyn} = v^2 R / G $ and the total gas mass as $\pi R^2 \Sigma_{gas}$ then equation \ref{eq:Q} can be rewritten as 
\begin{equation} \label{eq:fgas}
Q_{gas}=a \left(\frac{\sigma} {v}\right) \left( \frac{M_{dyn}}{M_{gas}} \right)
\end{equation}
Since we expect rapidly star-forming discs to be unstable and have $Q\sim 1$ we arrive at the {\it important result}:
\begin{equation}  \label{eq:fgas2}
\frac{v} {\sigma} \simeq  \frac{1}{f_{gas}} \
\end{equation}
i.e. that it is a high gas fraction that gives rise to these dynamically hot discs.  For a mixture of gas and young stars in a disc, if they share the same velocity, velocity
dispersion, and spatial distribution, then equations \ref{eq:Q} and \ref{eq:fgas2} are still valid with the substitutions $Q_{gas}\rightarrow Q_{young}$,
$\Sigma_{gas}\rightarrow \Sigma_{young}$, $f_{gas}\rightarrow f_{young}$.\footnote{Strictly the factor of $\pi$ in equation \ref{eq:Q} should be replaced by 3.36 to compute $Q$ for a stellar disc but this is a negligible difference at this level of detail.} Since young stars will form from the gas on timescales less than an
orbital time, it is natural to expect them to share the same distributions, and this is observed in the Milky Way \citep{Luna2006}.

For the range of $2<v/\sigma<4$ typically observed derived gas fractions are 25-50\%, which accords
with observations of molecular gas fractions at high-redshift \citep{Tacconi10,Tacconi12,Daddi-gas,CarriliWalter13}  A corollary of course is that since the gas/young stars fraction is not  close to 100\%  (except possibly 
in the case of the  `dispersion dominated galaxies' discussed in Section \ref{sec:DD}), there must be another component; the fractions defined in equation \ref{eq:fgas} are relative to the total {\it dynamical mass} (and noting that the
observational data from molecular gas surveys are usually relative to the total gas $+$ stellar mass which includes young stars). 
For multiple components in a disc the 
the approximation $Q_{eff}^{-1}=Q_{gas}^{-1}+Q_{stars}^{-1}$ \citep{WS94} is often used
\citep{Puech08,Gen11} though there are more sophisticated combinations 
 (e.g. \cite{Raf01,Romeo13}). If I assume the dynamical mass is dominated by an older stellar disc (i.e $f_{gas}<1$) then I can  show (using the Wang \& Silk
 approximation and similar working to equation~\ref{eq:fgas}) that:
  \begin{equation} 
Q = \frac{a \, \sigma_g}{v} \left(   \frac{\sigma_g}{\sigma_s} +f_{gas} \right)^{-1}
\end{equation}
Setting $Q\sim 1$ and $a\sim 1$ I then I get:
 \begin{equation} 
\frac{v} {\sigma} \simeq  \frac{1}{\left(  \frac{\displaystyle \sigma_{gas}\ }{\displaystyle \sigma_{stars}} \right) +  f_{gas}} \
\label{eq:more_fgas}
\end{equation}
which shows that the stellar dispersion needs to be several times higher than that of the gas to maintain the observed $v/\sigma > 1$ values. Alternatively, a dark matter or stellar spheroid could serve and we would expect theoretically baryon fractions of $\sim0.6$ within the disc radius \citep{DSC09}. In my view, it seems from the argument in Equation~\ref{eq:more_fgas}  that high-redshift discs will evolve in to local intermediate mass ellipticals or S0 galaxies (i.e. Fast Rotators), not local thick discs, as the implied stellar dispersions and masses are the right scale (100--150 \kms, $\sim 10^{11} \Msun$). This would also follow
from using the clustering properties of these high-redshift star-forming galaxies to trace descendants (e.g. \cite{Adel05,Hay07}). Local Slow Rotators are more massive and could not form by fading of these discs, single major mergers may not be enough and these objects likely require
multiple hierarchical mergers to achieve their kinematic state \citep{Burkert2008}.

\begin{figure}[t] 
\centering
\iftablet
   \includegraphics[width=13cm]{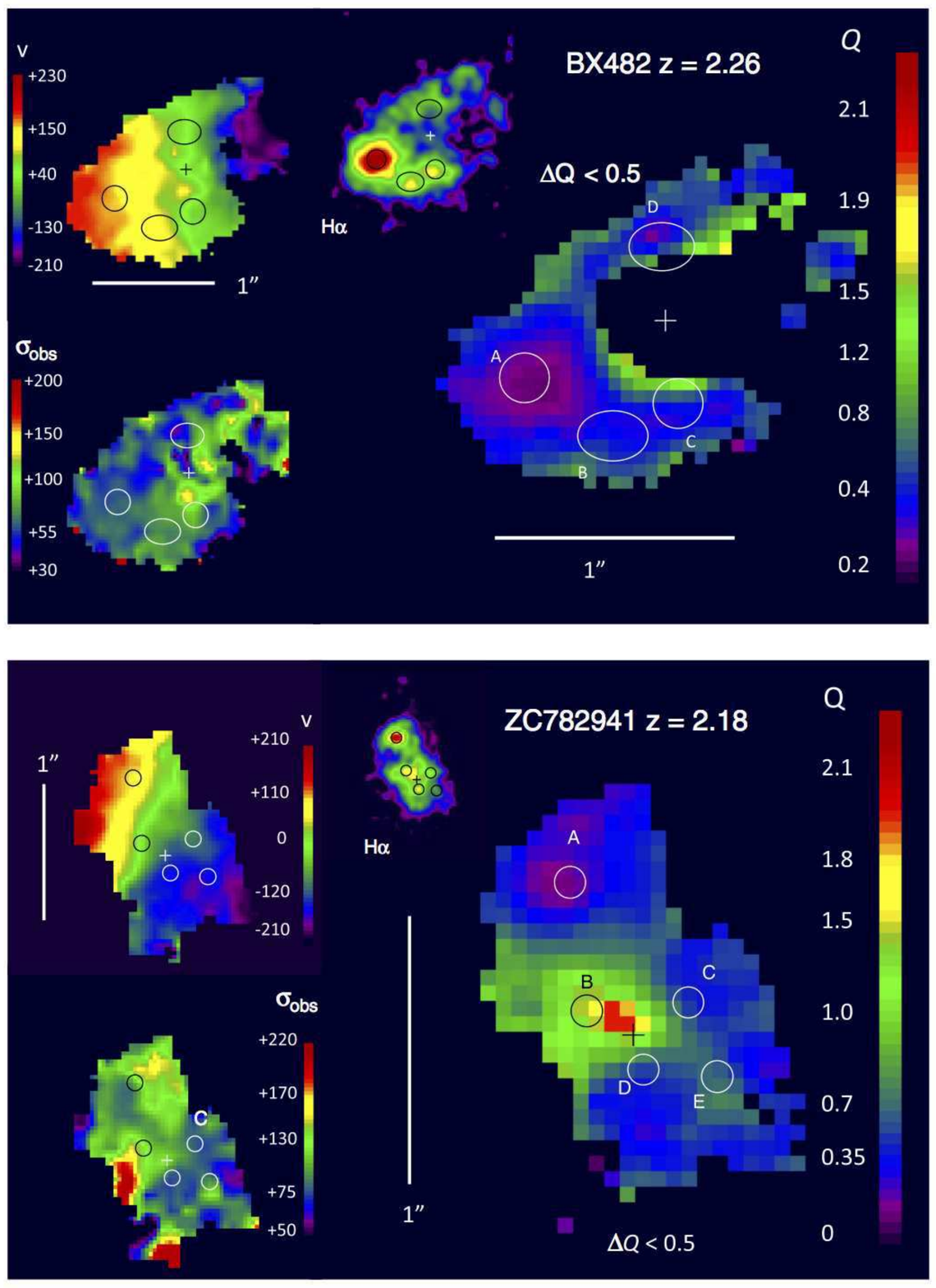}
\else
   \includegraphics[width=7.5cm]{figure-GenzelClumps.pdf}
\fi
\caption{\small Two clumpy $z \sim 2$ discs from the larger sample of \cite{Gen11} showing velocity, dispersion, \Ha\
 and {\it Q} maps. Data is AO at resolution 0.2 arcsec.  The circles denote the positions of clumps, note how these `disappear' in to the 
 velocity maps showing they are embedded in discs and occur in regions of $Q<1$. See also \cite{Wis12} for a similar
 finding (but also caveat in text here).  {\it Credit:  from Figures 4 \& 5 of \cite{Gen11}, reproduced by permission of the AAS.} } 
\label{fig:Genzel-clumps}
\end{figure}

\cite{Gen11} made maps of the $Q$ parameter in the SINS sample and found regions of $Q_{gas}<1$ corresponded to star-formation peaks providing some support for the idea that they are clumps generated by instability (see Figure~\ref{fig:Genzel-clumps}). However, it is critical to make the caveat that they calculated $\Sigma_{gas}$ as $\propto \Sigma_{SFR}^{0.73}$, i.e. using a Kennicutt-Schmidt type law, as $\Sigma_{gas}$ appears in the denominator, this naturally gives low derived values of $Q$ where the star-formation peaks. A critical test would be to repeat this using high-spatial resolution direct gas measurements.

The other important physical parameter that arises from this picture of the Jeans length and consequent Jeans mass which sets  the scale of collapsing gas clouds. 
The form of the Jeans length $L_J \simeq \sigma^2/(G\Sigma)$ is the same as that for the vertical scale height of a thin disc in 
gravitational equilibrium $h = 2\pi \sigma^2/(G\Sigma)$, thus we naturally expect the Jeans length to be similar to the disc thickness (this is also seen in simulations \cite{Bour10}). Putting in some numbers for high-redshift
discs ($\sigma = 70 \,$\kms, $M_{gas}=5\times 10^{10} M_\odot$, $R=3$ kpc), we obtain $L_J \simeq 1 $ kpc. 
The associated Jeans mass $\simeq \Sigma L_J^2$ is $\sim 10^9 M_\odot$. Once the clump collapses one would expect from general virial arguments that it
becomes and object of virial size, a factor of two less than the Jeans length and dispersion equal to the disc dispersion \citep{DSC09}. These
scales and masses match those of the giant clumps of star-formation commonly observed in high-redshift galaxies supporting this model. It is 
the large mass scale, which can be thought of as a cut-off mass of the HII region luminosity function \citep{Liv12},
and fundamentally arising from a high-gas fraction, that drives the clumpy appearance to the eye 
as a $\sim 10^{11} \Msun$ galaxy disc can only contain a handful of such
clumps.
For comparison, if we consider the Milky Way with $\sigma_{gas} = 5$ \kms\ and $M_{gas}=3 \times 10^9 M_\odot$ \citep{Combes},  we derive a Jeans length of $\simeq 100$ pc
and mass of  $\sim 10^6 M_\odot$ which correspond nicely to the scale height of the gas disc and the maximum mass of giant molecular clouds and regions. The scale height of 1 kpc  in $z\sim 2$ discs is
similar to the scale height of the thick disc of the Milky Way ($\simeq 1.4$ kpc, \cite{GR83}). Thick discs today tend to be old, red, and low surface brightness, however
they may contain as much mass again as  the bright thin
disc \citep{Com11}. An interesting suggestion is that these high-redshift discs could evolve in to modern thick discs if star-formation shuts down and gas is exhausted \citep{Genzel06}. There velocity dispersions are also in accord
with evolving in to lenticular galaxies today; or mergers could transform them in to massive ellipiticals.

\begin{figure}[t] 
\centering
\iftablet
   \includegraphics[width=13cm]{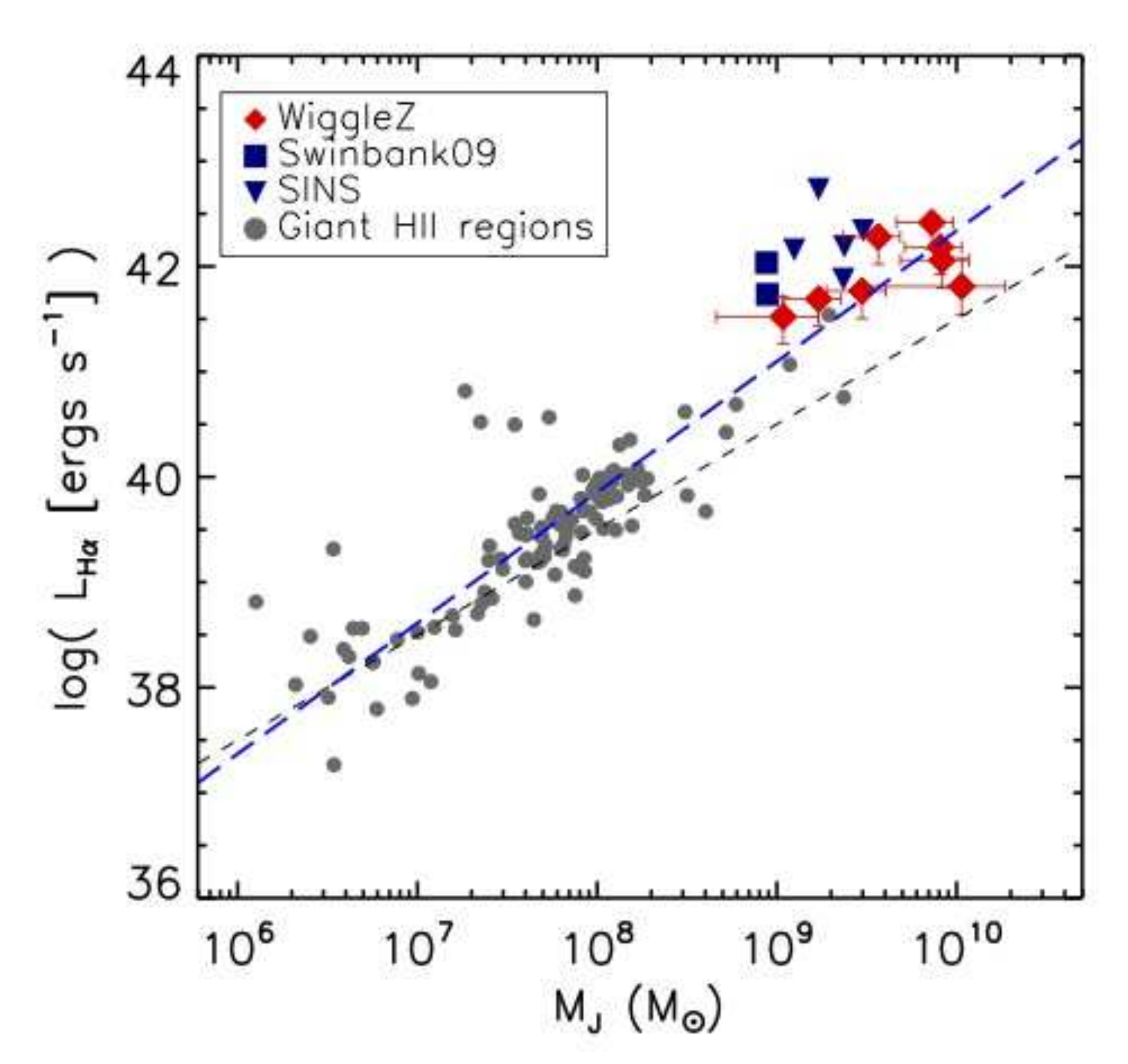}
\else
   \includegraphics[width=7.5cm]{figure-Jeans.pdf}
\fi
\caption{\small Scaling of \Ha\ luminosity (proxy for star-formation rate) with inferred clump Jeans mass $M_J=\pi^2 r \sigma^2 / 6G$
from local HII regions up to the most luminous $z>1$ clumps. The
correlation is quite tight and the slope close to unity (black dashed line). The blue dashed line is the best fit slope $M_J^{1.24}$ {\it Credit:  reproduced from 
Figure 5 of \cite{Wis12}.} } 
\label{fig:Jeans}
\end{figure}

The implication of all this is what we are observing at high-redshift are thick star-forming discs rich in molecular gas with very large star-formation complexes as I illustrated in Figure~\ref{fig:cartoon}.  Other support for this model comes from:
\begin{enumerate}
\item The axial ratio distribution of high-redshift `clump cluster' and `chain' galaxies suggest minimum disc thicknesses of  $\simeq $ 1 kpc \citep{Resh03,ELM-clump04,Elm-thick} (noting also that axial ratios may suggest  that some galaxies are  triaxial \citep{Law12}). 
\item The maximum sizes of clumps and clump scale height above the disc mid-plane match the disc thickness of $\simeq $ 1 kpc \citep{Elm-thick}.
\item The fact that star-forming clumps share the underlying rotational velocity and dispersion of the disc they are embedded in \citep{Gen11,Wis11}. 
\item That total star-formation rates seem to scale almost linearly with the inferred Jeans masses from HII regions up to giant clumps \citep{Wis12} as shown in Figure~\ref{fig:Jeans}.
\end{enumerate}

With typical star-formation rates of up to 50--100 $\Msun$ yr$^{-1}$ such high-redshift galaxies would exhaust their observed gas supply in 0.5--1 Gyr. The
preferred physical scenario has so far been that such galaxies are  continuously supplied  by cosmological `cold flows' \citep{Dek09,DSC09,Cev10}, with the term `cold'
denoting $\sim 10^4\,$K gas that has not been shocked and virialised on entering the galaxy halo and which can flow efficiently down to the centre of a young
galaxy. A $10^{11} M_\odot$ stellar mass galaxy could be smoothly assembled from star-formation in only 1--2 Gyr, a time scale comparable to the age of the Universe at $z\sim 2$. This is an attractive picture and also explains the tight star-formation rate--mass main sequence  but some health warnings are warranted. As recently discussed by \cite{Nelson2012} who compare hydrodynamical simulations using different kinds of codes (specifically the {\tt AREPO} moving mesh code and the {\tt GADGET-3} smoothed particle code), the distinction between `cold' and `hot' modes may be an over-simplification and can be dependent on definition and code type. Further, the delivery of large amounts of cold gas directly in to the centres of $z\sim 2$ galaxies may well be a numerical artefact of  {\tt GADGET-3}. 
Nevertheless, smooth accretion still dominates at $z\sim 2$ (compared to minor mergers) as the dominant mode of growth of large galaxies with 
accretion rates of up to $\sim 10 \,\Msun$ yr$^{-1}$ in large haloes.

The key physical detail needed to complete this picture is the energy source powering the observed velocity dispersion. The velocity dispersions are in the supersonic regime (i.e. $>12$ \kms) and
thus most likely arise from turbulent motions. However, turbulence will decay strongly on a disc crossing time 1 kpc $/$ 70 \kms\  which is only $\sim$  15 Myr. At $z\sim 2$, the Hubble time is 3 Gyr which is much greater than the crossing time and also of orbital timescales.
Since a large fraction of discs appear clumpy 
to maintain $Q\sim 1$ some sort of self-regulation is required. If $\sigma$ drops then $Q$ drops and the disc fragmentation increases, thus feedback associated
with this fragmentation operating on the same timescale is a good mechanism to self-regulate \citep{DSC09}.
In local galaxies, turbulence in the ISM is believed to be powered by star-formation feedback most likely by SNe feedback \citep{Dib06},
though stellar winds and radiation pressure from OB stars also contribute (see review by \cite{MacLow2004}).  At high-redshift, this
has also been suggested \citep{Lehnert2009,Tiran11}  which seems plausible given the connection between high dispersions and high star-formation rates; however, the absolute energetic coupling is difficult to calculate or simulate. Lehnert et al. found a correlation between {\it spatially resolved\/} star-formation rate {\em surface
density\/} and  velocity dispersion in the same spaxels
suggesting this mechanism; however \cite{Gen11} found a very poor correlation in much better resolved AO data. 
\cite{Green10} argued for a {\it global correlation} between {\it integrated} star-formation rates and mean dispersion. The relations between these findings is not yet clear. 
Other suggested mechanisms for generating high dispersions and thick discs are
(i) clump-clump gravitational interaction \citep{DSC09,Cev10}, (ii) accretion of cold flows \citep{EB2010,Aumer10}, (iii) disc instabilities and Jeans collapse  \citep{Im04,Bour10,Cev10,Aumer10}, and (iv) streams of minor mergers \citep{Bour09}.
No dominant energetic picture has emerged; rather a consistent theme of these papers are that  it is quite likely that there is 
more than one cause of high dispersion.
For example,  high turbulence may initially be set by the initial gas
accretion of the protogalaxy, then sustained by clump formation/interaction and/or star-formation
feedback. More observables to discriminate scenarios are desirable, for example the dependence of dispersion on star-formation rates \citep{Green10,Green13,Gen11} or galaxy inclination \citep{Aumer10}.

Another way forward in my view to further study of the energy sources powering turbulence may lie in the spatial structure that is apparent in dispersion maps, which in my view is seen consistently in all surveys (but needs AO to resolve). 
The dispersion varies often by factors of two across the disc. 
Notable examples are easy to find in the literature, for example simply  inspect the dispersion maps in Figures~\ref{fig:GenzelObject}, \ref{fig:Genzel-clumps} and \ref{fig:DD} of this review.  The dispersion shows distinct spatial correlations which do not seem to arise simply from random noise. For comparison, the models in Figure~\ref{fig:GenzelObject} shows the dispersion should be constant (apart from a central beam-smeared peak), however the two galaxies resolved by AO have a striking asymmetry of high-dispersion regions. This point is not commented on in any of the papers.
 Is it real or is it a numerical artefact of the line fitting 
 process used to create these maps?  If it is an artefact, why do some galaxies show such asymmetric dispersion (e.g. ZC782941 and D3a-15504 in Figuure~\ref{fig:GenzelObject})  despite the velocity map being very symmetrical? If it is real, I note that it is really interesting that the dispersion seems to be higher \emph{nearer} to the location
 of clumps but \emph{the dispersion peaks do not correspond to the star-formation peaks}. This is part of the reason why the $Q$-maps
 of Genzel et al. and Wisnioski et al. show minima on the clumps (the other is the increased star-formation density). It may also explain why there
 seem to be divergent findings between local and global dispersion star-formation rate correlations as mentioned above. This all
 in my view may point to non-uniform sources of energy powering dispersion associated with nearby clumps. Since the turbulent decay time
 is much less than an orbital time we would naturally expect this not to be well mixed. Demonstrating this effect is real and quantifying its spatial 
 relation to other galactic structures would make for interesting future work. 
 
Finally, let me end on a word of caution. Much of the work on clump properties has assumed that the dust extinction is constant across an individual galaxy. If extinction is patchy (e.g. \cite{Gen13}) then this could cause considerable scatter in clump properties (and even in clump identification). This is a particular problem for the rest-frame UV, \Ha\ is less affected but it is still a concern. Future resolved Balmer decrement studies combined with CO work (we expect dust to trace gas) would greatly improve our understanding of this issue.  
 
\subsection{Dispersion-dominated Galaxies}
\label{sec:DD}

\begin{figure*}[t] 
\centering
\iftablet
   \includegraphics[width=18cm]{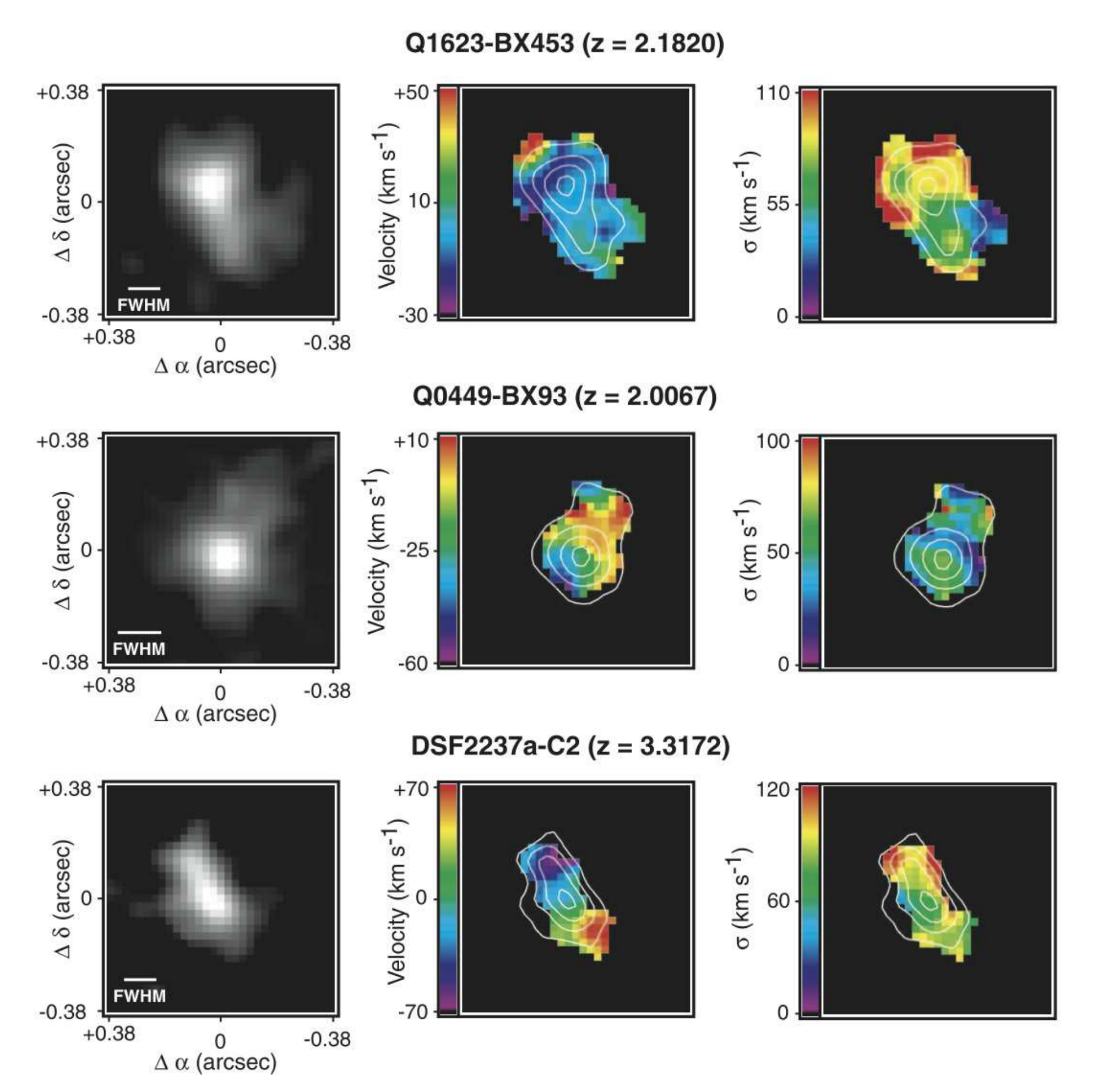}
\else
   \includegraphics[width=15cm]{figure-DD.pdf}
\fi
\caption{\small Sample UV-selected dispersion-dominated galaxies from \cite{Law07} observed with OSIRIS AO. The columns are \Ha\ intensity,
velocity and dispersion, in all cases $v\lesssim \sigma$.   {\it Credit:   from Figure~1 of \cite{Law07}, reproduced by permission of the AAS.} } 
\label{fig:DD}
\end{figure*}

\begin{figure*}[t] 
\centering
\iftablet
   \includegraphics[width=19cm]{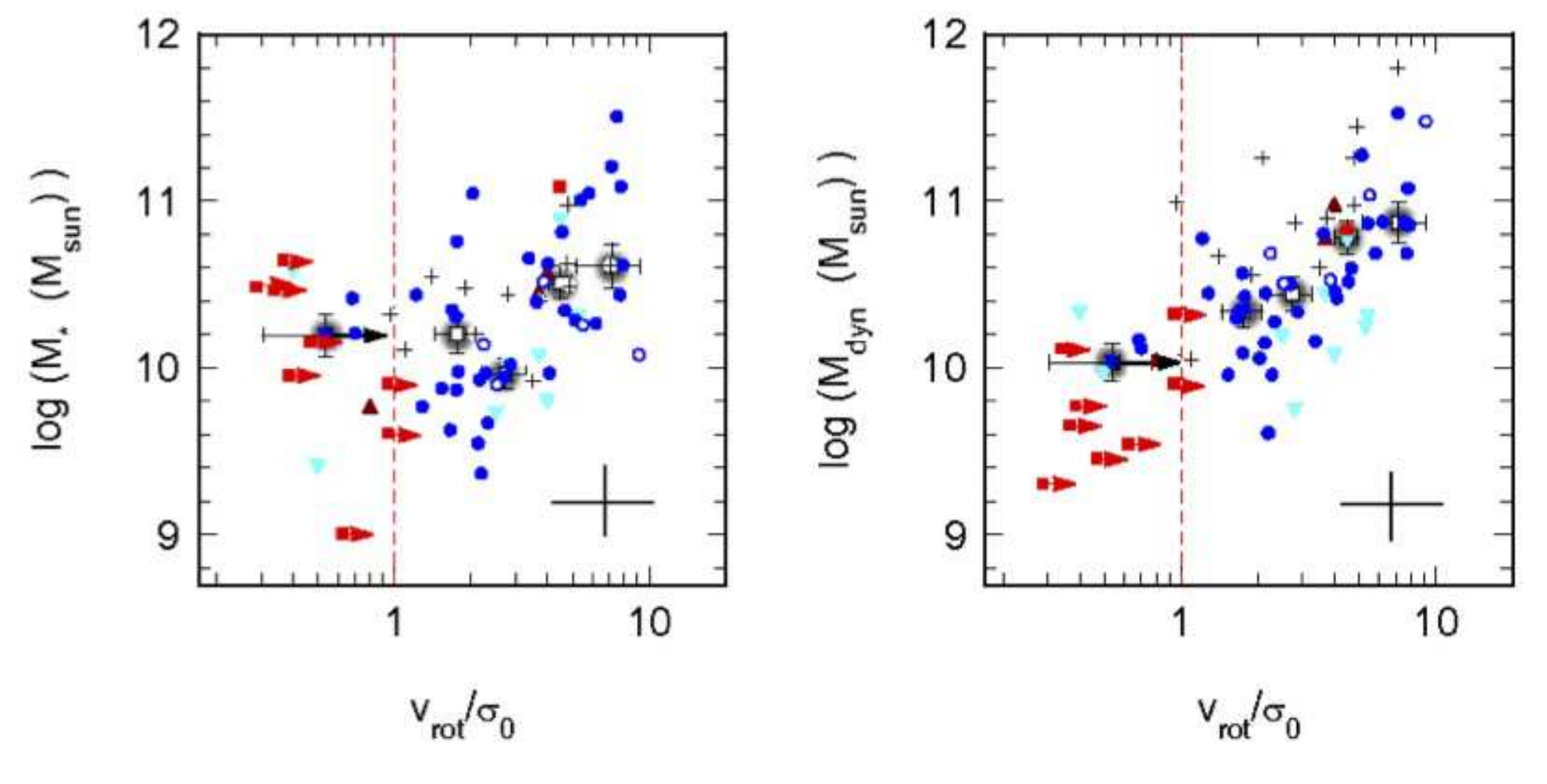}
\else
   \includegraphics[width=15cm]{figure-NewDD.pdf}
\fi
\caption{\small Dispersion-dominated galaxies ($v/\sigma<1$) tend to have smaller stellar and dynamical masses but the scatter
is large. (They also have smaller half-light radii not shown here). The samples are AO: red points \citep{Law09}, blue points  (SINS AO), cyan points \citep{SHiZELS1,SHiZELS2},  non-AO: black crosses  \citep{Lem10a,MASSIV2}.  Grey filled circles denote median values in bins.  Stellar masses
are corrected to the \cite{Chabrier-IMF} IMF. {\it Credit: from Figure 7 of \cite{NewDD}, reproduced by permission of the AAS.  } } 
\label{fig:NewDD}
\end{figure*}

Another surprise at high-redshift was the high fraction (30--100\% depending on sample definition) of galaxies which are very compact, are dominated by a single large star-forming clump, have large line widths  in integrated spectra but show very little evidence for systematic rotation. This was first noticed by \cite{Erb06} in a sample of $z>2$ Lyman Break galaxies (i.e. UV-selected) and later by \cite{Law07,Law09} in AO follow-up of a sub-sample (see Figure~\ref{fig:DD}). Objects with  $v/\sigma<1$  have been labeled 
as `Dispersion-Dominated Galaxies' though there is no evidence that this forms a distinct class, all surveys have
shown a continuous sequence of $v/\sigma$ (e.g. Figure~\ref{fig:NewDD}). This is usually measured with circular velocity and the isotropic resolved dispersion (e.g. as measured by disc fitting or data values beam-smearing corrected in some fashion) but an important caution is that 
the resulting values  may still depend substantially on spatial resolution due to beam-smearing \citep{NewDD}. 

However, the label is still useful in the sense that it is a population of galaxies 
that does not exist (at least in any abundance) in the local Universe but which seems to rise rapidly in number density with redshift \citep{Kassin12}. These
particular defining
physical characteristics are roughly a stellar mass of 1--5$\times 10^{10}\Msun$, effective half light radii $<$ 1--2 kpc, high star-formation rates and velocity dispersions of 
50--100 \kms\ \citep{Law07,Law09,MASSIV2,NewDD}. It is important to distinguish this population from that of compact red galaxies (sometimes called `red nuggets') also seen
at $z\sim 2$ (e.g. \cite{Daddi05,vdk08, Cimatti08,D09}), these have similar effective radii but have stellar masses up to a factor of ten higher ($>10^{11}\Msun$) and are quiescent. A popular observational and theoretical scenario is that they evolve in size via minor mergers on their outskirts
to become large elliptical galaxies today \citep{Bez09,Naab09,Newman10,Hopkins10}. Their ancestors at high-redshift ($2<z<3$) may be `blue nuggets' \citep{Barro12} of similar high mass; this population has yet to be probed in detail kinematically and its relation to the dispersion-dominated galaxies at lower redshifts and lower masses is an open question. 
Some of the dispersion-dominated samples do contain a few high stellar mass objects (e.g. \cite{Wis11} has two with $M>10^{11}\Msun$  with very large dispersions); though of course the  stellar mass signal may be coming from a different part of the
galaxy to that visible to the kinematics.

The dispersion-dominated galaxies  in general form a large proportion of UV-selected samples but the fraction appears to decline with higher stellar mass \citep{Law09,NewDD} as shown in Figure~\ref{fig:NewDD}; thus, they are less common in $K$-selected samples.
So what are dispersion-dominated galaxies physically? A simple interpretation might be that they are exactly as the observations suggest: high-star formation rate galaxies with negligible rotation and pressure supported. These would be star-forming analogs of modern day large elliptical galaxies (which also have $v/\sigma<1$ in their stellar kinematics \citep{Cap2007} but are a factor of ten more massive); perhaps they could have
formed from the collapse of a single gas cloud of low angular momentum? This is also suggestive of the  classical `monolithic collapse' picture of galaxy formation of \cite{ELS};
however, it is important to note that monolithic collapse-type processes still have important roles in modern hierarchical models in building initial seeds for galaxy growth \citep{Naab07}.
Maybe they could simply fade to make low mass ellipticals today?
One problem is we observe these galaxies to be highly star-forming, so we would suppose they are gas rich, and gas (unlike stars) dissipates very quickly. 
One would expect turbulent energy dissipation and settling of gas
in to a cold disc to occur on a crossing time of 1 kpc $/$ 70 \kms\  which is only $\sim$  15 Myr unless the turbulent energy is continuously refreshed. More generally,
one would expect gas to settle in to a disc on a dynamical time which is the same. Compared to the Hubble time at this redshift, this is short, so it is unlikely that we would observe galaxies during such a brief phase.

There are other more natural possibilities for such objects. Since the objects are known to be small, one might hypothesise that they are simply very small disc galaxies and that some
of the `dispersion' is in fact unresolved rotation. Another possibility is that they might be `clump cluster' disc galaxies but with only a single visible disc clump, 
perhaps due to their low mass. In
such a scenario, the stellar mass would not all come from the clump which simply dominates the UV/\Ha\  morphology (I note that many of the larger SINS galaxies in fact show one dominant clump sitting in an extended disc). A final possibility is that they
might be newly formed bulges, at the centre of clumpy discs, after clump coalescence. In some scenarios, clumps survive long enough to migrate to the centre and
merge via secular processes \citep{Nog99,NFS06,EBE08}. The stellar masses are in the right ball park for local bulges \citep{Graham-review}; however, the dynamical timescale argument for star-formation still applies
and it seems unlikely to find such a high fraction. 

 \cite{NewDD} present observations of 35 UV-selected $z\sim 2$ galaxies observed with AO as well as sources from the literature. They conclude that the `compact disc' hypothesis is the most plausible based on extrapolations of $v/\sigma$ which they find is strongly correlated with size. The stellar mass correlation is not so tight; and there are in fact
 some quite massive galaxies with $v/\sigma<1$.  The classification does depend on resolution in the sense that they find that if a source was classified as dispersion-dominated in natural seeing, it was quite likely to be reclassified as rotating by AO data. However, even at AO resolution there remains a substantial population of dispersion-dominated
 systems. The origin of the dispersion dominance is interpreted as arising partly from beam smearing in compact discs; but also as a genuine physical effect in the sense
 that their extrapolation of the velocity-size relation suggests that $v<50$ \kms\  half-light (\Ha) radii are $<$ 1.5 kpc whilst the dispersion remains `constant' ($\sim 50$--70 \kms) for galaxies of all sizes. (I note there is
some evidence for dispersion increasing in the smallest galaxies, e.g. Fig. 6 in Newman et al. and Fig. 6 in
\cite{MASSIV2}, though this may of course also be due to beam-smearing). In such a scenario, the Jeans length is comparable to the size of the galaxy disc both radially and vertically and the entire object is one large star-forming clump
as long as the observed level of turbulence can be sustained.

The scenario where there is a single clump is offset and embedded in a larger disc could be directly tested by searching for the extended galaxy. For example, HST imaging
may show diffuse disc emission or even a red bulge not coincident in centre with the clump. Morphological examples do in fact exist; this is the class of
galaxies known as `tadpole galaxies' \citep{vdB96,local-tadpoles,UDF-tadpoles}. If diffuse emission spectra can be stacked then one can test for differences in the velocity centroid as a function of surface brightness. A specific morphological comparison of galaxies labeled `dispersion dominated' with more general samples would be valuable. 
 
\subsection{Evolution of the scaling relations?}
\label{sec:TFR}

The evolution of the \TFR\ reflects the build-up of galaxy discs. In the framework of hierarchical clustering, \cite{MMW98} derived the following 
simple theoretical expression for the evolution in the disc mass $M_d$ and circular velocity $V_c$ in isothermal dark matter haloes;
\begin{equation}
M_d = \frac{m_d V_c^3}{10\, G H(z)} 
\end{equation}
where $m_d$ is the {\em fraction} of the total halo mass corresponding to the disc (typically $\sim 0.05$) and $H(z)$ is the cosmological
Hubble expansion rate. An assumption is that the disc vs halo mass fraction and angular momentum fraction are the same. The two pertinent
features of this result are (i) that if $m_d \simeq$ const., then one naturally expects a $M\propto V^3$ stellar mass \TFR\ and (ii) that at higher redshifts, galaxy discs could have a lower mass at fixed $V_c$ due to the increasing $H(z)$ factor.

So should there be an evolution in the zeropoint? One also expects the mass in the disc to be smaller at high-redshift as star-formation builds it up, so smaller $m_d$. We should also consider the effects of more realistic dark matter halo profiles \citep{NFW}  and interactions between baryons and dark matter, in particular angular momentum transfer which can give rise to disc expansion or contraction \citep{Dutton09}.
A more sophisticated treatment can arise by using semi-analytic models to estimate disc rotation  curves self-consistently \citep{Som2008,Dutton11,Tonini11,Benson12} or by running full hydrodynamical simulations with star-formation and feedback to follow disc galaxy evolution \citep{PSL,Sales2010}. Commonly it is found that the predicted evolution is {\it along\/} the stellar mass \TFR\ \citep{Dutton11,Benson12} so that the actual zero-point evolution is weak.
 
\begin{figure}[t] 
\centering
\iftablet
   \includegraphics[width=13cm]{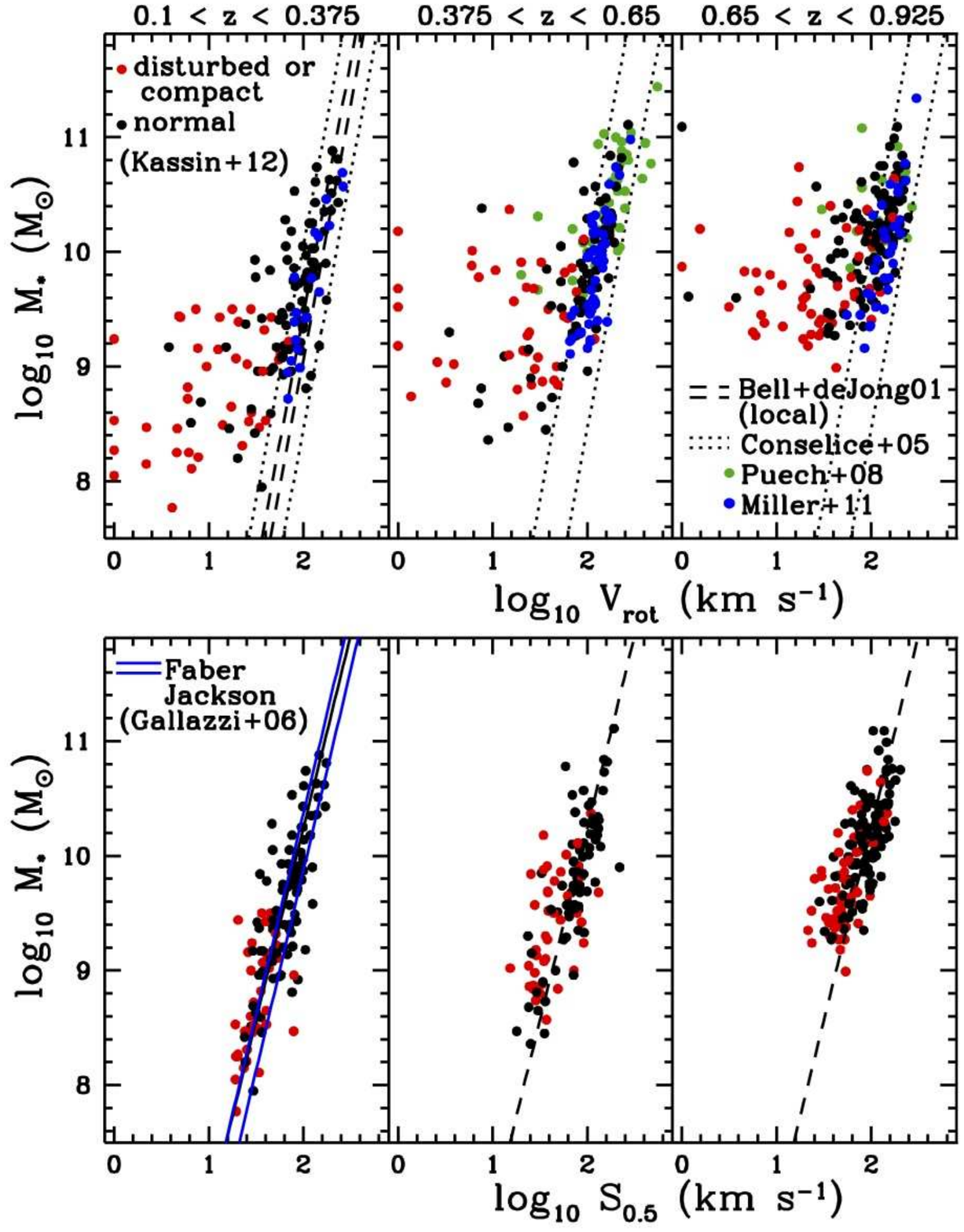}
\else
   \includegraphics[width=7.5cm]{figure-lowzTFR.pdf}
\fi
\caption{\small A comparison of \TFR\ findings at $z<1$ for the surveys mentioned in the main text. A much larger scatter in the $M$--$V$ relation 
is found in the sample of 
\cite{Kassin12} and  \cite{Puech08} than in that of \cite{Miller11} predominantly in objects with more disturbed morphologies. This scatter is considerably
reduced in Kassin et al.'s use of the $M$--$S_{0.5}$ relation and  is then brought on to the local Faber-Jackson relation  \citep{Gallazzi06}. 
The disagreements are likely due to some combination of sample selection, data quality and definition of kinematic quantities but the exact combination is not yet determined. See Sections \ref{sec:mslit} and \ref {sec:TFR} for further discussion of this. 
{\it Credit:  kindly provided by Susan Kassin (2013). } } 
\label{fig:lowzTFR}
\end{figure}

Despite the surveys outlined in Section~\ref{sec:surveys}, observationally, the situation is not clear. 
Even at low redshift, there is considerable disagreement  between surveys as illustrated in  Figure~\ref{fig:lowzTFR}.
First, there is the matter of   what fraction of star-forming galaxies at different redshifts can be usefully classified as discs and placed on a \TFR. Is it in the range
30--50\% at $ 0.5<z<3$ \citep{Yang08,MASSIV2,NFS09} or much closer to 100\% (e.g. \cite{Miller12} for $z<1.7$) if one obtains deeper data? This fraction may be increased by using the $S_{0.5}$ parameter which combines velocity and dispersion  instead of $V_c$; some authors have found that this allows all galaxies, including ones with anomalous kinematics, to be brought on to a tighter \TFR\  reducing scatter by large 
amounts \citep{Kassin07,Weiner06a,Puech10}. However, there
is debate over this true anomalous fraction and whether a tight \TFR\ for all star-forming galaxies can be produced conventionally \citep{Miller11,Miller12}.
Another issue is the choice of velocity parameter, for example $V_{2.2}$ may be more robust and produce less scatter than $V_c$ \citep{Dutton11,Miller12}. 

\begin{figure*}[t] 
\centering
\iftablet
   \includegraphics[width=19cm]{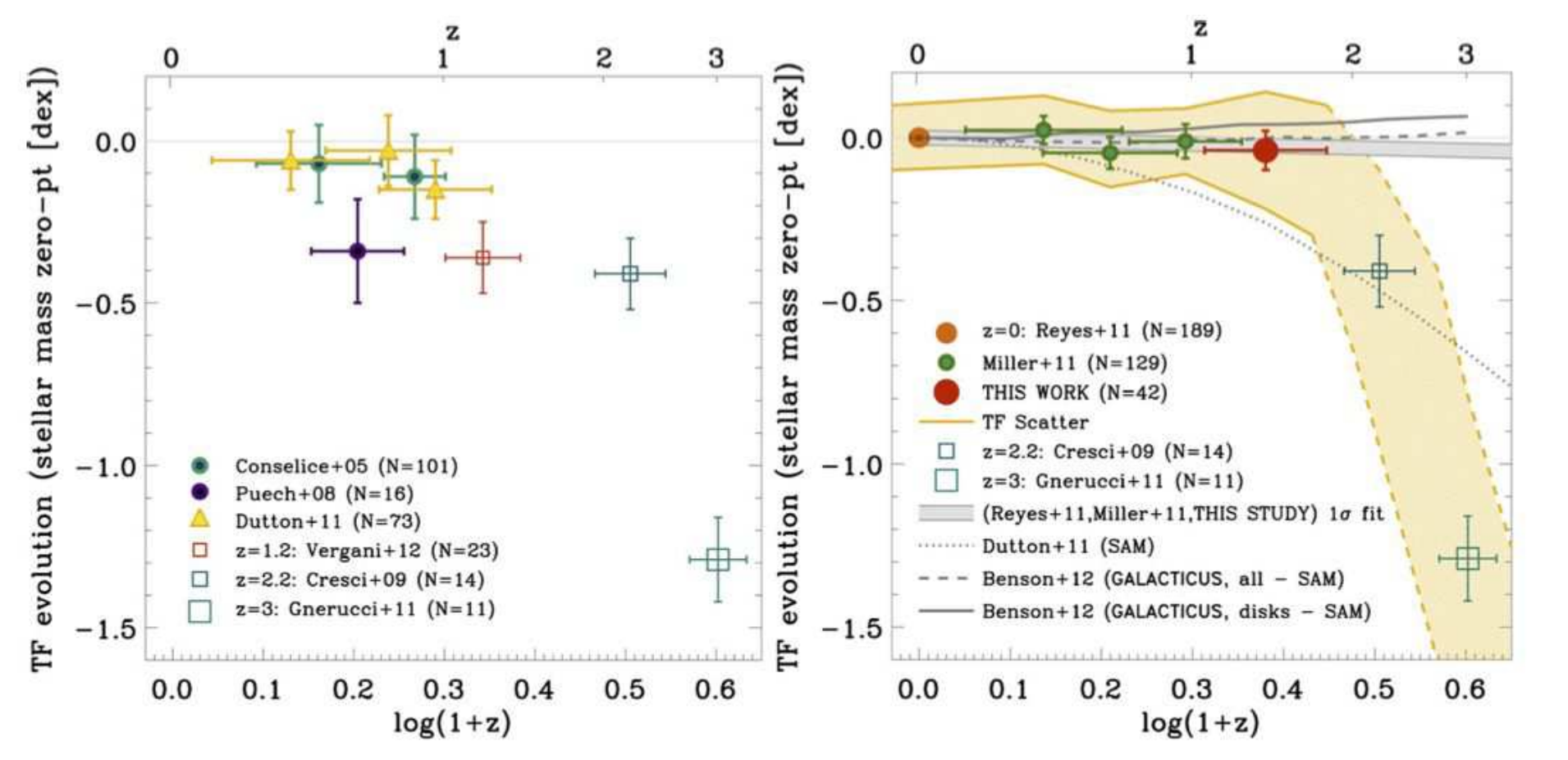}
\else
   \includegraphics[width=15cm]{figure-MillerTFR.pdf}
\fi
\caption{\small Evolution of the \TFR\ zeropoint  with redshift from Miller et al (2012).  Points show zeropoint and error bars show RMS scatter around the linear relations. The left panel shows mostly previous IFS results showing considerable disagreement. The right panel shows the results from Miller et al's very deep multi-slit work claiming no evolution in the zeropoint and very small scatter to $z\sim 2$. Some galaxy models and empirical fits are also shown (see Miller et al. for details). There is a clear inconsistency between with the (shallower) IFS results for $0.2<z<2$ where the redshift ranges overlap. Fast evolution at $z>2$ could be possible, however deeper surveys are needed to also verify the $z>2$ IFS results. The local relation is that of \cite{Reyes11} which is based on that of \cite{Pizagno07}.
{\it Credit:  from Figure 7 of \cite{Miller12}, reproduced by permission of the AAS. } } 
\label{fig:Miller-TFR}
\end{figure*}

Neither is it yet clear whether the \TFR\ zeropoint is observationally found to change with redshift. The range of findings is shown in Figure~\ref{fig:Miller-TFR} taken from \cite{Miller12}. There certainly appears to be a lack of consistency between surveys even at moderate redshifts ($\sim 0.5$). Some of this may be due to the 
methodological differences in deriving circular velocities (and perhaps stellar masses) discussed in Section~\ref{sec:disc-fitting}. It may
also be related to different choices of local relation to normalise evolution as discussed in Section~\ref{sec:IMAGES}. The local relations used have different slopes between them, this then will or will not cause an offset depending on the mass range probed. There is also no consistent local relation derived in a methodology which is consistent with that of high-redshift galaxies and that has been
tested via simulation against redshift effects.

Finally, it is interesting to note that some authors have tried to compute baryonic masses for high-redshift galaxies by adding to the stellar masses an estimate of gas mass using the observed star-formation rate surface density and the \KSR.  There is some evidence that this works in the sense that there is usually better agreement between dynamical masses and baryonic masses than with stelar masses \citep{Puech10,MASSIV4,AMAZE-TFR,Miller11}. In the 
local Universe, the `Baryonic \TFR'  is found to give a better linear relation, compared to stellar masses, down to low masses where galaxies become much more gas rich \citep{McG00,McG05}. This has been investigated
at high-redshifts; \cite{Puech10} found no evolution in the offset of the baryonic relation at $z\sim 0.6$ where the stellar relation
showed evolution and interpreted this as a conversion of gas in to stars from a fixed well over cosmic time. (Alternatively, one might suppose galaxies accrete gas and could move along the relation.) Similar lack of zero point evolution was found at $z>1$ by \cite{MASSIV4}. Given the range of results for the stellar mass zeropoint
evolution (Figure~\ref{fig:Miller-TFR}) and the uncertainty introduced by estimating gas masses from star-formation I would
argue it is premature to over-interpret the baryonic \TFR\ evolution. The local relation uses direct HI masses; it 
may be another decade before HI is available in normal galaxies at $z\gtrsim 1$, but it would be interesting in the near future to consider this with estimates of molecular gas masses from CO data.

The other kinematic scaling relation is the velocity--size one. \cite{Bou07} found that local spirals and $z \sim 2$ star-forming SINS galaxies 
overlap substantially in this plane and there is little evidence for evolution with the exception of sub-mm galaxies which were very compact. \cite{Puech07} found similar
results at $z \sim 0.6$. In both cases, the scatter was considerable and not as tight as the mass--velocity relation, a result that mirrors the local Universe. Puech et al. interpreted extra scatter as arising from their disturbed kinematic classes and it is also true that the SINS sizes were for all objects, not just well-modelled discs.  
MASSIV finds only $\simeq -0.1$ dex in size-mass and size-velocity relation (i.e. smaller discs at $z\simeq 1.2$ compared today). \cite{Dutt11} note that the lack of
evolution in size--velocity relations may be inconsistent with the sign of the observed \TFR\ evolution (including those derived within the same survey such as SINS). They also compare 
with data from DEEP2 and size--mass relations from photometric surveys and argue there is a consistent picture of early discs being smaller, as theoretically expected, and discrepancies can be attributed to (i) different methods and conversions of size measurements, (ii) selection biases in IFS surveys, and  (iii) possible differences in sizes between the young stellar populations probed by ionised gas compared to stellar mass.

\subsection{The merger rate}
\label{sec:merger-rate}

One key question that IFS surveys set out to address was the prevalence of mergers at high-redshift. Certainly one might have supposed they were common given the irregular structures of high-redshift 
galaxies (e.g. \cite{BCF96}) and an IFS is the instrument of choice for objects with unknown {\it a priori} kinematic axes. Merger rates can
be investigated by looking at galaxy image irregularities  \citep{Con03,Con08,Bluck12}
but as we have seen there are numerous examples of clumpy galaxies which are morphologically
irregular but kinematically regular.


Before the advent of IFS surveys the primary method of estimating the merger rate at high-redshift was via pair counts, starting with \cite{ZK89,Carlberg94,LeFevre2000,Lin2004}, and many papers since. By trying to estimate the fraction of galaxies that were `interacting pairs' $f_{int}$ (e.g. close on the sky, using redshift and tidal feature information if available) and then adding a merger timescale $T_{merge}$ (usually calibrated via simulations) one can estimate a merger {\it rate} $R$:
\begin{equation} \label{eq:merge}
R = \frac{f_{int}}{T_{merg}}
\end{equation}
(e.g. \cite{Bridge10}). This gives units of number of mergers per galaxy per Gyr (and can be further broken down by galaxy mass, merger mass ratio, etc.). 
Generally the merger rate is parameterised as evolving as $(1+z)^m$ where recent estimates are
$1<m<3$ \citep{Bundy2009,Kartaltepe2007,Bridge10,Lotz11,Xu2012}. The timescale $T_{merg}$ is set by dynamical friction and is around 1 Gyr \citep{Lotz08,KW08,Lotz11}; noting that
not all pairs identified in surveys
will eventually merge and this effect is often incorporated implicitly in to the calibration (the non-merging fraction may be
around 30--50\% for typical observational selections; \cite{KW08}). The star-formation rate in close pairs may even be enhanced as far as 150 kpc \citep{SFR-pairs}
 which at this distance are not likely future mergers;  but such effects do make clear the point that one should be careful of selection effects in pair
 catalogues especially in the rest-frame optical.

In determining the galaxy merger rare from kinematics, one arrives at a similar equation:
\begin{equation}
R = \frac{ f_{merg} }{T'_{merg}}  \label{eq:merge2}
\end{equation}
(e..g . \cite{Puech12,MASSIV5}) where $f_{merge}$ is the fraction of galaxies identified as mergers from
kinematics,  and $T'_{merge}$ is the timescale. Note this is a 
{\em different\/} timescale from equation \ref{eq:merge} as we are now considering  closer galaxies and a more
advanced merger stage.  However, $R$ should be the same (for an equivalent sample) as galaxies would be conserved
at all phases of the merging process \citep{CYP09}. One expects both timescales to be of order 1 Gyr \citep{Puech12}
but both can vary by factors of 2--3 \citep{Lotz08}.
 \cite{Chou2012} found that only $\sim 20\%$ of close pairs at $z<1$ were kinematically associated,
 that  the merging timescale was rather short ($<0.5$ Gyr), and that merging was dominated by blue-blue 
 pairs (i.e. opposed to red on red `dry mergers').

One approach to measuring merger rates more precisely 
is to take a close pair catalog and confirm the kinematic association spectroscopically using slit spectroscopy \citep{Chou2012}. An interesting hybrid approach was used by \cite{MASSIV5}, where they took advantage of the fact that their IFS maps were wider field than typical to count close kinematic pairs (within $\sim 20$ kpc and 500 \kms) as well as advanced ongoing mergers for star-forming galaxies with $10^{10-10.5}\Msun$.
Good agreement was found by L{\'o}pez-Sanjuan et al. between their kinematic and other's photometric surveys at $1<z<1.5$. Their `major merger fraction' (meaning 1:4 at least ratios) was $\sim 20\%$ over this redshift interval translating to a merger rate of $\sim 0.1$ Gyr$^{-1}$ (this is
equivalent for a typical $T'_{merg} \simeq 2$ Gyr which is what simulations typically indicate for pairs within 20 kpc \citep{KW08}). They found this to be
consistent with lower redshift photometric studies with an overall evolution of $(1+z)^4$ in the rate. 
Extrapolation of their power-laws would predict close to a 100\% 
merger fraction at $z=2.5$ (and a rate of $\sim 0.7$ Gyr$^{-1}$),  which  seems inconsistent with the results of the SINS \citep{NFS09} and AMAZE-LSD surveys \citep{AMAZE-TFR}
which both observe kinematic fractions of closer to $\sim 30\%$.  
One possibility is these surveys may be missing close pairs which have not yet progressed to the kinematic stage,
however  the MASSIV data by itself suggests a constant rate at $z>1$  so perhaps the power-law evolution could be flattening off at 
$z>1$. A direct comparison of merger fractions and rates from purely photometric pairs would be valuable at $2<z<3$. 
 \cite{Bluck09} looked at pairs (mass ratio 1:4) within 30 kpc of $>10^{11}\Msun$ galaxies at $1.7<z<3$, the inferred merger rate is $\sim 1$  Gyr$^{-1}$ (their Fig.3) which seems consistent with the higher MASSIV extrapolation; however, the mass range is different and would include a substantial fraction of quiescent non-star-forming galaxies than does MASSIV and SINS. Slit surveys of $z\sim 3$ LBGs seem to find a surprisingly high fraction of spectroscopic pairs (i.e. double lined) which could also imply a higher merger rate \citep{Cooke10}. Such a high merger/interaction rate at $z>2$ typically does not match modern $\Lambda$CDM model predictions,  \citep{Bertone09}, (though see Cooke et al. for a contrary view).
 
 Another possible tension is at $z\sim 0.6$. The IMAGES survey find a high-fraction of galaxies with anomalous IFS kinematics \citep{Neichel08,Yang08,Hammer09} which 
 is interpreted as a high merger fraction, 33\% involved in major mergers \citep{Puech12}.  The close pair studies mentioned above suggest the merger fraction is closer to 4\%
 at $z\sim 0.6$  (e.g. see Figure~24 of \cite{MASSIV5}) and if the timescales are similar then these should be comparable and clearly they are not.
  Puech et al. argued that
 a fraction of their objects were in what they called a `post-fusion' phase, these should be compared to close
 pairs at an earlier epoch ($z\sim 1.1$) which  lessens the tension due to the fast evolution in pairs with redshift. 
 
 However, the fact remains that
 if 33\% of star-forming galaxies are deeply kinematically disturbed at only $z\sim 0.6$ and one must ask the question how is this compatible with the fact that the majority of star-forming large galaxies today have thin, fragile discs. The traditional view is that a major merger quenches star-formation in
 a galaxy and forms a red-sequence elliptical; the morphological transformation is one-way and its star-forming life is then over. However, in CDM models it has long been supposed that such ellipticals could continue to accrete gas (either pristine or expelled during the merger) 
 and form new discs of young stars; they would then transform
 back in to a spiral albeit one with a large bulge-to-disc ratio \citep{Barnes02,SH05}. \cite{Hammer09} posit a 
 `Sprial Rebuilding Scenario' in which
 half of today's major spirals were in an active major merger phase 6 Gyr ago, and all have had a merger since $z=1$, (with the 
 Milky Way being exceptional) and the disc is then rebuilt by re-accreting the original gas. This may have more angular momentum than
 cosmological accretion as it retains that of the original disc.
 Puech et al. suggested that the high star-formation rate of  discs at $z\sim 0.6$ would permit them to regrow rapidly.
 Such a large recent merger rate is contingent on the results from the IMAGES survey being correct; it is possible they 
 have  overestimated the fraction of galaxies with anomalous kinematics --- deeper slit surveys have found a considerably smaller fraction of kinematically irregular galaxies at this epoch \citep{Miller11}. As well as being a shallower depth, the IMAGES data had a very coarse sampling of their kinematic maps
 making interpretation difficult (see Section~\ref{sec:IMAGES}). 

The main caveats in these comparisons of IFS-derived merger rates with other techniques (and indeed in general with inter-techniques comparisons) 
are the fact that (i) different surveys are probing different mass ranges, different galaxy populations with different selections even
if they are at the same redshift and (ii) the time scales for the different
stages of the merger process are a key model  uncertainty in converting observed fractions in to rates.
One final comment on this: 
comparing the form equations  \ref{eq:merge} and  \ref{eq:merge2}, I note that it is immediately obvious that if one wishes to establish the consistency of photometric and kinematic merger rates, one only needs to know the ratio of timescales $T_{merg} / T'_{merg}$ of the different phases. This
ratio may have a large range (ratio of 2--12, \citep{CYP09}) depending on the orbital parameters and is complicated by
non-merging pairs.  While the absolute values may be poorly constrained from simulations, it is interesting to speculate if the ratio might be better constrained, for example does
it vary strongly with mass ratio? The application of simulations to investigate this further would be interesting future work.

\section{Outstanding Questions and Future Directions}
\label{sec:wrap}

One thing that has become outstandingly clear in the course of this review is that we have only obtained cursory answers to the
questions that IFS surveys have set out to investigate. Certainly, we can see some definite discs and mergers at high-redshift and make plausible
physical models; however, the detailed abundance of these kinematics classes remains uncertain. The IFS surveys have been pioneering, however like
all pioneers they have set off in different directions, explored limited areas and different terrain. When they compare notes they find they have
all done things somewhat differently, they all agree on some of the major landmarks but the systematic detail is subject to a lot of difficult inter-comparisons. 

\subsection{Outstanding questions}
\label{sec:outstanding}

The outstanding questions remaining follow the themes of the section of this review. I will highlight some of the most important in my mind:

\mytopic{discs at high-redshift?} As I have discussed clearly many of the objects observed at high-redshift in IFS surveys 
are rotating with velocity fields that rise and turnover to a flat portion in a manner similar to local discs. 
It is not at all clear what fraction are discs at what redshift, the range is 30--100\% of star-forming galaxies and different surveys probe to different depths, sample different redshifts, and select different mass ranges. It is clear though that pure consideration of imaging surveys is not enough to establish 
the epoch at which discs arise in the Universe, as is often still done (e.g. \cite{Mort13}).
The scatter in the \TFR\ may be increasing at high-redshift, or it may not. Deeper IFS surveys are needed to probe the turnover in galaxy rotation curves at $z>1$. We also need greater overlap between broad-band HST surveys and kinematic AO surveys (only a few papers each with only a few objects).  Some objects may be too small to resolve as discs in natural seeing, and some may even be too small ($<1$ kpc) to resolve with AO data. The discs are certainly morphologically 
different to local ones: at a minimum they have much higher star-formation rates, have a high-velocity dispersion, and are physically thick. I note 
\cite{Law12} found they may be even more different: axial ratios provide some evidence that some may be triaxial ellipsoidal systems. Similar results have been found by  \cite{Chev12} for compact red galaxies. 
Nor is it clear whether the discs we see at high redshift are evolving in to the thick discs of today's spirals, 
or S0 galaxies, or massive ellipticals (via major mergers). These questions could perhaps be tested in the future by considering space density \citep{vDK2010} and clustering \citep{Adel05} of such objects as a means of tracing from high to low-redshift. Parent populations have been studied \citep{Hay07,Lin12} but current IFS 
sub-samples are not well-characterised in a mass-complete sense.

\mytopic{What are `Dispersion dominated galaxies'?} The existence of star-forming non-rotating galaxies is hard to explain. They start to appear at masses
$>10^{10}\Msun$ for $z>1$. High-star formation implies large amounts
of gas which naturally settles in to a disc on dynamical time. Unresolved (even at AO resolution) very 
compact discs would be one possibility as argued by \cite{NewDD}.  \cite{Law12} found that 
the morphology is not truly disc-like and suggested that may in fact indeed be transient structures, not in equilibrium, perhaps merger driven. 
If transient events are common and enhance star-formation, they may naturally populate UV-selected samples. More detailed comparisons of the
morphological axial ratios vs stellar mass and specific star-formation rate would be interesting especially given the wealth of new near-IR structural data
coming from HST in the CANDELS survey \citep{CANDELS,CANDELS2}.
The final answer to the question of the structure of these compact galaxies 
may depend on 30m class telescope AO resolution, though it is possible that the sub-resolution kinematics could be
tested with spectroastrometry.

\mytopic{The nature and driver of dispersion?} The large resolved velocity dispersion of high-redshift galaxies was an unexpected observation. It
is intrinsic and is not a beam-smearing effect. What is measured is the ionised gas dispersion as revealed by emission lines. One naturally then asks
is it coming from HII regions bound to a disc, in which case it reflects the gas disc dispersion, or from outflowing ionised gas? The consensus seems
to be the former, and in fact outflows are seen to be separately observed in even broader line wings of width 300--1000 \kms\ \citep{Gen11,Wis12}. 
A key test of this was measuring the dispersion of the cold molecular gas  \citep{Tacconi10,Tacconi12,Swin11-CO,Hodge12}, this will be improved to sub-galactic scales by the resolution and 
sensitivity of new telescopes such as ALMA. The 
resolved stellar kinematics of the young disc also ought to match the gas, however this has to be measured
from absorption lines which will likely require 30m class telescopes. If we interpret the dispersion as a turbulent
gas disc then the energy source powering it is not known. The picture of a $Q=1$ marginally stable discs requires but does not specify this.
Cosmic accretion, star-formation feedback, clump formation and stirring are all interesting candidates and progress will require a difficult quantitative estimate of these across large samples. Structure seems apparent in numerous dispersion maps of galaxies (see discussion in Section~\ref{sec:turbulent-discs}) but is
never commented on. Is it real and if so what does is correlate with? This may provide additional physical insight.

\mytopic{The physics of clumps?} We have seen a picture of the large clumps seen in high-redshift galaxies as Jeans mass objects embedded in galaxy discs. 
This explains the important observation of irregular morphology but smooth velocity fields without significant perturbations at the clump locations.
Their masses, luminosities, sizes and velocity dispersion seem to scale with star-formation rates, however it is not clear  if there are single scaling
relations connecting them with galactic HII regions today or two sequences. Part of this is that size measurements are particularly difficult to define even locally, HII regions clump together in complexes in spiral arms and the apparent size depends on the spatial resolution and image depth \citep{Rozas96}. More uniform and systematic approaches to comparing local and high-redshift galaxies (including quantitative artificial redshifting) are needed. 

We do know that the clumps we see in high-redshift galaxies are not just a resolution effect, i.e. the morphology is fundamentally
different from artificially redshifted local galaxies \citep{ELM-redshifting},  
however we do not know if these clumps will break up in to sub-clumps when viewed at even
higher angular resolution. If they do, then which clump scale is important for scaling relations? Are the clumps (or sub-clumps) single bound structures
with the velocity dispersion providing virial support? Do they also have rotational support \citep{Cev10}? How does metallicity, ionisation, stellar population age, and dust extinction
compare for young clumps vs the surrounding disc and radius within a galaxy? These could all be uncovered by
future IFS observations. One would presume that clumps also contain large amounts of cold molecular gas fuelling the star-formation. Do they contain super-Giant Molecular Clouds and what are their structure? One particularly clear and notable example of molecular clumps has been seen at $z=4.05$ in a sub-mm galaxy (albeit one of the most luminous)  by \cite{Hodge12}. 
Future  observations from ALMA and other facilities will produce
many more such observations of the general galaxy population extending to lower redshift. They are likely to test the
picture that clumps form in regions of $Q<1$ by measuring $Q$ properly using direct gas surface density measurements. Finally, one must ask
what is the fate of giant clumps, do they last a long time and gradually spiral in to the centre of a galaxy and form a bulge  or are they quickly
destroyed by intense feedback? 
Simulations support both short \citep{Genel2012} and long lifetime \citep{Cev12} scenarios depending on assumptions, 
observations  \citep{Gen11,Wuyts2012,Guo12} have yet
to settle the question.

\mytopic{The Star Formation Law?} We have seen that stars form in high-redshift galaxies in very different conditions than they do locally. The discs are more gas rich implying much greater pressure, the star-formation feedback is more intense especially within clumps and the dispersion of the disc
is much greater. Given the star-formation history of the Universe \citep{SFH}  most stars formed under these conditions. One must therefore ask basic questions,
for example is the star-formation law the same? Does a Kennicutt-Schmidt-like law apply or something different? 
The classical \KSR\ simply relates projected surface densities of gas and star-formation via a power law. 
The thickness of high-redshift 
discs would imply quite different results from  laws that depend on projected surface densities vs volumetric densities \citep{Krum12}. 
There are many other  proposed variations on this theme. For example, there may be `thresholds'  to star-formation (e.g. above
some critical density, \cite{Lada2010, Heiderman2010}). At high-redshift, it has been suggested that there are  in fact two relations --- a `sequence of starbursts' and a `sequence of discs' but which may be unified by introducing a dynamical time in to the formulation \citep{Daddi10}. Alternatively, it may simply reflect issues with CO
conversion factors \citep{SFL-review}.
Direct {\it resolved\/} tests of star-formation laws in high-redshift galaxies (see \cite{Freund13} for a first step towards this) are critical and will improve with the advent of high-resolution ALMA data in the next few years. Another related topic is the Initial Mass
Function (IMF) for star-formation. The possibility of IMF variations is important (see review of \cite{IMF-review}) and evidence for  variations in galaxies has attracted considerable recent interest and some tantalising results
(e.g. \cite{HG08,Meurer-IMF,vdK-IMF,CAP-IMF}). It seems plausible that the IMF could be different in high-redshift
discs and/or in clumps (e.g. \cite{clump-IMF}) and perhaps could be investigated by comparing colours and spectra as are done at low-redshift.

\mytopic{The Merger Rate?} As we have seen there seems to be some tension between some IFS results and those of other techniques. In particular,
at $z\sim 0.5$ deeper and higher resolution IFS observations are needed to determine if nearly half of all star-forming galaxies  have major
kinematic disturbances or whether this is just an artefact of low angular resolution or not being sensitive to galaxy outskirts. There is a clear tension with the estimates
of the merger rate by close pairs, and at these redshifts this technique is quite sensitive. At high-redshift ($z>2$) some estimates for Lyman Break
Galaxies put the merger fraction close to 100\%, this may be compatible with the existence of massive discs and the lower rates found in
IFS surveys  as the UV-selected objects are at the lower mass end. Can we define a consistent merger rate across the various merger phases
from close approach through to coalescence as a function of stellar mass, merger ratio, and redshift? This would be a powerful constraint on galaxy formation models. Deeper and more numerous IFS observations will help, but in my opinion it is equally important to 
find new techniques to extract time scales and mass ratios from IFS maps (which would obviously have to be calibrated
on simulations). 

In a sense, every galaxy at high-redshift is being subject to a continual accretion of matter of some degree of lumpiness, it is a question of
degree and how often. Every disc at high-redshift has probably had some sort of kinematic disturbance in it's recent past, likewise every major merger
remnant is probably busily regrowing a disc from new infall. Being able to quantify these effects continuously would be more helpful in comparison with models than the current somewhat artificial distinction between `disc' and `merger' which is predicated on the modern Universe where mergers are infrequent.  

\subsection{New surveys}
\label{sec:newsurveys}

Clearly, one next and  critical step at high-redshift is large scale IFS surveys of thousands of objects with uniform, homogenous selection functions.
Current surveys suffer from diversity --- selection is done using UV flux, near-IR flux, sub-mm flux, emission line flux, or some difficult to evaluate combination of all these. Even then selection from the parent sample is not necessarily homogenous. Of course the limiting factor in survey
size to date has been the necessity to observe one object at a time with IFS (with the notable exception of the IMAGES survey using the optical
FLAMES-GIRAFFE instrument). The instrument that is most likely to transform this is KMOS, recently 
commissioned at the end of 2012, on the VLT \citep{KMOS} which will work in the near-infrared and offer a 24-IFU multiplex (in natural seeing).  Multiplexed observations facilitate an improvement in numbers of course, but they also permit an improvement in depth as well as the telescope time is less expensive per object.
A number of groups are proposing IFS surveys of this scale with KMOS. It is remains desirable
to select galaxies for IFS from spectroscopic redshift surveys with prior information of the strength of the emission lines and their proximity to night sky lines. New redshift surveys using slitless spectroscopy in space will allow this \citep{3D-HST}; as will new near-IR redshift surveys using new
multi-slit instruments such as MOSFIRE on Keck \citep{MOSFIRE}. They will also provide well-defined {\it environments\/} for the IFS kinematic observations at high-redshift; a topic
that  so far has been completely unaddressed. If turbulent discs are fuelled by ongoing cosmic accretion, one might speculate on seeing strong environmental trends in their incidence and star-formation rates.

With this prospect, I also think it is critical to see a move to a uniformity of application of kinematic techniques; a good example is disc fitting where every group has developed their own bespoke code. Large surveys
need to develop a best practice with common codes and whole papers need to be devoted to describing and evaluating codes with full
treatments of errors, fit qualities, and degeneracies. This is the same transformation as
the photometric redshift community has gone through in the last decade as deep high-redshift imaging surveys have become industrialised. 

Another analogy is with the first deep imaging and spectroscopic surveys done with CCDs in the 1980's and 1990's, it was immediately 
apparent that the local comparison surveys done with photography were inadequate and this spawned the Sloan Digital Sky Survey \citep{SDSS}.
We seem to be in a similar position today with IFS surveys, they have been fruitful at handling objects with the complex morphological
structures common at high-redshift, however the majority of the local comparison to date is with traditional work done with long-slit spectroscopy.
This is changing rapidly as hundreds of local galaxies have been observed with wide field IFS instruments, notably the CALIFA \citep{CALIFA}, ATLAS3D \citep{ATLAS3D} and
DISKMASS surveys \citep{DISKMASS}. 
Low-redshift surveys are beginning with multi-IFU instruments; the MANGA survey in the U.S. and the SAMI survey
in Australia \citep{SAMI} will both observe several thousand galaxies in the next few years with well-defined selection 
and environments. They will provide
a `kinematic SDSS' and allow the statistical comparison of rotation, velocity dispersion, and angular momentum vs galaxy properties
across a range of environments from the field to rich clusters allowing fundamental tests of galaxy formation models. I predict we will always move from simple scaling relations
such as some measure of mass vs rotation towards {\it distribution functions}, for example the space density vs mass and angular momentum compared
to theoretical models.
These local surveys will also be extremely important for comparison with high-redshift; in particular the application of uniform 
techniques and the provision of large samples for artificial redshifting tests. This well-defined approach is necessary to settle the
question of the evolution in the \TFR\ ---  for example it is critical to test for selection biases to uncover the small amounts of 
evolution if any. Particularly important is these will provide high-quality baseline samples of galaxy mergers where kinematic features as well as low surface brightness photometric features (such as tidal tails) are available, confirming the merger nature but also providing approximate mass ratio
estimates by comparison with simulations. We will also likely see an increasing number of other rare objects
discovered that are similar to high-redshift galaxies (see Section~\ref{sec:analogues}) and whose close proximity will facilitate detailed
astrophysical observation, in particular multi-wavelength observations to measure gas content and it's role in shaping galaxy kinematics.

Future AO surveys will also be critical. It has been surprising how much progress has been made using natural seeing surveys given
how under-sampled the galaxies are are.   AO surveys
can deliver the kpc resolution required to resolve detailed internal structure and to make fundamental kinematic classifications of
compact galaxies. Detailed study of individual galaxies will remain an important complement to the large surveys of thousands of galaxies with
lower resolution. The main difficulty is that AO surveys remain small and 
 it is difficult to see how substantial progress will be made in increasing sample size in the near-future
given AO systems generally correct a small field-of view, hence no multiplexing of targets.
Another difficulty  of the current situation is
the lack of significant samples which have had AO and non-AO observations {\it of the same galaxies} for comparison. Even
groups who have done AO and non-AO observations {\it have not done so for the same objects} (a notable exception being \cite{NewDD}
however only limited comparisons have so far been made). Part of the reason for the limited size of AO overlap samples is the requirement for bright
guide stars --- even with laser guide star AO it is currently necessary to have a $R \lesssim 17$ mag tip-tilt correction star
and this has severely limited sample
selection to only 10--20\% of possible targets. The other issue is of course sensitivity --- at higher spatial resolution one has less photons per spaxel but 
also light is lost in the AO optical system and through the imperfect correction (i.e Strehl ratios well less than unity). Thus, more compact sources or those
with highly clumped high surface-brightness emission tend to be favoured and AO surveys have only had moderate completeness rates except when
very long integration times have been attempted.
Yet another restriction is the redshift coverage --- strong emission lines need to be used and AO works best at the redder near-IR wavelengths. We are currently subject
to an `AO redshift desert' at $0.3<z<1.2$ where we can not attain kpc resolution. 
The reddest strong emission line is \Ha\ which only achieves good Strehl in the $H$-band for $z>1.2$. The next reddest
strong star-formation line is Pa$\,\alpha$ but that redshifts in to the thermal infrared for $z>0.3$.

 Many of these issues are gradually being improved. Next generation AO systems will deliver higher throughput and higher Strehl at shorter wavelengths enabling AO observations of $z<1$ galaxies. Signal:noise is also improved by new near-IR detectors with lower readout noise (which is an issue due to the high spectral resolution of kinematic observations). Guide star availability is being improved through
more efficient wavefront sensors, near-IR wavefront sensors, which helps because so many faint stars are red M-stars \citep{NGAO}, the development
of compromise `no tip tilt' laser AO modes \citep{noTT}  and the development of
`Adaptive Optics Deep Fields' with low galactic extinction and high stellar density \citep{AODF}. Multiple object integral field
AO observations (denoted `MOAO') may also become possible due to the development of compact deployable wavefront sensors \citep{MOAO} allowing greater
number of objects and longer integration times. AO work will extend down in to the optical as technology improves, for example the next generation MUSE instrument  
on VLT \citep{MUSE} will offer a diffraction limited visible imaging mode with a 7.5 arcsec IFU (as well as a contiguous 1 arcmin wide field mode). 
Finally, the advent of 20-40m class telescopes in the 2020's will increase both AO resolution (from the diffraction limit) and light gathering power.
Ultimately in my view large and complete AO IFS surveys will have greater  impact on our physical understanding, but will take longer to arrive, than large seeing-limited surveys.

Even AO surveys can be under-sampled when some galaxy sizes approach a kpc at high-redshift \citep{vdk08, D09, Wuyts2011}. The future prospects are also very bright for taking advantage of the extra spatial resolution boost from gravitational strong-lensing which coupled with AO has allowed us to probe sub-kpc  scales \citep{Stark08,Jones11,Liv12}. New sky surveys such as the {\it Dark Energy Survey} \citep{DES}, the Hyper Suprime-Cam survey \citep{HSC}, and the {\it Large Synoptic Sky Telescope} \citep{LSST} survey will produce thousands to tens of thousands of new strong-lens candidates allowing a greater diversity of objects to be studied and statistics to be assembled. As such targets only have a sky density of order one per deg$^2$ they do not suffer a relative disadvantage from the single-object nature of AO and there ought to be ample with suitable tip-tilt stars. 

I predict the most important developments in the immediate future (the next five years) will not be at optical wavelengths. The 
Atacama Large Millimeter / sub-millimeter Array (ALMA) \citep{ALMA} is being commissioned in Chile and is being officially inaugurated
this year and is likely to dominate the near-future of high-redshift galaxy kinematics. Why do I  make this statement? Today, high-redshift is dominated by optical and near-IR observations which are mainly sensitive to stars and hot ionised gas (e.g. from star-formation or AGN). However, it is important
to consider `the fuel as well as the fire'. We have seen from existing sub-mm observations that high-redshift galaxies are rich in molecular gas \citep{Daddi10,Tacconi10}.
Current sub-mm telescopes barely resolve high-redshift galaxies with their beams of 0.5--1 arcsec and require many hours of integration per galaxy. However, integration time performance of radio telescopes scales much faster with increased area ($\propto A^2$) than do background-limited optical telescopes ($\propto A$).
ALMA will have three times more collecting area  and baselines up to 16 km and hence will improve resolution and integration times by factors of ten. 
In the northern hemisphere, upgrades to the  Plateau de Bure Interferometer (the `NOEMA' project) 
will double the number of dishes (increasing the collecting area to 40\% of full ALMA) and maximum baselines (allowing
sub-arcsec resolution) by 2018. Upgrades to lower frequency radio interferometers may enable such studies
to be extended to even higher redshifts. 
These new facilities will enable kpc-resolution morphology and kinematics of molecular gas and dust in normal star-forming galaxies to be routinely made.  The `turbulent clumpy disc model' predicts galaxies to be gas rich and thick. Will we see thick cold molecular gas discs co-rotating and aligned with the young stars seen by the near-IR IFS observations? Will we see {\it super-giant molecular clouds} associated with the bright giant star-forming regions see in the UV? I predict we will! 
It should be noted though that the observations are likely to be even  more time-consuming than optical/near-IR. Even with the full ALMA of 50 dishes I calculate that 0.3 arcsec/50 \kms\ resolution CO(3-2) observations of a $z=2.0$ galaxy with
$10^{11}\Msun$ of molecular hydrogen would take 20 h.\footnote{\ I use equation 1 of \cite{Tacconi12}, which
represented normal $z\sim 2$ star-forming galaxies, to relate H$_2$ masses to total CO fluxes and the ALMA Sensitivity Calculator at
\href{http://almascience.eso.org/proposing/sensitivity-calculator}{http://almascience.eso.org/proposing/sensitivity-calculator}} On the other hand the $\sim$ one
arcmin field-of-view and wide bandwidth of ALMA could allow multiple targets at similar redshifts to be observed simultaneously, somewhat offsetting this.

Another  extremely important question for these high-resolution sub-mm  facilities is the nature of the star-formation law relating gas density to star-formation rate, a critical theoretical ingredient of numerical galaxy formation simulations (this is often referred to as the `sub-grid physics').  Around 80\% of the stars in the Universe formed at $z>1$ but we have seen throughout this review that  galaxies in the the high-redshift Universe are physically very different from today's galaxies. Will the star-formation law be the same as in today's galaxies or quite different? Future facilities will bring a highly superior set of data to bear on this important problem and I will predict some surprises! Finally one interesting new prediction that could perhaps be tested by ALMA is the possible existence of {\em dark\/} turbulent discs \citep{EB2010}. The prediction is that turbulence in gas discs starts initially in a dark accretion-driven phase lasting for $\sim 180$ Myr before star-formation turns on and renders the galaxy optically visible. The gas would be cold and molecular --- the actual visibility of such objects to ALMA has not yet been calculated, but would make for an interesting paper.

\subsection{Final words}

I am fortunate in the timing of this review as I sense that in 2013 we are now at the end of the first major phase of high-redshift IFS kinematic studies
which started around 2005. My impression of the topic  is that in the next few years, we are going to see a phase change in the field and an avalanche of new data from
large  surveys with instruments such as KMOS and the first sub-mm wavelength kinematic studies at high angular resolution.
Large surveys with consistent selection will allow us to firmly address the statistical questions about the incidence of kinematic structures that have been identified at high-redshift and longer wavelength observations will allow us to view the cold molecular gas, both before and after forming stars,  directly. The combination of improved AO instruments and sub-mm telescopes will allow us to test the detailed physics of internal star-formation and probe galactic structure at high-redshift.  I will look forward to seeing some of the outstanding physical questions raised by the first generation of surveys answered.

\section*{Acknowledgments}

I would like to thank Bridget Glazebrook for her love and support in the long weekends required to write this review and the various little Glazebrooks for not creating too much disruption. I would also like to thank
Bryan Gaensler for inviting me to write the very first Dawes review and continual reminders about my progress! Special thanks for providing
useful useful references along the way go to Jeff Cooke, Andy Green, Luc Simard and Emily Wisnioski. I would like to thank Roberto Abraham for a careful and critical review of the draft manuscript. Thanks also go to Reinhard Genzel, Richard Ellis,
Sarah Miller, Susan Kassin, Joss Hawthorn, Kar\'in Men\'endez-Delmestre,  Enrica Bellocchi, Tucker Jones and Francois Hammer for providing useful comments and suggestions on the submitted manuscript and I have tried to reflect most of them to the best
of my ability. I would like to specially thank Cara Faulkner of Ivanhoe Girls' Grammar School for her valuable assistance in correcting the final arXiv version of this review.
The final responsibility and/or blame for the content and opinions of this review remain with me alone.
Finally I would like to extend great thanks to the  referee, Natascha F{\"o}rster Schreiber, for many useful comments and suggestions which have greatly added to this review.



\def\apj{The Astrophysical Journal}
\def\apjl{The Astrophysical Journal Letters}
\def\aj{The Astronomical Journal}
\def\aap{Astronomy \& Astrophysics}
\def\pasp{Publications of the Astronomical Society of the Pacific}
\def\apjs{The Astrophysical Journal Supplement}
\def\araa{Annual Reviews of Astronomy \& Astrophysics}
\def\mnras{Monthly Notices of the Royal Astronomical Society}
\def\nat{Nature}
\def\nar{New Astron. Rev.}
\def\na{New Astron.}
\def\physrep{Phys. Reports}



\end{document}